\documentclass{llncs}

\usepackage{amsmath}
\usepackage{amsfonts}

\usepackage{xspace}
\usepackage{times}

\usepackage{multicol}

\usepackage{color}
\usepackage{graphicx}
\usepackage{varioref}
\usepackage{enumerate}

\usepackage{algorithmic}
\usepackage{algorithm}
\usepackage{caption}
\usepackage{capt-of}

\usepackage{float}

\usepackage{amssymb}
\usepackage{subfigure}
\usepackage{multirow}

\usepackage{ifthen}
\newboolean{IsFinal}
\setboolean{IsFinal}{false} 

\newcommand{\maxatt}{{\mathtt{maxAttempts}}}

\newcommand{\fxcostdelay}{{\mathtt{cdelay}}}
\newcommand{\fxcostmul}{{\mathtt{cmul}}}
\newcommand{\fxcost}{{\mathtt{cost}}}
\newcommand{\maxErr}{\mathtt{maxError}}
\newcommand{\fxtype}{{\mathbf {fx\tau }}}
\floatname{algorithm}{Procedure}
\renewcommand{\algorithmicrequire}{\textbf{Input:}}
\renewcommand{\algorithmicensure}{\textbf{Output:}}
\newcommand{\fixedpoint}{fixed-point\xspace}
\newcommand{\infeasible}{{\sc{infeasible}}\xspace}

\newcommand\iwl{\mathtt{IWL}}
\newcommand\fwl{\mathtt{FWL}}
\newcommand{\wl}{{\mathtt{WL}}}
\newcommand{\signbit}{{\mathtt{Signedness}}\xspace}

\newcommand{\optV}{\ensuremath{{\mathcal{O}}_V}}
\newcommand{\optS}{\ensuremath{{\mathcal{O}}_S}}

\newcommand {\extended}[1]{}

\newcommand {\forcav}[1]{}

\usepackage{ifthen}
\newboolean{TENPAGE}
\setboolean{TENPAGE}{false}
\newcommand{\swVers}[2]{\ifthenelse{\boolean{TENPAGE}}{#1}{#2}}

\newcommand{\comment}[1]{}

\newcounter{myctr}

\newenvironment{myitemize}{\begin{list}{$\bullet$}
{\setlength{\topsep}{1mm}\setlength{\itemsep}{0.25mm}
\setlength{\parsep}{0.1mm}
\setlength{\itemindent}{0mm}\setlength{\partopsep}{0mm}
\setlength{\labelwidth}{15mm}
\setlength{\leftmargin}{4mm}}}{\end{list}}

\newcommand\ignore[1]{}

\newcommand{\x}{x}

\makeatletter
{\catcode`\/\active\catcode`\_\active\catcode`\.\active
\gdef\URLslash{\futurelet\next\@@URLslash}%
\gdef\@@URLslash{\ifx\next\URLslash\char`\/\else\slash\fi}%
\gdef\URLdot{\char`\.\penalty\exhyphenpenalty}%
\gdef\URLprepare{\catcode`\/\active\catcode`\_\active\catcode`\.\active
        \let/\URLslash\let.\URLdot\def~{\char`\~}\def_{\char`\_}}}%
\def\URL{\bgroup\URLprepare\realURL}%
\def\realURL#1{\tt #1\egroup}%

\newcommand{\optsynth}{$\mathtt{optInduce}$}
\newcommand{\optsynthm}{\mathtt{optInduce}}
\newcommand{\verifyerr}{$\mathtt{testErr}$}
\newcommand{\verifyerrm}{\mathtt{testErr}}

\newcommand{\swati}{\mathtt{swati}}

\newcommand {\noop}[1]{}

\begin{document}
\sloppy



\ifthenelse {\boolean{IsFinal}}
{\title{From Verification to Synthesis of Fixedpoint Implementation of Numerical Software Routines}}
{\title{SWATI: Synthesizing
Wordlengths Automatically Using Testing and Induction}}

%


\author{Susmit Jha\inst{1,2} \and Sanjit A. Seshia\inst{2}}

\institute{Strategic CAD Labs, Intel \and UC Berkeley}


\vspace{-1cm}
\maketitle

\begin{abstract}
In this paper, we present an automated technique
\textit{$\swati$: \underline{S}ynthesizing \underline{W}ordlengths \underline{A}utomatically Using \underline{T}esting and \underline{I}nduction},
which uses a combination of Nelder-Mead optimization based testing, and induction from examples
to automatically synthesize optimal fixedpoint implementation of numerical routines.
The design of numerical software is commonly done using floating-point
arithmetic in design-environments such as Matlab.
However, these designs are often implemented using fixed-point
arithmetic for speed and efficiency reasons especially in embedded systems. The fixed-point
implementation reduces implementation cost, provides better performance,
and reduces power consumption.
The conversion from
floating-point designs to fixed-point code is subject to two
opposing constraints:
(i) the word-width of fixed-point types must be minimized, and
(ii) the outputs of the fixed-point program must be accurate.
In this paper, we propose a new solution to this problem.
Our technique takes the floating-point program, specified accuracy
and an implementation cost model and provides the \fixedpoint
program with specified accuracy and optimal implementation cost.
We demonstrate the effectiveness of our approach on a set of
examples from the domain of automated control, robotics and digital signal
processing.
\noop{
In this paper, we propose a new solution to this problem based
on inductive synthesis. Our technique
takes the floating-point program, specified accuracy and an
implementation cost model as input, and provides the \fixedpoint
program with specified accuracy and optimal implementation cost.
The core idea
is to perform inductive synthesis
from input-output examples using optimization oracles.
These optimization oracles are iteratively invoked to generate
candidate programs that are optimal for a set of examples
as well as to find new input-output examples. Perfect optimization
oracles that can find globally-optimal solutions guarantee
that we will find the optimal fixed-point program meeting
the specified accuracy.
We demonstrate the effectiveness of our technique on programs
drawn from the domains of digital signal processing and control implementations.
}
\end{abstract}




\section{Introduction}
\label{sec:intro}

Numerical software forms a critical component of embedded systems
such as robotics, automated control and digital signal processing.
These numerical routines have two important characteristics.
First, these routines are procedures that compute some
mathematical functions designed ignoring precision issues
of fixed-point arithmetic. Design environments such as
Simulink/Stateflow and LabVIEW allow design and simulation
of numerical routines using floating-point arithmetic that
closely resembles the more intuitive real arithmetic.
Second, the implementation of these numerical routines run in
resource-constrained environments, requiring their optimization
for low resource cost and high performance.
It is common for embedded platforms to have processors without
floating-point units due to their added cost and performance penalty.
The signal processing/control engineer must thus
redesign her floating-point program to instead use {\em fixed-point arithmetic}.
Each floating-point variable and operation
in the original program is simply replaced by
 a corresponding fixed-point variable and operation,
so the basic structure of the program does not change.
The tricky part of the redesign process is to
find the {\em optimal fixed-point types},
viz., the optimal wordlengths (bit-widths) of fixed-point variables, so that
the implementation on the platform is optimal --- lowest cost and
highest performance --- {\em and} the resulting fixed-point program
is sufficiently accurate.
\noop{.
Thus, to summarize, the conversion from floating-point to
fixed-point is subject to two opposing constraints:
(i) the word-width of fixed-point types must be minimized, and
(ii) the outputs of the fixed-point program must be accurate,
lying within a specified distance of those of the floating-point version.
}
\noop{
Our technique
takes the floating-point program, specified accuracy, and an
implementation cost model as input, and generates the \fixedpoint
program with specified accuracy and optimal implementation cost.
The core idea
is to perform inductive synthesis
from input-output examples using optimization oracles.
The optimization oracles are iteratively invoked to generate
candidate programs that are optimal for a set of examples
as well as to find new input-output examples to drive the synthesis process.
}
The following novel contributions are made in this paper to address this problem:
\begin{itemize}
\item
We present a new approach for inductive synthesis of fixed-point
programs from floating-point versions. The novelty stems in part from our
use of optimization: we not only use optimization routines
to minimize fixed-point types (bit-widths of fixed-point variables),
as previous approaches have, but also show
how to use an optimization oracle to 
systematically test the program and generate input-output examples for inductive synthesis.


\item
We illustrate the practical effectiveness of our technique on programs
drawn from the domains of digital signal processing and control theory.
For the control theory examples, we not only exhibit the synthesized
fixed-point programs, but also show that these programs, when integrated
in a feedback loop with the rest of the system, perform as accurately as the original
floating-point versions.

\end{itemize}
\noop{
The paper is organized as follows.
We begin in Sec.~\ref{pldi12:sec:prelim} with some background on floating-point and fixed-point arithmetic.
Sec.~\ref{pldi12:sec:probdef} presents a formal definition of the problem and a running example.
Our approach is described in Sec.~\ref{pldi12:sec:approach}, and
experimental results provided in Sec.~\ref{pldi12:sec:exp}.
Related work is discussed in Sec.~\ref{pldi12:sec:related}
and we conclude in Sec.~\ref{pldi12:sec:conc}.
}

\noop{
Along with a floating-point implementation of numerical
computation, the user can provide additional specification
regarding the desired fixed-point implementation.
The floating-point program takes inputs from a given {\em input domain}.
This input domain can be provided as a logical condition on the
inputs.
Numerical computation is expected to have some
{\em specified threshold of numerical accuracy}
which can be provided as a correctness condition for the fixed-point
program.
Further, the fixed-point implementation has a cost associated
with it and an optimal implementation is
expected to take minimal cost.
The user can provide a cost model for the implementation.
In this paper, we present an automated approach to synthesize
optimal fixed-point implementation from floating-point code
that satisfies the correctness condition and is of minimal cost.
PAPER ORGANISATION IF SPACE PERMITS
}

\section{Preliminaries}
\label{pldi12:sec:prelim}

Floating-point arithmetic~\cite{goldberg-ACM91} is a system for
approximately representing real numbers that
supports a wide range of values. It approximates a real number using
a fixed number of significant digits
scaled using an exponent.
The floating-point system is so called because
the radix point can {\em float} anywhere relative to the
significant digits of the number.
This is in contrast to fixed-point arithmetic~\cite{yates-09}
in which there are a fixed number of digits and the radix point
is also fixed.
Due to this feature, a floating-point
representation can represent a much wider range of
values with the same number of digits.
The most common
floating-point representation used in computers is
that defined by the IEEE 754 Standard~\cite{IEEE754}.
\ifthenelse {\boolean{IsFinal}}
{}
{
The storage layout of the floating-point numbers consist of
three basic components:
the sign, the exponent, and the mantissa.
The storage layout of the single-precision and double-precision floating
point numbers is presented in Table~\ref{prog:prel:floatlayout}
\begin{itemize}
\item
The {\em sign bit} is $0$ for a positive number and $1$ for
a negative number. Flipping the value of this bit flips the sign of the number.
\item
The {\em mantissa}, also known as the significand, represents the precision
bits of the number. It is composed of an implicit leading bit and the
fraction bits.
In order to maximize the quantity of representable numbers, floating-point
numbers are typically stored with the radix point after the first non-zero
digit. In base $2$, the only possible non-zero digit is $1$.
Thus, we can just assume a leading digit of 1, and don't need to represent
it explicitly. As a result, the mantissa has effectively $24$ bits of
resolution, by way of $23$ fraction bits in single-precision floating-point
numbers, and $53$ bits of resolution, by way of $52$ fractional bits in
double-precision.
\item
The {\em exponent}
field needs to represent both positive and negative exponents.
To do this, a bias is added to the actual exponent in order to get the
stored exponent. For IEEE single-precision floats, this value is $127$.
Thus, an exponent of $0$ means that $127$ is stored in the exponent field.
A stored value of $200$ indicates an exponent of $(200-127)$, or $73$.
Exponents of $-127$ (all $0$s)
and $+128$ (all $1$s) are reserved for special numbers.
For double precision, the exponent field is $11$ bits,
 and has a bias of $1023$.
\end{itemize}
\begin{table}
\centering
\begin{tabular}{|c|c|c|c|c|}
\hline
& Sign & Exponent & Fraction & Bias \\
\hline
Single Precision & $1 \;[31]$ & $8 \;[30-23]$ & $23 \;[22-00]$ & $127$ \\
\hline
Double Precision & $1 \;[63]$ &	$11 \;[62-52]$ & $52 \;[51-00]$ & $1023$\\
\hline
\end{tabular}
\caption{Floating-point Number Layout}
\label{prog:prel:floatlayout}
\end{table}
Floating-point solves a number of representation problems.
Fixed-point has a fixed window of representation, which limits
it from representing very large or very small numbers.
Floating-point, on the other hand, employs a sort of
``sliding window'' of precision appropriate to the scale of the number.
The range of positive floating-point numbers can be split into
normalized numbers (which preserve the full precision of the mantissa),
and denormalized numbers.
The denormalized
numbers do not have an implicit leading bit of $1$ and
allow representation of really small numbers but with
only a portion of the fraction's precision.
The exponent of all $0$s ($-127$) and all $1$s ($128$)
are reserved for denormalized numbers and
representing infinity respectively.
A complete discussion on the semantics of floating-point
operations can be found in the
 IEEE 754 Standard~\cite{IEEE754}.
A floating-point unit (FPU) is used to carry out
operations on floating-point numbers such as addition, subtraction,
multiplication, division and square root.
FPUs are integrated with CPUs in computers but most embedded
processors do not have hardware support for floating-point
operations. Emulation of floating-point operations
without hardware support can be very slow.
Inclusion of FPUs also increases the power consumption of the processors.
This has made the use of fixed-point arithmetic very common in
embedded systems.
}
In spite of the benefits of floating-point arithmetic,
embedded systems often use fixed-point
arithmetic to reduce resource cost and improve performance.
A fixed-point number consists of a sign mode bit,
an integer part and a fractional
part. We denote the fixed-point type of a variable $x$ by $\fxtype(x)$.
Formally, a fixed-point type is a triple:
$$\langle \signbit, \iwl, \fwl \rangle.$$
The sign mode bit $\signbit$ is $0$ if the data is unsigned and is $1$ if the
data is signed.
The length of the integer part
is called the integer wordlength ($\iwl$) and the length of the
fractional part
is called the fractional wordlength ($\fwl$).
The fixed-point wordlength
($\wl$) is the sum of the integer wordlength
and fractional wordlength; that is, $\wl = \iwl + \fwl$.
\ifthenelse {\boolean{IsFinal}}
{}
{
A fixed-point number with fractional word length ($\fwl$)
is scaled by a factor of ${1}/{2^{\fwl}}$.
For example, a fixed point number $01110$
with $0$ as $\signbit$ , integer wordlength of 3
and fractional wordlength of 2 represents $14\times {1}/{2^{2}}$,
that is, $3.5$.
Converting a fixed-point number with scaling
factor R to another type with scaling factor S,
requires multiplying the underlying integer by R
and dividing by S; that is, multiplying by the ratio R/S.
For example, converting $01110$
with $0$ as $\signbit$, integer wordlength of 2
and fractional wordlength of 2 into a fixed-point
number with $0$ as $\signbit$, integer wordlength of 2
and fractional wordlength of 3 requires
multiplying with ${2^3}/{2^{2}}$ to obtain
$011100$. If the scaling factor is to be reduced,
the new integer will have to be rounded. For example,
converting the same fixed-point number $01110$
to a fixed-point number with fractional wordlength
of $0$ and integer wordlength of $2$ yields
$011$, that is, $3$ which is obtained by rounding down
from $3.5$.
}
The operations supported by fixed-point arithmetic
are the same as those in the floating-point arithmetic standard~\cite{IEEE754}
but the semantics can differ on custom hardware.
For example, the rounding mode for arithmetic
operations could be different, and the
result could be specified to saturate or overflow/underflow
in case the wordlength of a variable is not sufficient to store
a computed result. One complete semantics of
fixed-point operation is provided with the Fixed-point Toolbox
in Matlab~\cite{fixed-matlab}.
The range of the fixed-point number is much smaller compared
to the range of floating-point numbers for the same number
of bits since the radix point is fixed and no dynamic
adjustment of precision is possible.
\ifthenelse {\boolean{IsFinal}}
{}
{
\begin{table}
\centering
\begin{tabular}{|c|c|c|}
\hline
Number systems with $\wl = 32$  & Range & Precision\\
\hline
\hline
Single-precision Floating-point &  $\pm$ $ ~10^{-44.85}$ to $~10^{38.53}$ & {\em Adaptive}\\
\hline
Fixed-point type: $\langle 1, 8, 24 \rangle$& $~$ $-10^{2.11}$ to $10^{2.11}$ &  $~10^{-7.22}$  \\
\hline
Fixed-point type: $\langle 1, 16, 16 \rangle$ &  $~$ $-10^{4.52}$ to $10^{4.52}$ & $~10^{-4.82}$ \\
\hline
Fixed-point type: $\langle 1, 24, 8 \rangle$ &   $~$ $-10^{6.93}$ to $10^{6.93}$ & $~10^{-2.41}$  \\
\hline
\end{tabular}
\caption{Range of $32$ bit fixed-point and floating-point numbers}
\label{prog:prel:fxrange}
\end{table}
}
Translating a floating-point program into fixed-point program
is non-trivial and requires careful consideration of loss of precision
and range. The integer wordlengths and fractional wordlengths of
the fixed-point variables need to be carefully selected to ensure that
the computation remains accurate to a specified threshold.
\ifthenelse {\boolean{IsFinal}}
{Please refer to the extended version of the paper available as technical report~\cite{extended}
for more background discussion.}
{}

\noop{In Chapter~\ref{fpfx}, we present an automated synthesis technique
to derive fixed-point code from floating-point code.
Fixed-point arithmetic is performed on custom hardware
and FPGA is often a natural choice for implementing
these algorithms.
}

\section{Problem Definition}
\label{pldi12:sec:probdef}
\label{icse12:sec:probdef}

We introduce a simple illustrative example to explain the
problem of synthesizing an optimal fixed-point program
from a floating-point program, and then present the formal
problem definition.\\
\\
\ifthenelse {\boolean{IsFinal}}
{\noindent \textbf{Floating-point Implementation: }}
{\subsection{Floating-point Implementation}}
Given a floating-point program,
we need to synthesize fixed-point type for each floating-point variable.\\
\\
\noindent \textbf{Example 1:}
The floating-point
program in this example~\ref{icse12:proc:ch1:fpcircle}
takes $\mathtt{radius}$ as the input, and
computes the corresponding $\mathtt{area}$ of the circle.
Notice that the fixed-point program is essentially identical
to the floating-point version, except that the
fixed-point types of variables $\mathtt{mypi, radius, t}$
and $\mathtt{area}$ must be identified.
Recall that the fixed-point
type is a triple
$\mathtt{\langle s_j,iwl_j,fwl_j \rangle}$ for $j$-th variable
where $\mathtt{s_j}$ denotes the $\signbit$ of the variable,
$\mathtt{iwl_j}$ denotes the integer wordlength and $\mathtt{fwl_j}$ denotes
the fraction wordlength.
\ifthenelse {\boolean{IsFinal}}
{
\begin{minipage}[t]{\textwidth}
\noindent\begin{minipage}[t]{0.48\textwidth}
\small
\begin{algorithm}[H]
\caption{Floating-point program to compute circle area}
\label{icse12:proc:ch1:fpcircle}
\begin{algorithmic} [style=tt]
\REQUIRE $\mathtt{radius}$
\ENSURE $\mathtt{area}$
\STATE \textbf {double} $\mathtt{mypi, radius, t, area}$
\STATE $\mathtt{mypi=3.14159265358979323846}$
\STATE $\mathtt{t=radius \times radius}$
\STATE $\mathtt{area=mypi \times t}$
\RETURN $\mathtt{area}$
\end{algorithmic}
\end{algorithm}
\normalsize
\end{minipage}
\noindent\begin{minipage}[t]{0.48\textwidth}
\small
\begin{algorithm}[H]
\caption{Fixed-point program to compute circle area}
\label{icse12:proc:ch1:fxcircle}
\begin{algorithmic}[style=tt]
\REQUIRE $\mathtt{radius}$,
$\mathtt{\langle s_j,iwl_j,fwl_j \rangle}$ for $\mathtt{j=1,2,3,4}$
\ENSURE $\mathtt{area}$
\STATE $\mathbf{fx \langle s_1,iwl_1,fwl_1 \rangle}$ $\mathtt{mypi}$
\STATE $\mathbf{fx \langle s_2,iwl_2,fwl_2 \rangle}$ $\mathtt{radius}$
\STATE $\mathbf{fx \langle s_3,iwl_3,fwl_3 \rangle}$ $\mathtt{t}$
\STATE $\mathbf{fx \langle s_4,iwl_4,fwl_4 \rangle}$ $\mathtt{area}$
\STATE $\mathtt{mypi=3.14159265358979323846}$
\STATE $\mathtt{t=radius \times radius}$
\STATE $\mathtt{area=mypi \times t}$
\RETURN $\mathtt{area}$
\end{algorithmic}
\end{algorithm}
\normalsize
\end{minipage}
\end{minipage}
}
{
\small
\begin{algorithm}
\caption{Floating-point program to compute circle area}
\label{icse12:proc:ch1:fpcircle}
\begin{algorithmic} [style=tt]
\REQUIRE $\mathtt{radius}$
\ENSURE $\mathtt{area}$
\STATE \textbf {double} $\mathtt{mypi, radius, t, area}$
\STATE $\mathtt{mypi=3.14159265358979323846}$
\STATE $\mathtt{t=radius \times radius}$
\STATE $\mathtt{area=mypi \times t}$
\RETURN $\mathtt{area}$
\end{algorithmic}
\end{algorithm}
\normalsize
\small
\begin{algorithm}
\caption{Fixed-point program to compute circle area}
\label{icse12:proc:ch1:fxcircle}
\begin{algorithmic}[style=tt]
\REQUIRE $\mathtt{radius}$,
$\mathtt{\langle s_j,iwl_j,fwl_j \rangle}$ for $\mathtt{j=1,2,3,4}$
\ENSURE $\mathtt{area}$
\STATE $\mathbf{fx \langle s_1,iwl_1,fwl_1 \rangle}$ $\mathtt{mypi}$
\STATE $\mathbf{fx \langle s_2,iwl_2,fwl_2 \rangle}$ $\mathtt{radius}$
\STATE $\mathbf{fx \langle s_3,iwl_3,fwl_3 \rangle}$ $\mathtt{t}$
\STATE $\mathbf{fx \langle s_4,iwl_4,fwl_4 \rangle}$ $\mathtt{area}$
\STATE $\mathtt{mypi=3.14159265358979323846}$
\STATE $\mathtt{t=radius \times radius}$
\STATE $\mathtt{area=mypi \times t}$
\RETURN $\mathtt{area}$
\end{algorithmic}
\end{algorithm}
\normalsize
}
We use $F_{fl}(X)$ to
denote the floating-point program with inputs $X = \langle x_1,x_2,\ldots,x_n \rangle$.
$F_{fx}(X,\fxtype)$ denotes the fixed-point version of the program,
where the fixed-point type of a variable $x \in X$ is $\fxtype(x)$.
Note that the fixed-point types in $F_{fx}(X,\fxtype)$ are defined
by the mapping $\fxtype$.\\
\\
\ifthenelse {\boolean{IsFinal}}
{\noindent\textbf{Input Domain:}}
{\subsection{Input Domain}}
The context in which a fixed-point program
$F_{fx}(X,\fxtype)$ is executed
often provides a precondition that must be satisfied
by valid inputs $\langle x_1,x_2,\ldots,x_n \rangle$.
This defines the input domain
denoted by  $Dom(X)$.\\
\\
\noindent \textbf{Example 2:} In our example of computing area of a circle,
suppose that we are only interested in the radii in the range $[0.1,2.0)$.
Then, the input domain $Dom(\mathtt{radius})$
is $$\mathtt{radius} \geq 0.1 \wedge \mathtt{radius} < 2.0$$

\ifthenelse {\boolean{IsFinal}}
{\noindent\textbf{Correctness Condition for Accuracy: }}
{\subsection{Correctness Condition for Accuracy}}
The correctness condition specifies
an error function $Err(F_{fl}(X),F_{fx}(X,\fxtype))$, and
a maximum error threshold $\maxErr$.
The error function and error threshold together define a bound
on the ``distance'' between outputs generated by the
floating-point and fixed-point programs respectively.
An {\em accurate} fixed-point program is one whose error function
lies within the error threshold for all inputs in the input domain.
Some common error functions are:
\begin{myitemize}
\item Absolute difference between the floating-point function
and fixed-point function: $\left |F_{fl}(X) - F_{fx}(X,\fxtype) \right|$
\item Relative difference between the floating-point function
and fixed-point function: $\left |\frac{F_{fl}(X) - F_{fx}(X,\fxtype)}{F_{fl}(X)} \right|$
\item Moderated relative difference: $\left |\frac{F_{fl}(X) - F_{fx}(X,\fxtype)}{F_{fl}(X)+ \delta} \right|$.
This approaches the relative difference
for $F_{fl}(X)>> \delta$ and approaches a
weighted absolute difference for $F_{fl}(X) << \delta$.
When $F_{fl}(X)$ can be zero
for some values of $X$, the moderated relative difference remains bounded unlike
the relative difference which becomes unbounded.

\end{myitemize}

The {\em correctness condition for accuracy}
requires that for all inputs in the
provided {\em input domain} $Dom(X)$, the error function
$Err(F_{fl}(X),F_{fx}(X,\fxtype))$ is below
the specified threshold $\maxErr$; i.e.,
$$ \forall X \in Dom(X) \;. \;
Err(F_{fl}(X),F_{fx}(X,\fxtype)) \leq \maxErr$$
\noindent \textbf{Example 3:} In our running example of
computing area of a circle,
the error function is chosen to be relative difference, the
error threshold $0.01$, and thus the
correctness condition is $\forall \mathtt{radius}, \; s.t. \;
\mathtt{radius} \geq 0.1 \wedge \mathtt{radius} < 2.0$
$$\frac{F_{fl}(\mathtt{radius}) - F_{fx}(\mathtt{radius},\fxtype)}{F_{fl}(\mathtt{radius})}  \leq 0.01$$

\ifthenelse {\boolean{IsFinal}}
{\noindent\textbf{Implementation Cost Model: }}
{\subsection{Implementation Cost Model}}
\noop{
Recall that the fixed-point program only differs from the floating-point
version in the types: floating-point types are replaced by fixed-point types.
Thus, the cost of the fixed-point program depends on
the fixed-point types of its variables.
Recall that a fixed-point type is
triple
$\mathtt{\langle s_j,iwl_j,fwl_j \rangle}$ for $j$-th variable
where $\mathtt{s_j}$ denotes the $\signbit$ of the variable,
$\mathtt{iwl_j}$ denotes the integer wordlength and $\mathtt{fwl_j}$ denotes
the fraction wordlength.}
The {\em cost model} of the fixed-point program
is a function mapping fixed-point types to a real number.
For a given fixed-point program $F_{fx}(X,\fxtype)$,
let $T = \{t_1, t_2, \ldots, t_k\}$ be the set of
fixed-point program variables with corresponding
types $\{\fxtype(t_1), \fxtype(t_2), \ldots, \fxtype(t_k)\}$.
Then the cost model (or simply {\em cost}) of $F_{fx}$ is a function
$$\fxcost: (\fxtype(t_1),\fxtype(t_2),\ldots,\fxtype(t_k)) \rightarrow \mathbb{R} $$
In practice, $\fxcost$ is often just a function
of the total wordlengths ($\wl = \iwl + \fwl$) of the variables.
It can incorporate hardware implementation metrics such as area, power and delay.
A number of cost models are available in the
literature~\cite{chien-vlsi97,macii-iccad98,clarke-ISCAS06,constantinides-FCCM02},
and all of these can be used in our approach.\\
\noindent \textbf{Example 4:} The cost model proposed
by Constantinides et al~\cite{constantinides-FCCM02}
for the running example yields the following cost function.
We use this cost model in all our examples.
\small
$$\fxcost (\fxtype(\mathtt{mypi}),\fxtype(\mathtt{radius}),\fxtype(\mathtt{t}), \fxtype(\mathtt{area})) = $$
$$\fxcostdelay(\wl(\mathtt{mypi})) + \fxcostmul(\wl(\mathtt{radius}), \wl(\mathtt{radius}), \wl(\mathtt{t}))$$
$$+ \fxcostmul(\wl(\mathtt{mypi}), \wl(\mathtt{t}), \wl(\mathtt{area}))\;, \mathtt{where}$$
$$\fxcostdelay(l) = l+1 \; \mathtt{ and } \; \fxcostmul(l_1,l_2,l) = 0.6 \times (l_1+1) * l_2 - 0.85 * (l_1 + l_2 - l)$$
\normalsize
The area of a multiplier
grows almost linearly with both the coefficients and the data
wordlength.
The first term in the Constantinides model represents
this cost. The second term represents the area cost of
computational elements required only for carry propagation.
The coefficients $0.6$ and $0.85$ were obtained
through least-squared fitting
to area of several hundred multipliers of different coefficient
value and width~\cite{constantinides-FCCM02}.\\

\ifthenelse {\boolean{IsFinal}}
{\noindent\textbf{\underline {Problem Definition} }}
{\subsection{Problem Definition}}
\begin{definition}[{\em Optimal Fixed-point Types Synthesis}]
The optimal fixed-point types synthesis problem
is as follows. Given
a floating-point program $F_{fx}(X,\fxtype(T))$ with variables $T$,
an input domain $Dom(X)$,
a correctness condition $Err(F_{fl}(X),F_{fx}(X,\fxtype(T))) \leq \maxErr$,
and a cost model $\fxcost(\fxtype(t_1),\fxtype(t_2),\ldots,\fxtype(t_k))$,
\noop{
\begin{enumerate}
\item  a floating-point program $F_{fx}(X,\fxtype(T))$ with variables $T$,
\item  an input domain $Dom(X)$,
\item  a correctness condition $Err(F_{fl}(X),F_{fx}(X,\fxtype(T))) \leq \maxErr$,
\item  a cost model $\fxcost(\fxtype(t_1),\fxtype(t_2),\ldots,\fxtype(t_k))$
\end{enumerate}
}
the optimal fixed-point types synthesis problem is to discover
fixed-point types
$$\fxtype^*(T) = \{ \fxtype^*(t_1), \fxtype^*(t_2), \ldots, \fxtype^*(t_k) \}$$
such that the fixed-point program $F_{fl}(X)$ with the above types for
fixed-point variables satisfies the correctness condition for accuracy,
that is,
\small
$$ (a) \; \forall X \in Dom(X) \;.\; Err(F_{fl}(X),F_{fx}(X,\fxtype^*(T))) \leq \maxErr$$
\normalsize
and has minimal
cost with respect to the given cost function among all fixed-point types
that satisfy condition (a), that is,
$$ (b) \; \fxtype^* = \underset{\fxtype \mathtt{\; satisfies \; (a)}}{\operatorname{argmin}} \; \fxcost(\fxtype(T))$$
\noop{
\begin{enumerate}[(a)]
\item satisfies the correctness condition for accuracy,
that is,
$$ \forall X \in Dom(X) \;.\; Err(F_{fl}(X),F_{fx}(X,\fxtype^*(T))) \leq \maxErr$$
\item has minimal
cost with respect to the given cost function among all fixed-point types
that satisfy condition (a), that is,
$$\fxtype^* = \underset{\fxtype \mathtt{\; satisfies \; (a)}}{\operatorname{argmin}} \; \fxcost(\fxtype(T))$$
\end{enumerate}
}
\end{definition}
Our goal is to automated this search for optimal fixed-point types.
We illustrate this problem using the running example below.\\
\\
\noindent \textbf{Example 5: }
In our running example of computing the {\em area of a circle},
we need to discover $\fxtype^*(\mathtt{mypi}),$ $\fxtype^*(\mathtt{radius}),$ $\fxtype^*(\mathtt{t})$
and $\fxtype^*(\mathtt{area})$
such that
\begin{enumerate}[(a)]
\item the fixed-point program with the given
fixed-point types satisfies the correctness condition; that is, $\forall \mathtt{radius},  \; s.t., \;
\mathtt{radius} \geq 0.1 \wedge \mathtt{radius} < 2.0$
\small
$$\;\;
\frac{F_{fl}(\mathtt{radius}) - F_{fx}(\mathtt{radius},\fxtype^*)}{F_{fl}(\mathtt{radius})}  \leq 0.01$$
\normalsize
\item and the cost is minimized; that is,
$$\;\; \fxtype^* = \underset{\fxtype \mathtt{\; satisfies \; (a)}} {\operatorname{argmin}} \; \fxcost(\fxtype(\mathtt{mypi}, \mathtt{radius},\mathtt{t}, \mathtt{area}))$$
\end{enumerate}
We use this example to illustrate the trade-off between cost and error
and how a human might use trial and error to discover the correct
wordlengths. We vary the wordlength of the variables. The integer wordlength
is selected to avoid overflow and the remaining bits are used for fractional
wordlength.

\ifthenelse {\boolean{IsFinal}}
{}
{
\noindent \textbf{Case 1} (Figure~\ref{icse12:fig:circlearea8}):
$WL = 8$ for all variables. $\fxtype(\mathtt{mypi}) = \langle 0, 2, 6\rangle, \fxtype(\mathtt{radius}) = \langle 0,1, 7\rangle, \fxtype(\mathtt{t}) = \langle 0, 2, 6\rangle, \fxtype(\mathtt{area}) = \langle 0, 4,4 \rangle$.
Cost is $81.80$.\\
}
\ifthenelse {\boolean{IsFinal}}
{
\noindent \textbf{Case 1} (Figure~\ref{icse12:fig:circlearea12}):
}
{
\noindent \textbf{Case 2} (Figure~\ref{icse12:fig:circlearea12}):
}
$WL = 12$ for all variables. $\fxtype(\mathtt{mypi}) = \langle 0, 2, 10\rangle,
\fxtype(\mathtt{radius}) = \langle 0,1, 11\rangle, \fxtype(\mathtt{t}) = \langle 0, 2, 10\rangle, \fxtype(\mathtt{area}) = \langle 0, 4,8 \rangle$.
Cost is $179.80$.\\
\ifthenelse {\boolean{IsFinal}}
{
\noindent \textbf{Case 2} (Figure~\ref{icse12:fig:circlearea16}):
}
{
\noindent \textbf{Case 3} (Figure~\ref{icse12:fig:circlearea16}):
}
$WL = 16$ for all variables. $\fxtype(\mathtt{mypi}) = \langle 0, 2, 14\rangle, \fxtype(\mathtt{radius}) = \langle 0, 1, 15\rangle, \fxtype(\mathtt{t}) = \langle 0, 2, 14\rangle, \fxtype(\mathtt{area}) = \langle 0, 4, 12 \rangle$.
Cost is $316.20$.\\

\ifthenelse {\boolean{IsFinal}}
{

\begin{minipage}[t]{\textwidth}
\noindent \begin{minipage}[t]{0.5\textwidth}
\begin{figure}[H]
\includegraphics[width=2.5in]{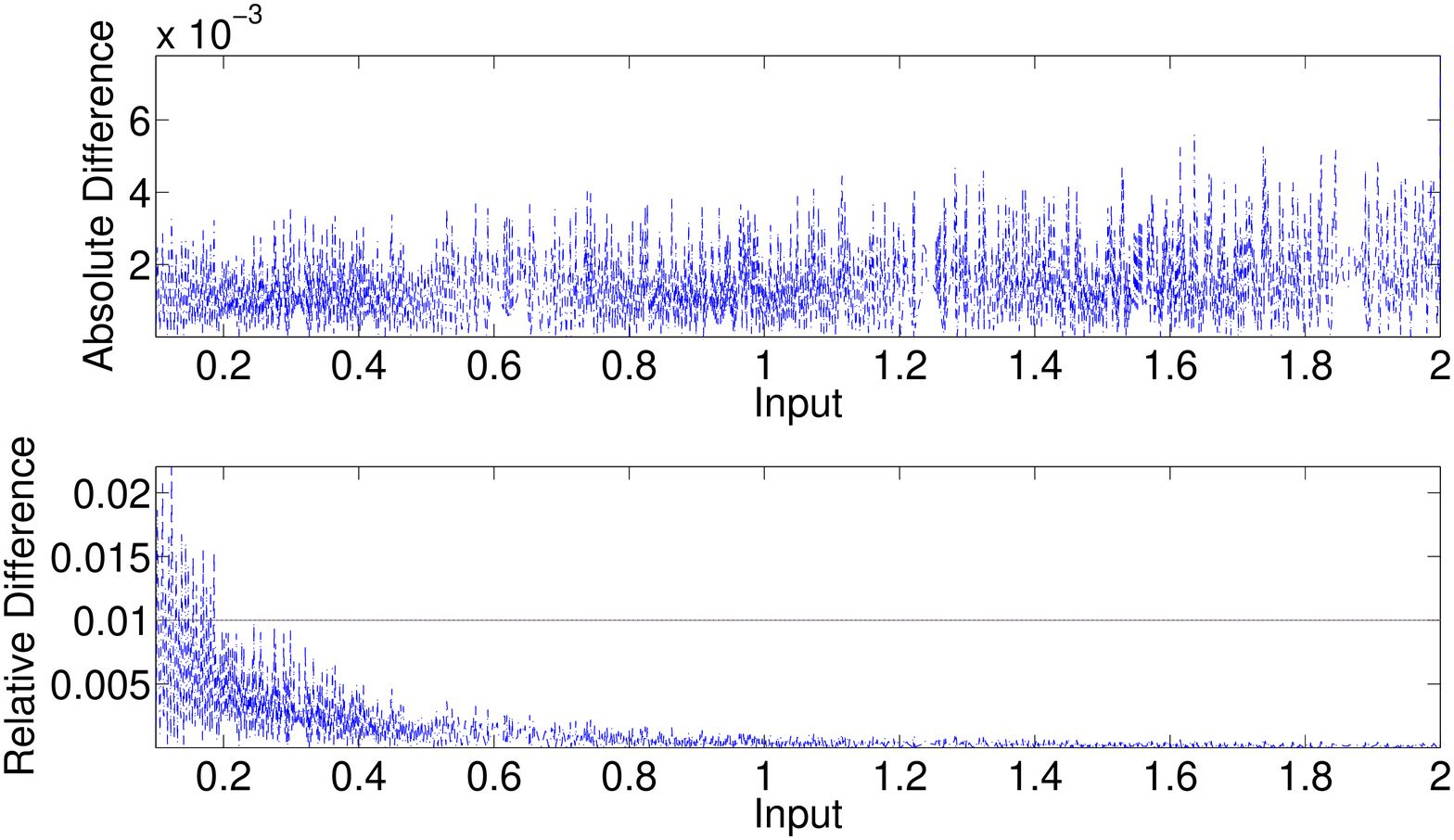}
\caption{$\wl=12$. {\footnotesize{Error threshold at $0.01$ is violated}}.}
\label{icse12:fig:circlearea12}
\end{figure}
\end{minipage}
\noindent \begin{minipage}[t]{0.5\textwidth}
\begin{figure}[H]
\includegraphics[width=2.5in]{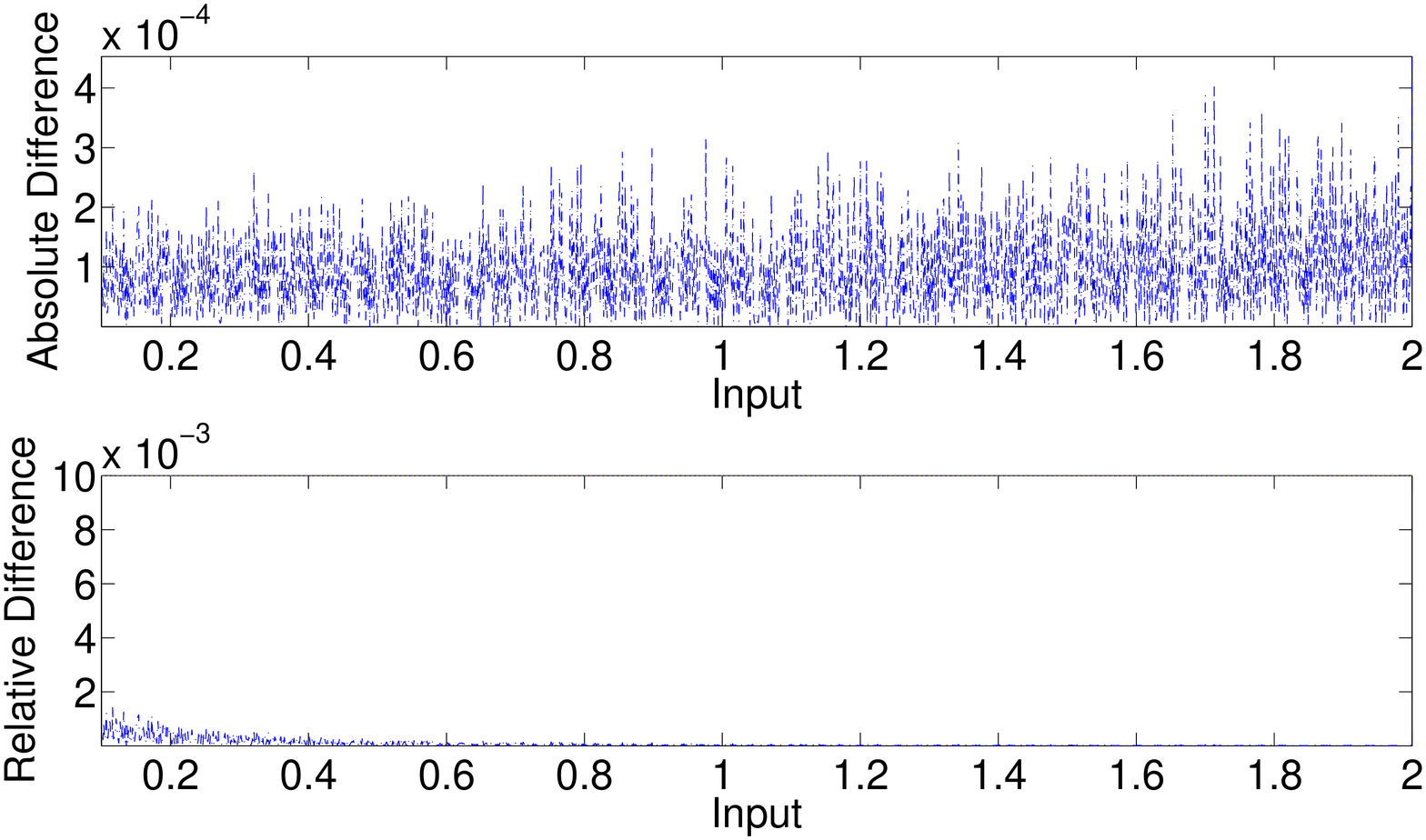}
\caption{$\wl=16$. {\footnotesize{Error threshold at $0.01$ is not violated}}.}
\label{icse12:fig:circlearea16}
\end{figure}
\end{minipage}

\end{minipage}
}
{
\begin{figure}[!ht]
\centering
\includegraphics[width=4in]{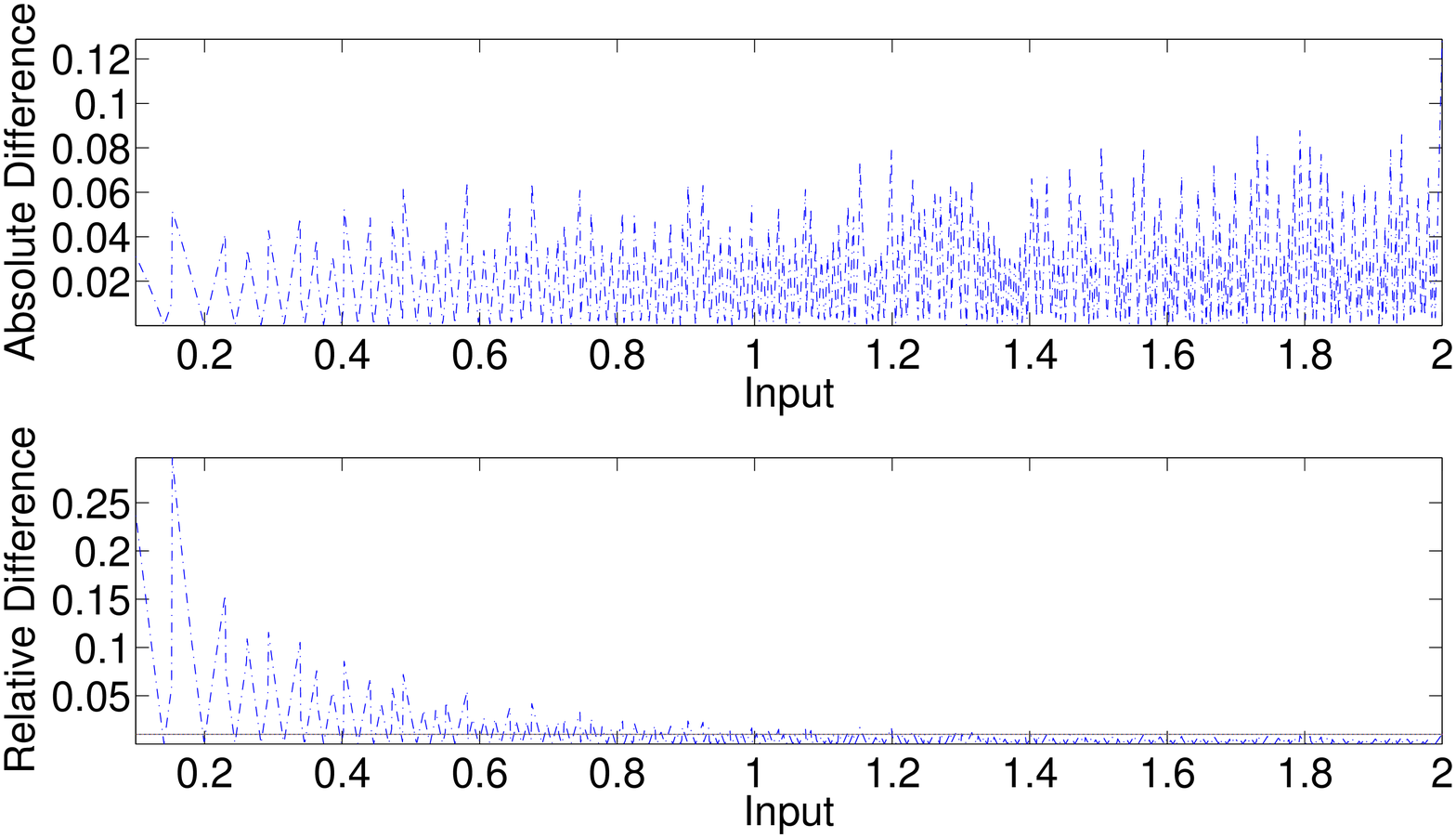}
\caption{Error for $\wl=8$. {\footnotesize{Error threshold at $0.01$ is violated}}.}
\label{icse12:fig:circlearea8}
\end{figure}

\begin{figure}[!ht]
\centering
\includegraphics[width=4in]{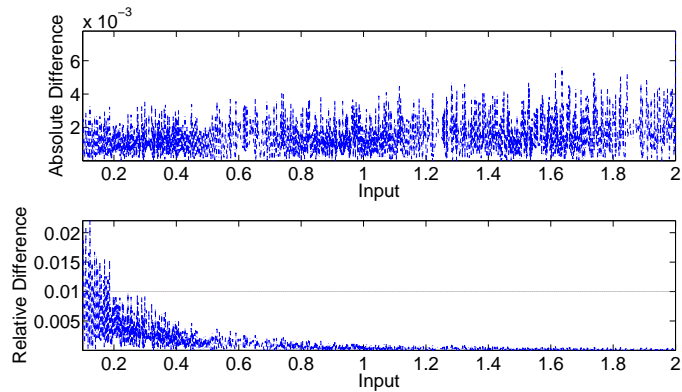}
\caption{$\wl=12$. {\footnotesize{Error threshold at $0.01$ is violated}}.}
\label{icse12:fig:circlearea12}
\end{figure}

\begin{figure}[!ht]
\centering
\includegraphics[width=4in]{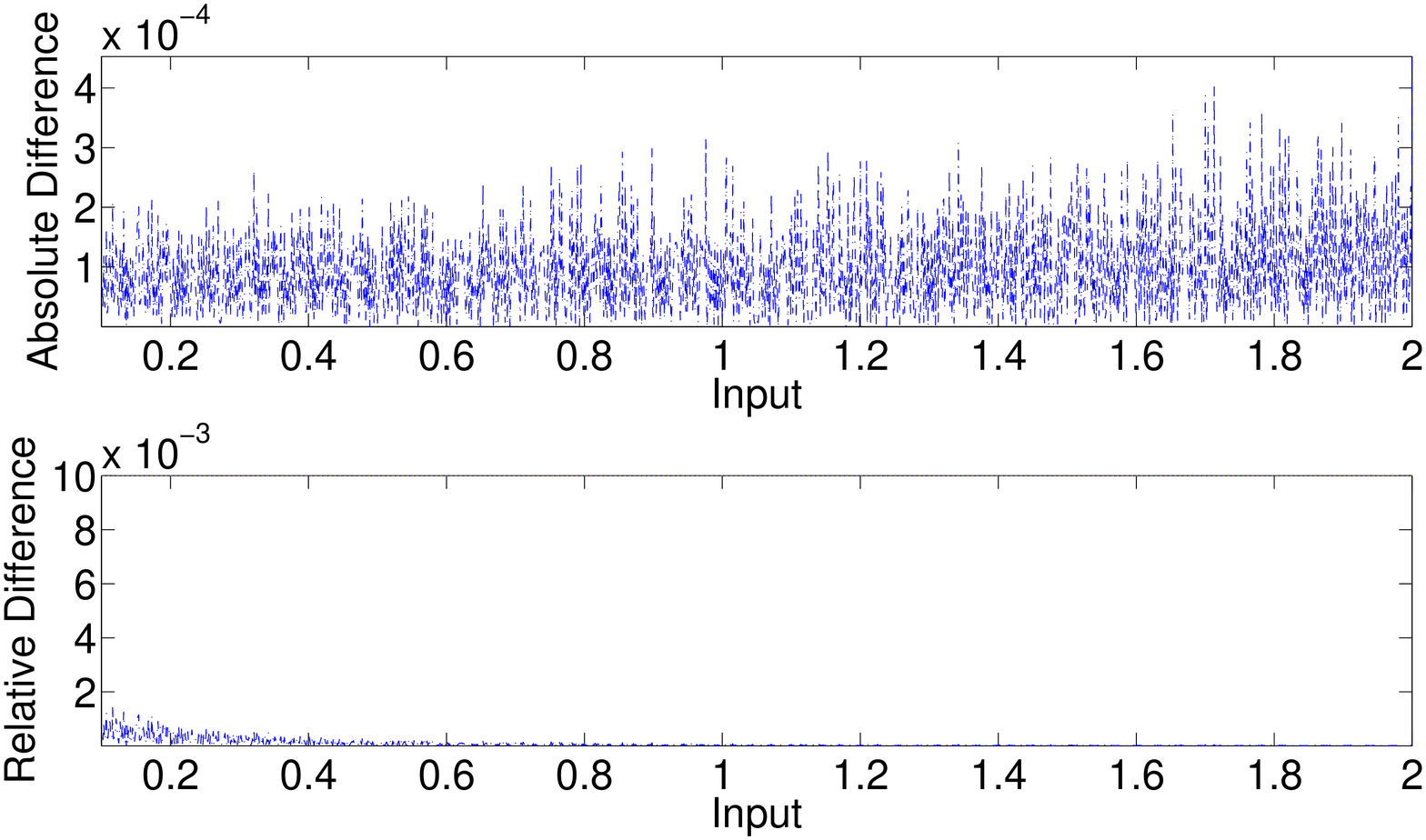}
\caption{$\wl=16$. {\footnotesize{Error threshold at $0.01$ is not violated}}.}
\label{icse12:fig:circlearea16}
\end{figure}
}

As we will show in the next section, our approach computes
fixed-point types that meet the accuracy threshold and
yield a cost of only $104.65$, which, while being less than the cost in Case 1,
satisfies the correctness criterion like Case 2.
In the following
section, we  discuss our automated approach to solve this problem.

\noop{
Manual synthesis of types typically requires
the developer to run a number of experiments with varying
fixed-point types, identify the cases where
the correctness condition is satisfied, and then
find the fixed-point types that minimize the cost from amongst these cases.
Such a manual trial-and-error process is time-consuming and
does not guarantee correctness of the resulting fixed-point code;
since the input domain can be very large,
it is not possible to simulate the program for all possible inputs.
One thus needs a systematic approach to selecting inputs to use in
the optimization process.
In the following
section, we present an automated approach to solve this problem.
}

\section{Our Approach}
\label{pldi12:sec:approach}
\label{icse12:sec:approach}

A central idea behind our approach, $\swati$ is to
identify a small set of {\em interesting} inputs $S(X)$ using testing
from the input domain $Dom(X)$ such that
the optimal implementation found using induction that satisfies the correctness condition for the
inputs in $S(X)$ will be optimal and correct for all inputs in
the given input domain $Dom(X)$. 
\ifthenelse {\boolean{IsFinal}}
{
\tiny
\begin{algorithm}
\caption{Overall Synthesis Algorithm: $\swati$}
\label{icse12:algo:floatfix}
\begin{algorithmic}
\REQUIRE Floating-point program $F_{fp}$, Fixed-point program $F_{fx}$ with
fixed-point variables $T$,
Domain of inputs $Dom$, Error function $Err$, maximum error threshold
$\maxErr$, Cost Model $\fxcost$, maximum wordlengths $\wl_{max}$
\ENSURE Fixed-point type $\fxtype$ for variables $T$  or \infeasible
\STATE $S^0$ = random sample from $Dom$, $Bad^0 = S^0$, $i = 0$
\WHILE{$Bad^i \not = \emptyset$}
\STATE $i = i+1$, $S^{i} = S^{i-1} \cup Bad^{i-1}$, $\fxtype^i = \optsynthm{}(F_{fp},F_{fx}, Dom, Err, \maxErr,$\\$\;\;\;\;\;\;\;\;\;\;\;\;\;\;\;\;\;\;\;\;\;\;\;\;\;\;\;\;\;\;\;\;\;\;\;\;\;\;\fxcost, \wl_{max}, S^i)$
\IF {$\fxtype^i = \bot$} \RETURN \infeasible
\ENDIF
\STATE $Bad^i = \verifyerrm(F_{fp},F_{fx}, \fxtype^i, Dom, Err, \maxErr)$
\ENDWHILE
\RETURN $\fxtype^* = \fxtype^i $
\end{algorithmic}
\end{algorithm}
\normalsize
}
{
\small
\begin{algorithm}
\caption{Overall Synthesis Algorithm: $\swati$}
\label{icse12:algo:floatfix}
\begin{algorithmic}
\REQUIRE Floating-point program $F_{fp}$,\\ Fixed-point program $F_{fx}$ with
fixed-point variables $T$,\\
Domain of inputs $Dom$, Error function $Err$,\\ maximum error threshold
$\maxErr$, Cost Model $\fxcost$,\\ maximum wordlengths $\wl_{max}$
\ENSURE Fixed-point type $\fxtype$ for variables $T$\\
\hspace{1cm} or \infeasible
\STATE $S^0$ = random sample from $Dom$, $Bad^0 = S^0$, $i = 0$
\WHILE{$Bad^i \not = \emptyset$}
\STATE $i = i+1$
\STATE $S^{i} = S^{i-1} \cup Bad^{i-1}$
\STATE $\fxtype^i = \optsynthm{}(F_{fp},F_{fx}, Dom, Err, \maxErr,$\\$\;\;\;\;\;\;\;\;\;\;\;\;\;\;\;\;\;\;\;\;\;\;\;\;\;\;\;\;\;\;\;\;\;\;\;\;\;\;\fxcost, \wl_{max}, S^i)$
\IF {$\fxtype^i = \bot$} \RETURN \infeasible
\ENDIF
\STATE $Bad^i = \verifyerrm(F_{fp},F_{fx}, \fxtype^i, Dom, Err, \maxErr)$
\ENDWHILE
\RETURN $\fxtype^* = \fxtype^i $
\end{algorithmic}
\end{algorithm}
\normalsize
}
The top-level synthesis
algorithm is presented in Procedure~\ref{icse12:algo:floatfix}.
$\wl_{max}$ is an upper bound on wordlengths
beyond which it is non-optimal to use the fixed-point version.
The algorithm starts with a randomly selected set of examples
$S^0$ from the given input domain.
Then, a fixed-point implementation that satisfies
the accuracy condition for each of these inputs
and is of minimal cost is synthesized using the
routine \optsynth{}.
If no such implementation is found,
the algorithm reports \infeasible. Otherwise,
the testing routine \verifyerr{}
checks whether the implementation fails the
correctness condition for any input.
If so, a set of
inputs $Bad^i$ on which the implementation violates
the correctness condition are added to the set
$S^i$ used for synthesis, and the process is repeated.
If the correctness condition is satisfied,
the resulting fixed-point types are output.
In the rest of this section, we describe the main components
of our approach in detail, including the theoretical result.

\subsection{Synthesizing Optimal Types for a Finite Input Set}

The \optsynth{} function
(see Procedure~\ref{icse12:algo:synthMinCost})
is used to obtain optimum fixed-point types such that
the fixed-point program with these types satisfies
the correctness condition for a finite input set $S$
and has minimal cost.
First, the floating-point program $F_{fl}$ is executed for all the inputs
in the sample $S$ and the range of each variable $t_i$ as well
as its $\signbit$ is recorded by the functions $\mathtt{getRange}$
and $\mathtt{isSigned}$ respectively. Then, the integer wordlength $\iwl$
sufficient to represent the computed range
is assigned to each variable $t_i$ and the $\signbit$ is $1$ if
the variable takes both positive and negative values, and $0$ otherwise.
If the fixed-point program with maximum wordlengths $\wl_{max}$
fails the correctness condition, we conclude that the synthesis
is not feasible and return $\bot$.
If not, we search for the wordlength
with minimum cost satisfying the correctness condition using
our optimization oracle $\optS$. The result is used to compute
the fractional wordlengths, and the resulting fixed-point types
are returned.

More precisely, $\optS$ solves
the following optimization problem over $\fxtype$:
$\mathtt{Minimize\;\;} \fxcost(\fxtype) \mathtt{\;\; s.t. \;} $
\begin{align} \label{fxopt}
\displaystyle \bigwedge_{x \in S} Err(F_{fx}(x,\fxtype), F_{fl}(x)) \leq \maxErr
\end{align}
Let us reflect on the nature of the above optimization problem.
The overall synthesis algorithm might make several
calls to $\optS$ for solving the optimization
problem for different sets of inputs and hence,
$\optS$  must be a fast procedure.
But it is a discrete optimization problem
with a non-convex constraint space,
a problem class that is known to be computationally hard~\cite{fletcher-book}.
This rules out any computationally efficient algorithm to implement
$\optS$ without sacrificing correctness guarantees.
Since the space of possible types grows exponentially with the number
of variables, brute-force search techniques will not scale beyond a few
variables.
Satisfiability solvers can also not be directly
exploited to search for optimal wordlengths since the existential
quantification is over the types and not the variables.
The arithmetic operators have different semantics when operating on
operands with different types and hence, the only way to
encode this search problem as a satisfiability problem is to case-split
{\em{exhaustively}} on all possible types (word-lengths),
where each case encodes the fixed-point program with one
possible type. The number of such cases is
exponential in the number of the variables in the program
under synthesis and hence, SAT problems will be themselves
exponentially large in size. Further, one would need to
invoke SAT solvers multiple times in order to optimize
the cost function. Thus, satisfiability solving would be a wrong choice
to address this problem.
Further, the space of possible types is also not totally ordered and hence,
binary search like techniques would also not work. For a binary search like
technique to work, we will need to define a
domination ordering over the types which has three properties.
Firstly, it is a total ordering relation.
Secondly,
if a particular type assignment satisfies the correctness condition
for all inputs then all dominating types satisfy
the correctness condition for all inputs.
Thirdly,
the cost function is monotonic with respect to the
domination ordering relation.
In general, this may not be feasible for any given floating-point
program and cost function.
\noop{
In spite of this, since floating-point and fixed-point programs
can ultimately be encoded to Boolean satisfiability (SAT),
due to the dramatic practical progress in SAT solving, one might be
hopeful of an efficient SAT-based pseudo-Boolean optimization procedure
to solve this problem (e.g.,see~\cite{een-jsat06,chai-dac03}).
However, we note two main factors going against such a SAT-based approach.
First, the optimization is over wordlengths, rather than bits encoding
variable values; thus, one must do an explicit case-split over the large space of
possible wordlengths for each vector of variables $X$.
Second, SAT solving is known to be notoriously difficult for
reasoning about programs with arbitrary floating-point arithmetic operations
involving non-linear arithmetic.
Additionally, the size of the constraint that
encodes correctness grows linearly in the number of input examples.
While a SAT-based approach may work for programs with a small number
of variables and no non-linear floating-point operators, for the general
case, these factors effectively rule out such an approach. }
Hence, we implement $\optS$ using a greedy procedure $\mathtt{getMinCostWL}$
presented in Procedure~\ref{icse12:algo:getMinCostWL}.
\noop{
The wordlength of each variable is increased or decreased by $1$
independently to construct a set of
candidate wordlength assignments $cand\wl$.
We select a set of
valid candidate wordlength assignments $valcand\wl$ from
$cand\wl$ which satisfy
the correctness condition. From these,
the least-cost assignment is then selected as our next greedy
choice. This process is continued till no further reduction in
cost is possible and we have reached a local minimum.

This method can be used for any floating-point program
because it relies only on executing the program and does not
require any explicit modeling of the operations.
However, it only guarantees finding
a wordlength with a locally-minimum cost
and not globally-minimum cost.
}

\ifthenelse {\boolean{IsFinal}}
{
\small
\begin{algorithm}
\caption{Optimal Fixed-Point Types Synthesis: \optsynth{}}
\label{icse12:algo:synthMinCost}
\begin{algorithmic}
\REQUIRE
 Floating-point program $F_{fp}$, Fixed-point program $F_{fx}$ with fixed-point variables $T$,
Domain of inputs $Dom$, Error function $Err$, maximum error threshold
$\maxErr$, Cost Model $\fxcost$, max wordlengths $\wl_{max}$, Input $S$
\ENSURE Optimal wordlengths $\wl$ for inputs $S$ or $\bot$
\FORALL{fixed-point variable $t_i$ in $F_{fx}$}
\STATE $\iwl(t_i) = \lceil \log(\mathtt{getRange}(t_i, F_{fl},S)+1)  \rceil$, $\signbit(t_i) = \mathtt{isSigned}(t_i, F_{fl},S)$
\ENDFOR
\IF {$\wl_{max} < \iwl$}
\RETURN $\bot$
\ENDIF
\STATE $\fxtype = \langle \signbit, \iwl, \wl_{max} - \iwl \rangle$
\IF {$Err(F_{fp}(x), F_{fx}(x,\fxtype)) > \maxErr$}
\RETURN $\bot$
\ENDIF
\STATE $\wl = \mathtt{getMinCostWL}(F_{fp},F_{fx}, Dom, Err, \maxErr, fxcost, \wl_{max}, S^i,\iwl, \signbit)$
\RETURN $\fxtype = \langle \signbit, \iwl , \wl - \iwl   \rangle$
\end{algorithmic}
\end{algorithm}
\normalsize

\small
\begin{algorithm}
\caption{$\mathtt{getMinCostWL}$}
\label{icse12:algo:getMinCostWL}
\begin{algorithmic}
\REQUIRE
 Floating-point program $F_{fp}$, Fixed-point program $F_{fx}$ with fixed-point variables $T$,
Domain of inputs $Dom$, Error function $Err$, maximum error threshold
$maxErr$, Cost Model $\fxcost$, max wordlengths $\wl_{max}$, Input $S$
\ENSURE Optimal wordlengths $\wl$
\STATE $valcand\wl = \{\wl_{max}\}$
\WHILE{$valcand\wl$ is not empty}
\STATE $\wl = \underset{vc\wl \in valcand\wl}{\operatorname{argmin}}cost(vc\wl)$, $\fxtype = \langle \signbit, \iwl, \wl - \iwl\rangle$, $cand\wl = \emptyset$, $valcand\wl = \emptyset$
\FORALL{fixed-point variable $t_i$ in $F_{fx}$}
\STATE $\wl^{i-}(j) = \wl(j) \; \forall j \not = i$, $\wl^{i-}(i) = \wl(i)-1$, $\wl^{i+}(j) = \wl(j) \; \forall j \not = i$, $\wl^{i+}(i) = \wl(i)+1$, $cand\wl = cand\wl \cup \{\wl^{i-},\wl^{i+}\} $
\ENDFOR
\FORALL{$cand$ in $cand\wl$}
\STATE $cand\fxtype = \langle \signbit, \iwl, cand\wl - \iwl\rangle$
\IF {$Err(F_{fp}(x), F_{fx}(x,cand)) \leq maxErr \; \forall x \in S$\\
$\mathtt{and}\; \fxcost(cand\fxtype) < \fxcost(\fxtype)  $}
\STATE $valcand\wl = valcand\wl \cup \{cand\}$
\ENDIF
\ENDFOR
\ENDWHILE
\RETURN $\fxtype$
\end{algorithmic}
\end{algorithm}
\normalsize
}
{
\small
\begin{algorithm}
\caption{Optimal Fixed-Point Types Synthesis: \optsynth{}}
\label{icse12:algo:synthMinCost}
\begin{algorithmic}
\REQUIRE
 Floating-point program $F_{fp}$,\\ Fixed-point program $F_{fx}$ with fixed-point variables $T$,
Domain of inputs $Dom$, Error function $Err$,\\ maximum error threshold
$\maxErr$, Cost Model $\fxcost$,\\ max wordlengths $\wl_{max}$, Input $S$
\ENSURE Optimal wordlengths $\wl$ for inputs $S$ or $\bot$
\FORALL{fixed-point variable $t_i$ in $F_{fx}$}
\STATE $\iwl(t_i) = \lceil \log(\mathtt{getRange}(t_i, F_{fl},S)+1)  \rceil$
\STATE $\signbit(t_i) = \mathtt{isSigned}(t_i, F_{fl},S)$
\ENDFOR
\IF {$\wl_{max} < \iwl$}
\RETURN $\bot$
\ENDIF
\STATE $\fxtype = \langle \signbit, \iwl, \wl_{max} - \iwl \rangle$
\IF {$Err(F_{fp}(x), F_{fx}(x,\fxtype)) > \maxErr$}
\RETURN $\bot$
\ENDIF
\STATE $\wl = \mathtt{getMinCostWL}(F_{fp},F_{fx}, Dom, Err, \maxErr,$\\$\;\;\;\;\;\;\;\;\;\;\;\;\;\;\;\;\;\;\;\;\;\;\;\;\;\;\;\;\;\;\;\;\fxcost, \wl_{max}, S^i,\iwl, \signbit)$
\RETURN $\fxtype = \langle \signbit, \iwl , \wl - \iwl   \rangle$
\end{algorithmic}
\end{algorithm}
\normalsize

\small
\begin{algorithm}
\caption{$\mathtt{getMinCostWL}$}
\label{icse12:algo:getMinCostWL}
\begin{algorithmic}
\REQUIRE
 Floating-point program $F_{fp}$,\\ Fixed-point program $F_{fx}$ with fixed-point variables $T$,\\
Domain of inputs $Dom$, Error function $Err$,\\ maximum error threshold
$maxErr$, Cost Model $\fxcost$,\\ max wordlengths $\wl_{max}$, Input $S$
\ENSURE Optimal wordlengths $\wl$
\STATE $valcand\wl = \{\wl_{max}\}$
\WHILE{$valcand\wl$ is not empty}
\STATE $\wl = \underset{vc\wl \in valcand\wl}{\operatorname{argmin}}cost(vc\wl)$
\STATE $\fxtype = \langle \signbit, \iwl, \wl - \iwl\rangle$
\STATE  $cand\wl = \emptyset$, $valcand\wl = \emptyset$
\FORALL{fixed-point variable $t_i$ in $F_{fx}$}
\STATE $\wl^{i-}(j) = \wl(j) \; \forall j \not = i$, $\wl^{i-}(i) = \wl(i)-1$
\STATE $\wl^{i+}(j) = \wl(j) \; \forall j \not = i$, $\wl^{i+}(i) = \wl(i)+1$
\STATE $cand\wl = cand\wl \cup \{\wl^{i-},\wl^{i+}\} $
\ENDFOR
\FORALL{$cand$ in $cand\wl$}
\STATE $cand\fxtype = \langle \signbit, \iwl, cand\wl - \iwl\rangle$
\IF {$Err(F_{fp}(x), F_{fx}(x,cand)) \leq maxErr \; \forall x \in S$\\
$\mathtt{and}\; \fxcost(cand\fxtype) < \fxcost(\fxtype)  $}
\STATE $valcand\wl = valcand\wl \cup \{cand\}$
\ENDIF
\ENDFOR
\ENDWHILE
\RETURN $\fxtype$
\end{algorithmic}
\end{algorithm}
\normalsize
}

\subsection{Verifying a Candidate Fixed-Point Program}

In order to verify that the fixed-point program $F_{fx}(X,\fxtype)$
satisfies the correctness condition, we need to check if the
following logical formula is satisfiable.
\small
\begin{align} \label{fxverify}
\exists X \in Dom(X) \;\; Err(F_{fx}(X,\fxtype), F_{fp}(X)) > \maxErr
\end{align}
\normalsize
If the formula is unsatisfiable, there is no input on which
the fixed-point program violates the correctness condition.

For arbitrary (possibly non-linear)
floating-point and fixed-point
arithmetic operations, it is extremely difficult to solve such a problem
in practice with current constraint solvers.
Instead, we use a novel optimization-based approach to verify the
candidate fixed-point program.
The intuition behind using an optimization-based approach is that the
error function is continuous in the inputs
or with very few discontinuities~\cite{majumdar-rtss09,chaudhuri-popl10},
and hence, optimization routines can easily find inputs which maximize
error function by starting from some random input and gradually adjusting
the output to increase the value of the error function.
The optimization oracle $\optV$
is used to maximize the error function $Err(F_{fx}(X,\fxtype), F_{fp}(X))$
over the domain $Dom(X)$.
If there is no input $X \in Dom(X)$ for which the error
function exceeds $\maxErr$,
the fixed-point program is correct and we terminate.
Otherwise, we obtain an example input on which the fixed-point program
violates the correctness condition. Multiple inputs can also be generated
where they exist.

In practice, with the current state-of-the-art optimization routines,
it is difficult to implement $\optV$ to find a global optimum.
Instead, we use a numerical optimization routine based on the
Nelder-Mead method~\cite{nelder-compj65} which can handle arbitrary
non-linear functions and generates local optima.
Procedure~\ref{icse12:algo:verify} defines $\verifyerrm$ which
invokes the Nelder-Mead routine (indicated by ``${\operatorname{argmaxlocal}}$'').
This routine requires
one to supply a starting value of $X$, which we generate randomly.
To find multiple inputs, we invoke the routine from
from different random initial points and record all example inputs on which
the fixed-point program violates the correctness condition. Since a global
optimum is not guaranteed, we repeat this search
$\maxatt$ times before declaring that the fixed-point program is correct.
%
\ifthenelse {\boolean{IsFinal}}
{
\small
\begin{algorithm}
\caption{Verification Routine \verifyerr{}}
\label{icse12:algo:verify}
\begin{algorithmic}
\REQUIRE
 Floating-point program $F_{fp}$, Fixed-point program $F_{fx}$,
Fixed-point type $\fxtype$, Domain of inputs $Dom$, Error function $Err$, maximum error threshold
$\maxErr$
\ENSURE Inputs $Bad$ on which $F_{fx}$ violates correctness condition
\STATE $Bad = \emptyset$
\WHILE {$i \leq \maxatt$}
\STATE $i=i+1$, $X_0 =$ random sample from $Dom$, $X_{cand} = \underset{X}{\operatorname{argmaxlocal}}(Err(F_{fp}(X),F_{fx}(X,\fxtype)), X_0)$
\IF {$Err(F_{fp}(X_{cand}), F_{fx}(X_{cand},\fxtype)) > \maxErr$ and $X \in Dom$}
\STATE $Bad = Bad \cup \{X\}$
\ENDIF
\ENDWHILE
\end{algorithmic}
\end{algorithm}
\normalsize
}
{
\small
\begin{algorithm}
\caption{Verification Routine \verifyerr{}}
\label{icse12:algo:verify}
\begin{algorithmic}
\REQUIRE
 Floating-point program $F_{fp}$,\\ Fixed-point program $F_{fx}$,
Fixed-point type $\fxtype$, \\Domain of inputs $Dom$, Error function $Err$, \\maximum error threshold
$\maxErr$
\ENSURE Inputs $Bad$ on which $F_{fx}$ violates correctness condition
\STATE $Bad = \emptyset$
\WHILE {$i \leq \maxatt$}
\STATE $i=i+1$, $X_0 =$ random sample from $Dom$
\STATE $X_{cand} = \underset{X}{\operatorname{argmaxlocal}}(Err(F_{fp}(X),F_{fx}(X,\fxtype)), X_0)$
\IF {$Err(F_{fp}(X_{cand}), F_{fx}(X_{cand},\fxtype)) > \maxErr$ and $X \in Dom$}
\STATE $Bad = Bad \cup \{X\}$
\ENDIF
\ENDWHILE
\end{algorithmic}
\end{algorithm}
\normalsize
}
\ifthenelse {\boolean{IsFinal}}
{}
{
\subsection{Illustration on Running Example} \label{icse12:sec:example}

\begin{figure}[ht]
\centering
\includegraphics[width=4in]{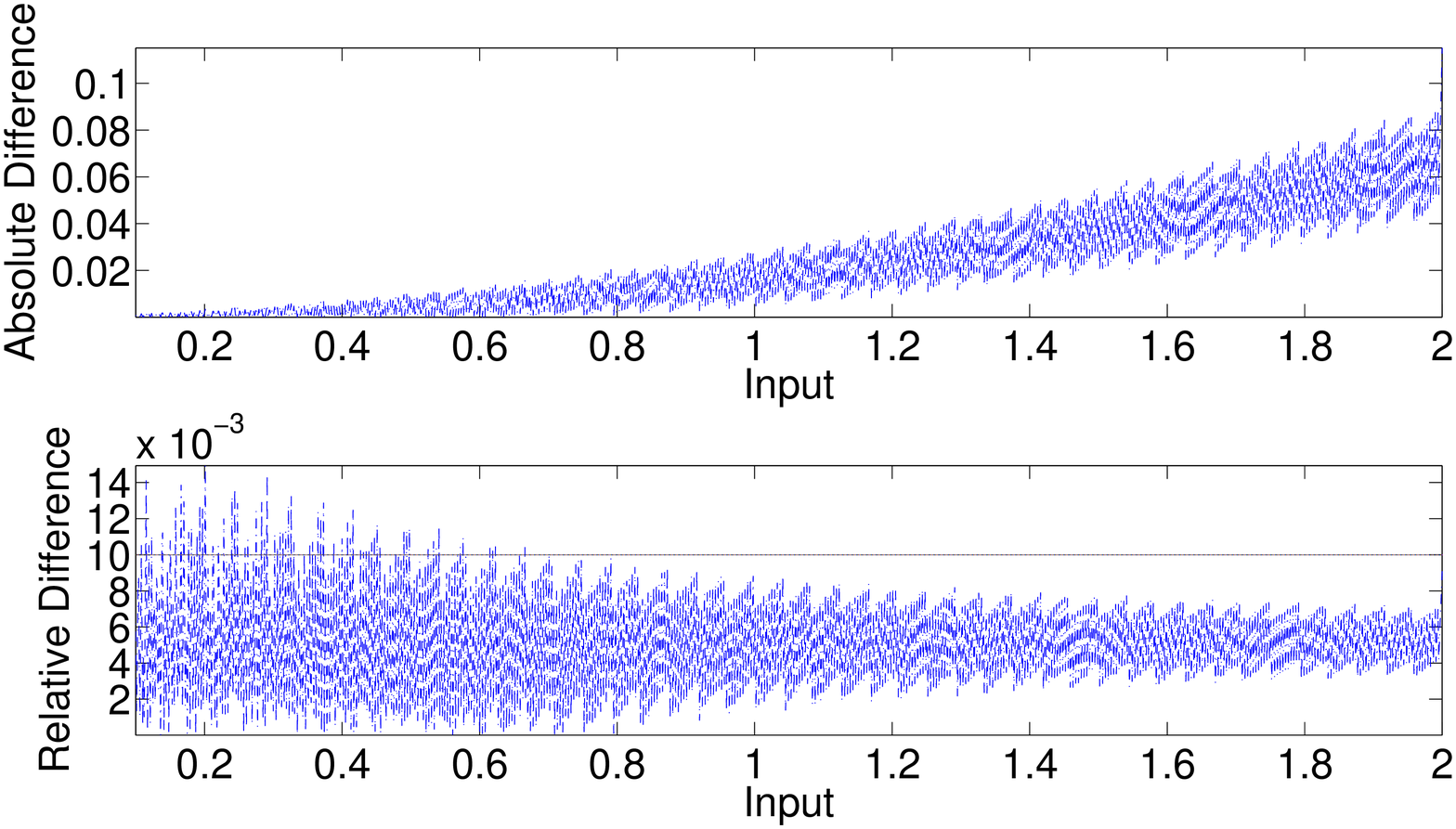}
\caption{Iteration 2}
\label{icse12:fig:approach}
\end{figure}

\begin{figure}[ht]
\centering
\includegraphics[width=4in]{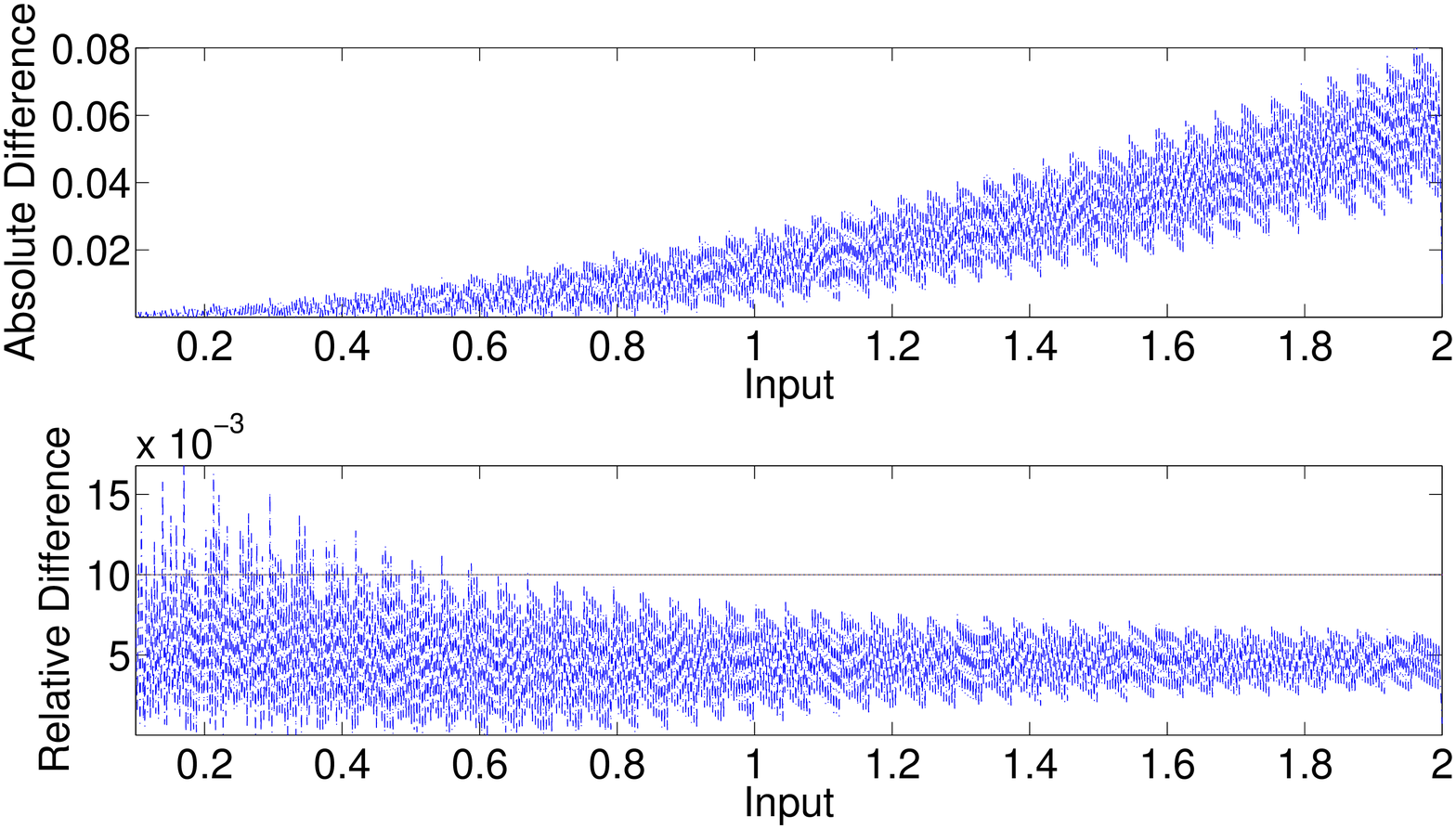}
\caption{Iteration 3}
\label{icse12:fig:approach1}
\end{figure}

\begin{figure}[ht]
\centering
\includegraphics[width=4in]{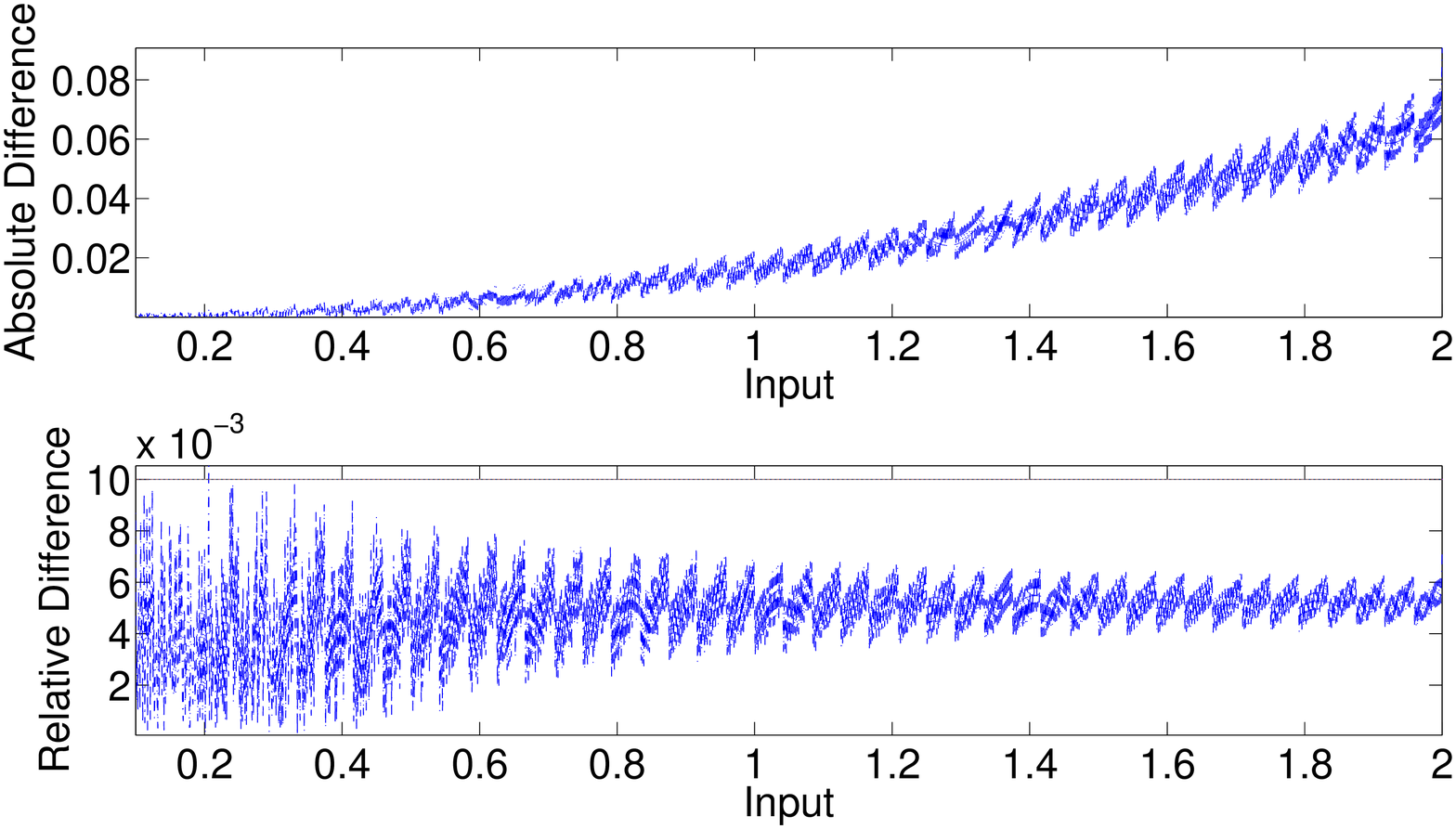}
\caption{Iteration 4 (last)}
\label{icse12:fig:approach2}
\end{figure}

The initial sample $S^0$ is of size $10$.
$\maxatt$ in the verification routine was also set to $10$.
The number of
Our algorithm took 4 iterations. We record the intermediate
implementations produced by $synthMinCost$ in
each of the last 3 iterations taken by our algorithm.
The initial selected random sample is of size
$10$ and the $MAXATTEMPTS$ was set to $10$.
The number of samples used in each subsequent step of iteration
after adding examples discovered by \verifyerr{}
procedure was $18, 22$ and $34$.
Figure~\ref{icse12:fig:approach},\ref{icse12:fig:approach1} and \ref{icse12:fig:approach2} illustrates
the error using the intermediate implementations and the final
implementation produced by our approach.
The wordlengths in Figure~\ref{icse12:fig:approach}
are $mypi(2,3)$, $radius(1,8)$, $t(2,11)$, $area(4,11)$,
those in Figure~\ref{icse12:fig:approach1}
are $mypi(2,5)$, $radius(1,8)$, $t(2,11)$, $area(4,12)$
and in Figure~\ref{icse12:fig:approach2}
are  $mypi(2,3)$, $radius(1,9)$, $t(2,11)$, $area(4,10)$.
We observe that the
number of inputs violating correctness constraint reduces after
each iteration. This illustrate how our algorithm works by identifying
a few representative inputs violating correctness condition
in each iteration and adding that to the set of examples for which we
synthesize the least cost implementation.
} 

\ifthenelse {\boolean{IsFinal}}
{}
{\subsection{Theoretical Results}}

The following theorem summarizes the correctness and optimality guarantees
of our approach.
\ifthenelse {\boolean{IsFinal}}
{Proof is presented in extended version~\cite{extended}.}
{}
\begin{theorem}
The synthesis procedure presented in Procedure~\ref{icse12:algo:floatfix}
is guaranteed to
synthesize the fixed-point program which is of minimal cost and satisfies
the correctness condition for accuracy if optimization oracles
$\optS$ and $\optV$ find globally-optimal solutions (when they exist).
\noop{
\begin{itemize}
\item the optimization engine $\optS$ finds
the global optimum of the problem in Equation~\ref{fxopt}, and
\item the verification engine is a decision procedure for
the satisfiability problem in Equation~\ref{fxverify}.
\end{itemize}
}
\end{theorem}

\ifthenelse {\boolean{IsFinal}}
{}
{
\begin{proof}
We first prove correctness and then optimality of the obtained
solution. Consider Equation~\ref{fxverify}:
$$\exists X \in Dom(X) \;\; Err(F_{fx}(X,\fxtype), F_{fp}(X)) > \maxErr$$
If $\optV$ finds globally-optimal solutions, it will find the $X \in Dom(X)$
that maximizes $Err(F_{fx}(X,\fxtype), F_{fp}(X))$, and hence determine
where or not Equation~\ref{fxverify} is satisfiable.
Thus, the correctness condition for accuracy is satisfied.

Next, let $\fxtype^*$ ($\neq \bot$) be the fixed-point type returned by
Procedure~\ref{icse12:algo:floatfix}. Let us assume that there
exists a fixed-point type $\fxtype'$  with a cost lower than
$\fxtype^*$ which also satisfies the correctness condition:
\begin{eqnarray*}
& \forall X \in Dom(X) \;\; Err(F_{fx}(X,\fxtype'), F_{fp}(X)) \leq \maxErr \\
\land & \fxcost(\fxtype') < \fxcost(\fxtype^*)
\end{eqnarray*}
Hence, for any $D \subseteq Dom(X)$,
\begin{eqnarray*}
& \forall X \in D \;\; Err(F_{fx}(X,\fxtype'), F_{fp}(X)) \leq \maxErr \\
\land & \fxcost(\fxtype') < \fxcost(\fxtype^*)
\end{eqnarray*}
But $\fxtype^*$ is the solution generated by applying $\optS$ to the optimization
problem of Equation~\ref{fxopt}:
\begin{align}
\mathtt{Minimize\;\;} \fxcost(\fxtype) \mathtt{\;\; s.t.} \nonumber \\
\displaystyle \bigwedge_{x \in S} Err(F_{fx}(x,\fxtype), F_{fl}(x)) \leq \maxErr
\end{align}
Since $\optS$ is guaranteed to generate globally-optimal solutions, setting
$D = S$, we obtain a contradiction. Hence, there
exists no fixed-point type $\fxtype'$  with a cost lower than
$\fxtype^*$ which also satisfies the correctness condition. Hence,
$\fxtype^*$ is the optimal correct solution.
\end{proof}
As noted earlier, it is difficult to implement ideal $\optS$ and $\optV$
(that find global optima) with current SAT and optimization methods for
arbitrary floating-point programs. Nonetheless, our experience with
heuristic methods that find local optima has been very good.
Also, improvements in optimization/SAT methods
can directly be leveraged with our inductive synthesis approach.
In contrast, the current techniques for synthesizing fixed-point versions
of floating-point programs perform heuristic optimization over a
randomly selected set of inputs (see Sec.~\ref{pldi12:sec:related} for
a detailed discussion).
Such techniques do not provide any correctness
guarantees and the number of inputs needed could be much larger.
Our
approach systematically discovers a small number of example
inputs such that the optimal fixed-point program for this set
yields that for the entire input domain.
}

\noop{
In practice,
\begin{itemize}
\item Optimality of the obtained solution depends on the
used optimization routine $\mathtt{getMinCostWL}$. Using a better optimization
engine could yield fixed-point program with lower costs. We use a greedy
heuristic search method which only guarantees finding a locally optimum
solution.
\item Correctness of the obtained solution with respect to accuracy
constraint depends on the testing routine \verifyerr{}.
Using a better testing routine improves the correctness guarantee
of the obtained fixed-point program.
We use numerical optimization routine which is
repeated with a number of initial points for
verification. Its accuracy can be improved by increasing the number
of initial points.
\end{itemize}

An alternative naive approach would be to randomly select a large number
of sample inputs and then use the  \optsynth{} routine
to synthesize a fixed-point program which is correct for all these inputs
and is of minimal cost. Such an approach would not have any correctness
guarantees and the number of inputs needed could be very large
as illustrated in Section~\ref{icse12:sec:example}. We use
our
approach to discover a small number of example
inputs which would induce the correct fixed-point program.
}

\section{Experiments}
\label{pldi12:sec:exp}
\label{icse12:sec:exp}

Apart from the running example,
we present case studies from DSP and control systems to illustrate
the utility of the presented synthesis approach.
Our technique was implemented in Matlab, and
Nelder-Mead implementation available in Matlab as
$\mathtt{fminsearch}$ function was used for numerical optimization.
We use the Constantinides et al~\cite{constantinides-FCCM02}
cost model.

\subsection{Running Example}
We illustrate the synthesis approach (more details in~\cite{extended}) presented in
Section~\ref{icse12:sec:approach} using the running example.
Our algorithm used $34$ examples and needed $4$ iterations.
To evaluate our approach, we exhaustively simulated the
generated fixed-point program
on the given domain ($0.1 \leq \mathtt{radius} < 2$)
at intervals of $0.0001$.
The is presented in Figure~\ref{fig:fxex1single}.

\ifthenelse {\boolean{IsFinal}}
{
\begin{minipage}[t]{\textwidth}
\noindent \begin{minipage}[t]{0.5\textwidth}
\begin{figure}[H]
\includegraphics[width=2.5in]{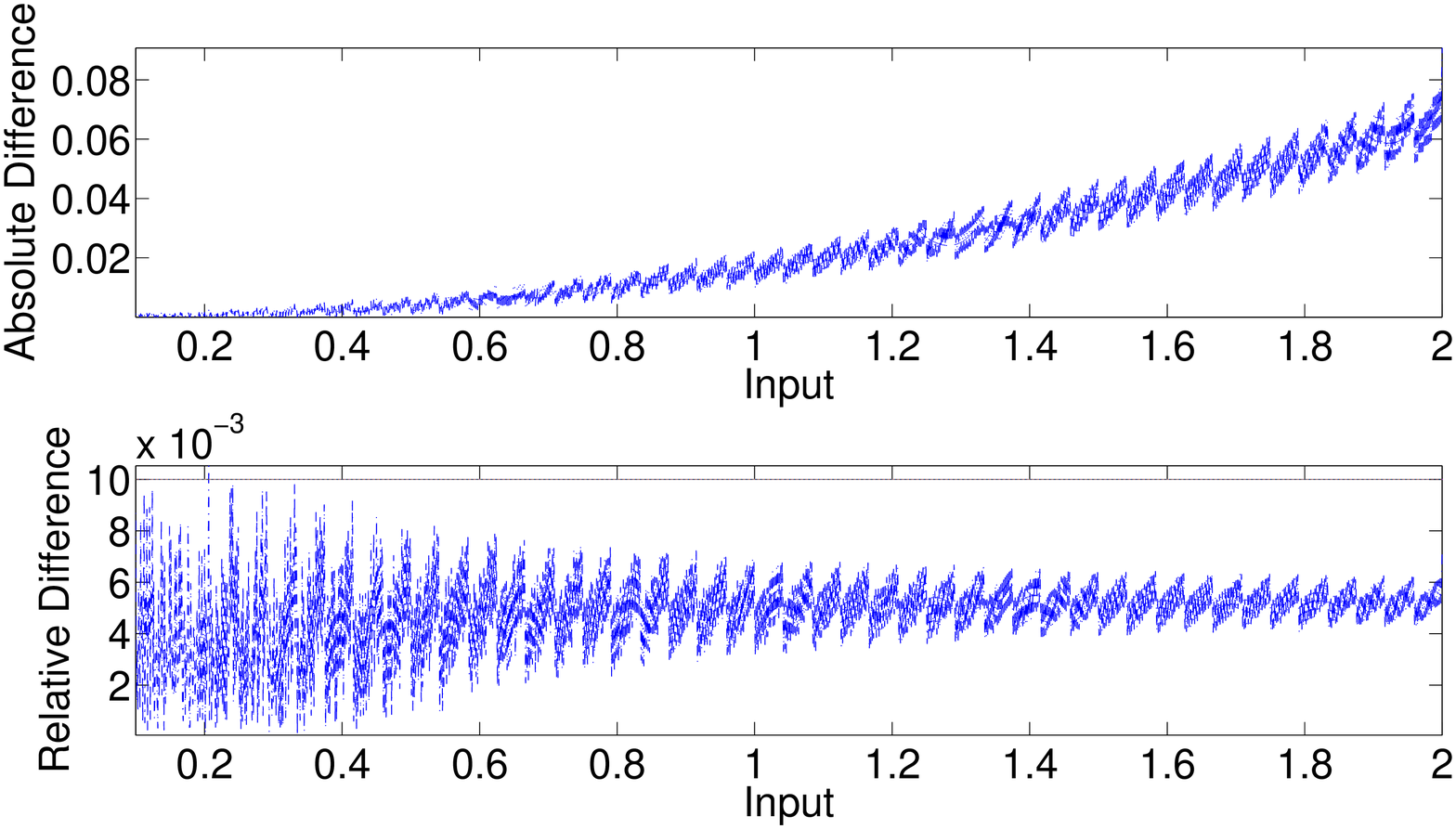}
\label{fig:fxex1single}
\caption{Our Approach on Running Example}
\end{figure}
\end{minipage}
\noindent \begin{minipage}[t]{0.5\textwidth}
\begin{figure}[H]
\includegraphics[width=2.5in] {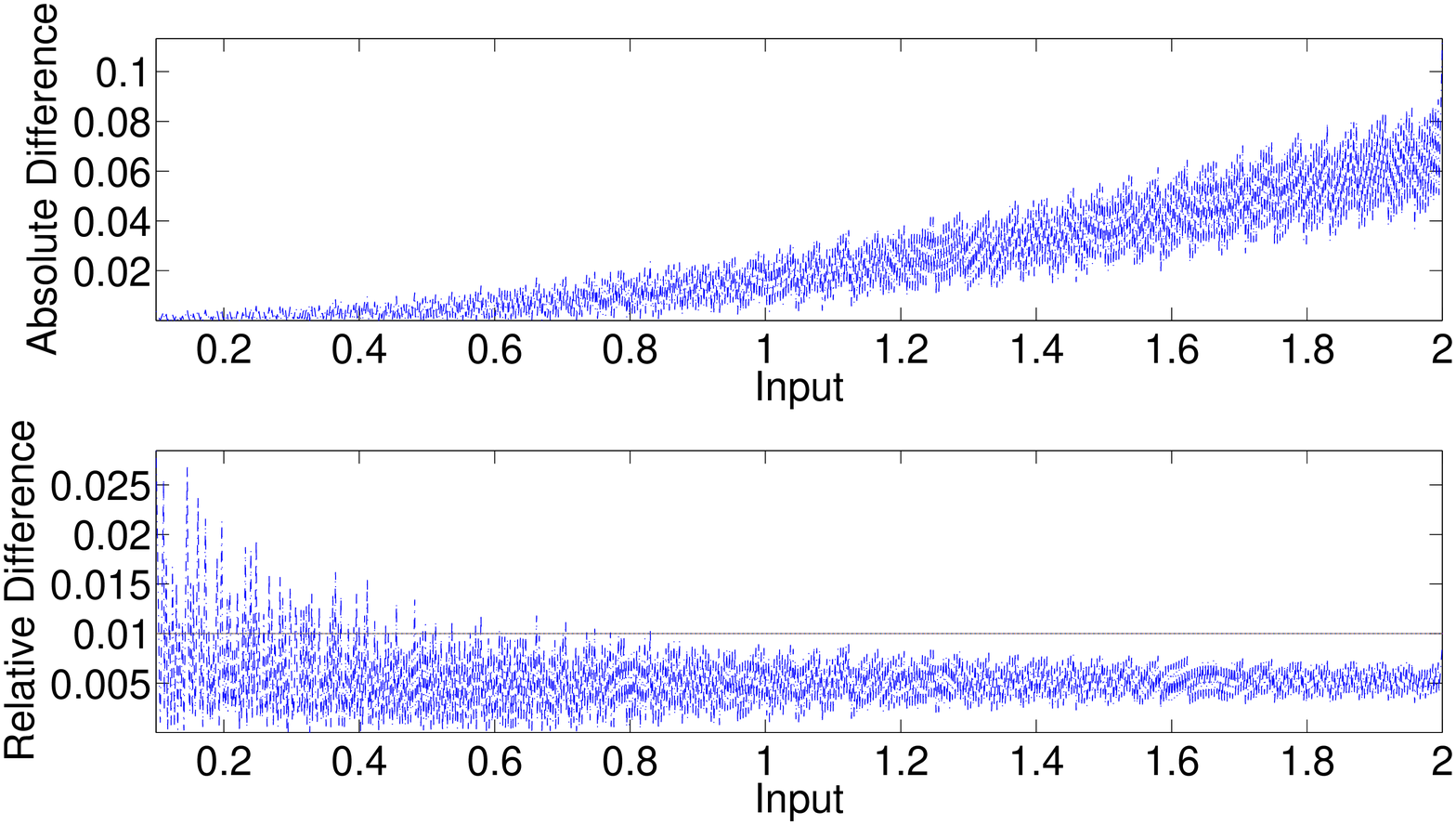}
\label{fig:fxex2}
\caption{Running Example Using Random Inputs.}
\end{figure}
\end{minipage}
\end{minipage}
}
{
\begin{figure}[!ht]
\centering
\includegraphics[width=3in]{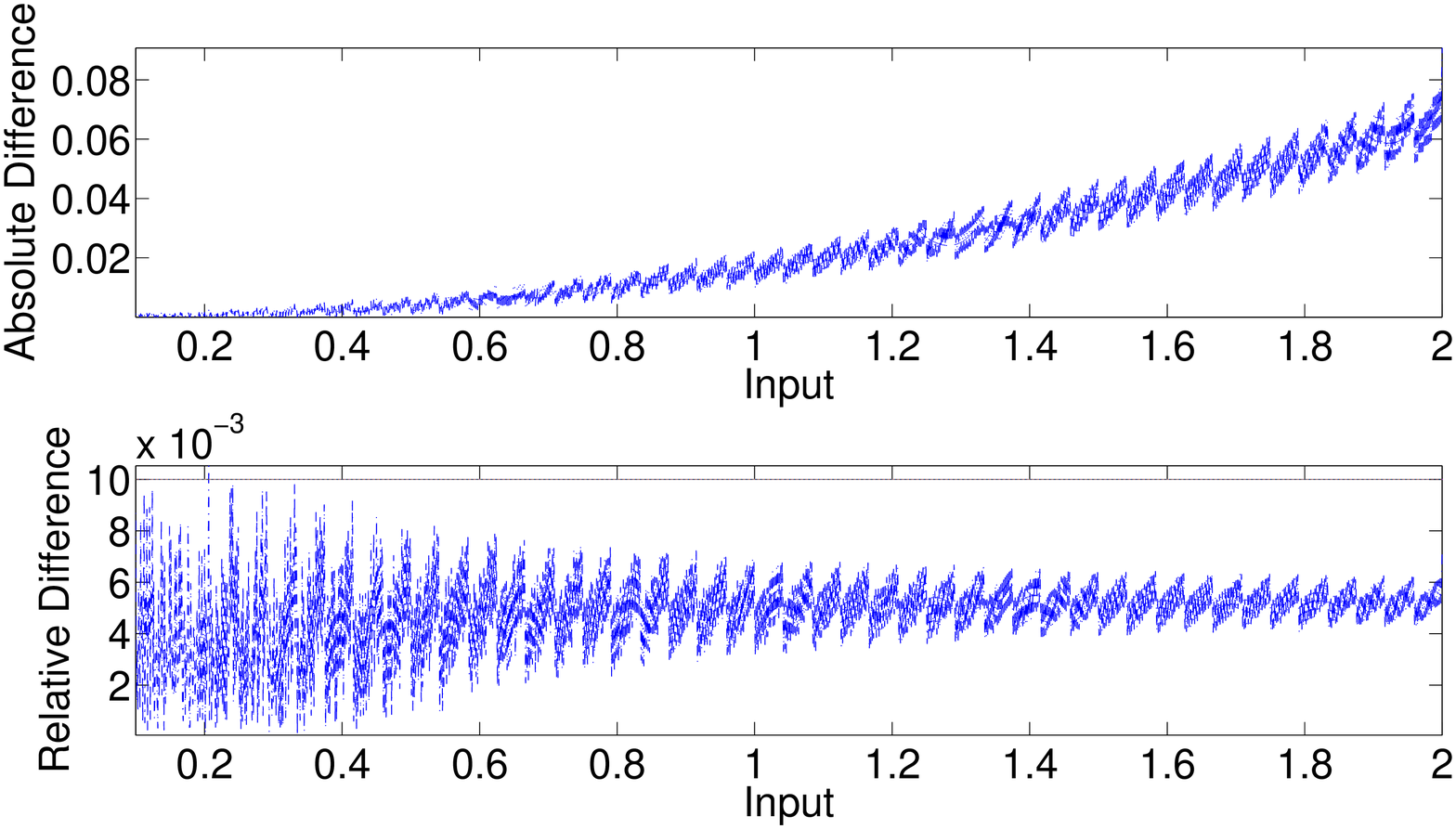}
\label{fig:fxex1single}
\caption{Our Approach on Running Example. }
\end{figure}

\begin{figure}[!ht]
\centering
\includegraphics[width=3in] {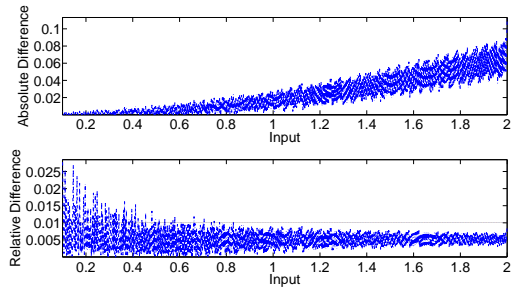}
\label{fig:fxex2}
\caption{Running Example Using Random Inputs. }
\end{figure}
}

As a point of comparison, we also show the result of synthesizing a
fixed-point program
using the \optsynth{} routine with $100$ inputs ($3$ times as many
as our approach) selected uniformly at random (Figure~\ref{fig:fxex2}).
The horizontal line in the plots denotes the maximum
error threshold of $0.01$ on the relative difference error function.
The cost of the fixed-point program synthesized with random sampling is
$89.65$, and the
fixed-point types of the variables are
$\fxtype(\mathtt{mypi}) = \langle 0,2,3 \rangle$,
$\fxtype(\mathtt{radius}) = \langle 0,1,8 \rangle$,
$\fxtype(\mathtt{t}) = \langle 0, 2, 10\rangle$ and
$\fxtype(\mathtt{area}) = \langle 0, 4, 8 \rangle$.
Notice, however, that it is incorrect for a large number of inputs.
In contrast, the cost of the implementation
produced using our technique is $104.65$, and
the fixed-point types of the variables are
$\fxtype(\mathtt{mypi}) = \langle 0,2,3 \rangle$,
$\fxtype(\mathtt{radius}) = \langle 0,1,9 \rangle$,
$\fxtype(\mathtt{t}) = \langle 0, 2, 11\rangle$ and
$\fxtype(\mathtt{area}) = \langle 0, 4, 10 \rangle$.

\ifthenelse {\boolean{IsFinal}}
{
\subsection{Infinite Impulse Response (IIR) Filter}

The first case study is a
first-order direct form-II IIR filter (see extended version~\cite{extended} for details).
We use our synthesis technique to discover the appropriate fixed-point
types of the coefficients of the filter.
The input domain used in synthesis is $-2 < input < 2$.
The correctness condition for accuracy is to ensure that the relative
error between the floating-point and fixed-point program is less
than $0.1$.

\begin{minipage}[t]{\textwidth}
\begin{minipage}[t]{0.5\textwidth}
\begin{figure}[H]
\centering
\includegraphics[width=2.5in]{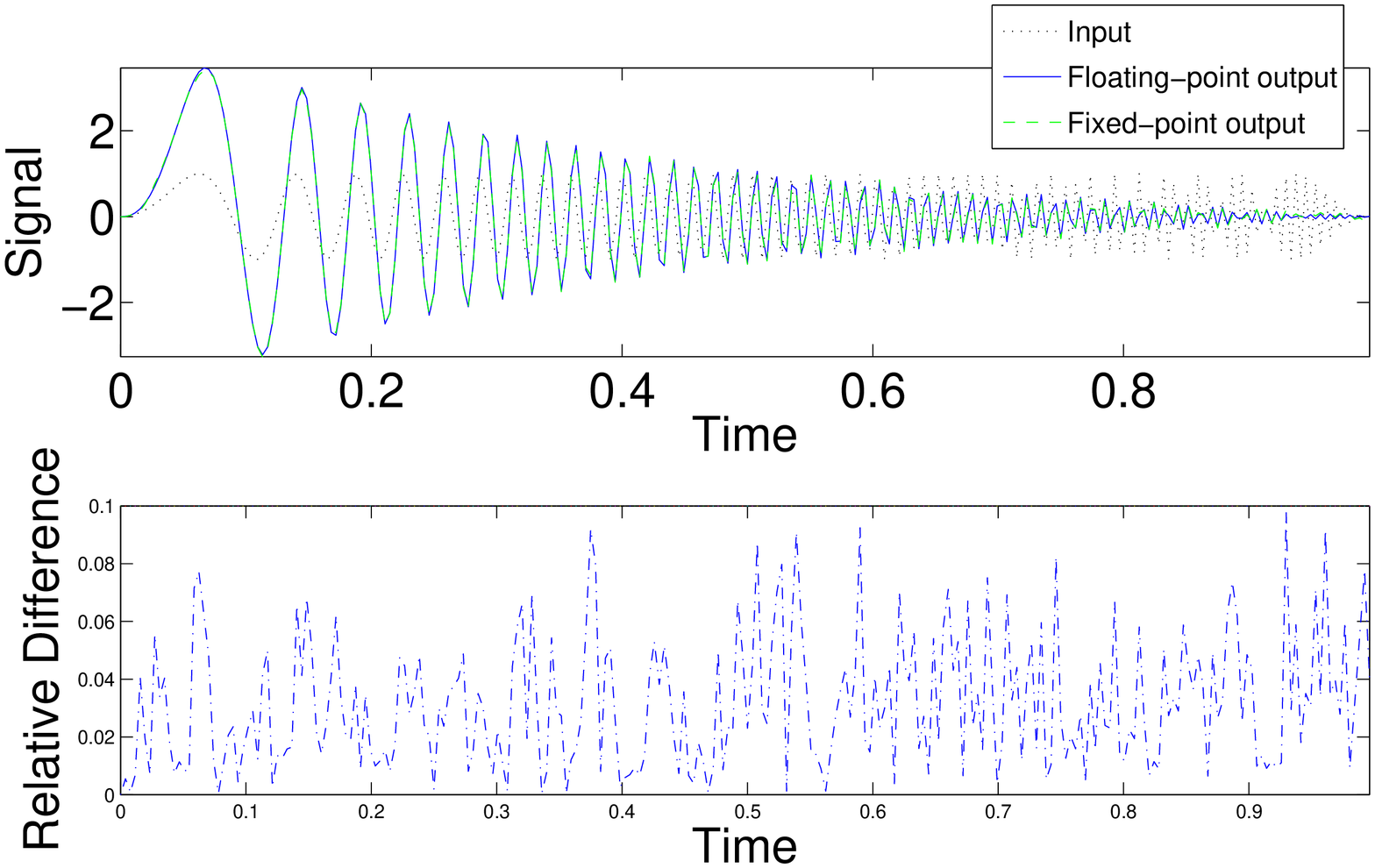}
\caption{IIR Filter}
\label{fig:fx:iirsim}
\end{figure}
\end{minipage}
\begin{minipage}[t]{0.5\textwidth}
\begin{figure}[H]
\centering
\includegraphics[width=2.5in]{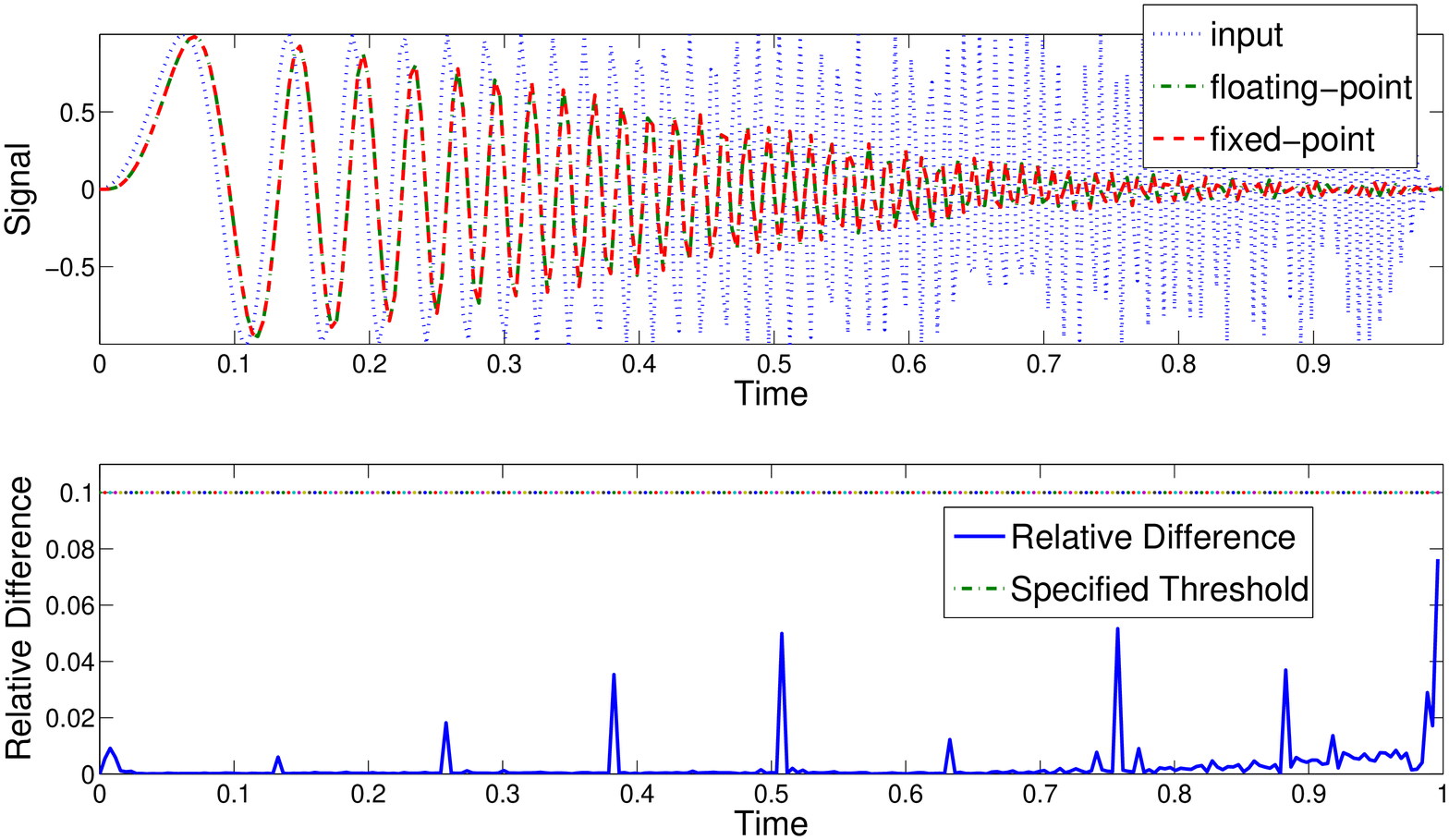}
\caption{FIR Filter}
\label{fig:fx:firsim}
\end{figure}
\end{minipage}
\end{minipage}

In order to test the correctness of our implementation,
we feed a common input signal to both the IIR filter implementations:
floating-point version and the fixed-point version obtained by our
synthesis technique.
The input signal is a linear chirp from $0$ to $\frac{Fs}{2}$ Hz in $1$ second.
$$input = (1-2^{-15}) \times sin(\pi \times \frac{Fs}{2}\times t^2)$$
where $Fs = 256$  and $t=0$ to $1-\frac{1}{Fs}$ and is
sampled at intervals of $\frac{1}{Fs}$.
Figure~\ref{fig:fx:iirsim} shows the
input,
outputs of both implementations and the relative error
between the two outputs. We observe that the implementation
satisfies the correctness condition and the relative error
remains below $0.1$ throughout the simulation.

\subsection{Finite Impulse Response (FIR) Filter}

The second case study is a low pass FIR filter of order 4
with tap coefficients $0.0346,0.2405, 0.4499, 0.2405$ and $0.0346$.
The input domain, correctness condition and input signal to test the
floating-point implementation and synthesized fixed-point program
are same as the previous case study.
Figure~\ref{fig:fx:firsim} shows the
input,
outputs of both implementations and the relative error
between the two outputs. We observe that the implementation
satisfies the correctness condition and the relative error
remains below $0.1$ throughout the simulation.

\subsection{Field Controlled DC Motor}

The next case study is a field controlled DC Motor.
It is a classic non-linear control example from
Khalil~\cite{khalil-book}.
A detailed discussion of this example is presented in
the extended version~\cite{extended}. The goal in this work
was to find an optimal fixedpoint implementation of
the control law computed mathematically for DC motor.
The computed control law can be mathematically shown to be
correct by designers who are more comfortable in reasoning with
real arithmetic but not with finite precision arithmetic.
Its implementation using floating-point computation
also closely mimics the arithmetic in reals
 but the control algorithms are often implemented using
fixed-point computation on embedded platforms.
We use our synthesis technique to automatically
derive a low cost fixed-point implementation of the control law
computing input $u$.
The input domain is $0 \leq i_a,i_f,\omega \leq 1.5$
where $i_a$ is armature current and $i_f$ is field current.
The correctness condition for accuracy is that the absolute
difference between the control input $u$ computed by fixed-point program and the
floating-point program is less than $0.1$.

Figure~\ref{fig:fx:dcmotorfx} shows the simulation of the system
using the fixed-point implementation of the controller
and the floating-point implementation.
This end-to-end
simulation shows that fixed-point program generated by our technique
can be used to control the system as effectively as the floating-point program.
This illustrates the practical utility of our technique.
Figure~\ref{fig:fx:motorerror} plots the difference between
the control input computed by the fixed-point program and the floating-point
program. It
shows that the fixed-point types
synthesized using our approach
satisfy the correctness condition, and the
difference between the control input computed by
the fixed-point and floating-point program is within
the specified maximum error threshold of $0.1$.
The number of inputs needed in our approach was $127$.
In contrast, the fixed-point types found using $635$(5X our approach)
randomly selected inputs violate
the correctness condition for a large number of inputs.

\begin{minipage}[t]{\textwidth}
\begin{minipage}[t]{0.5\textwidth}
\begin{figure}[H]
\centering
\includegraphics[width=2.7in]{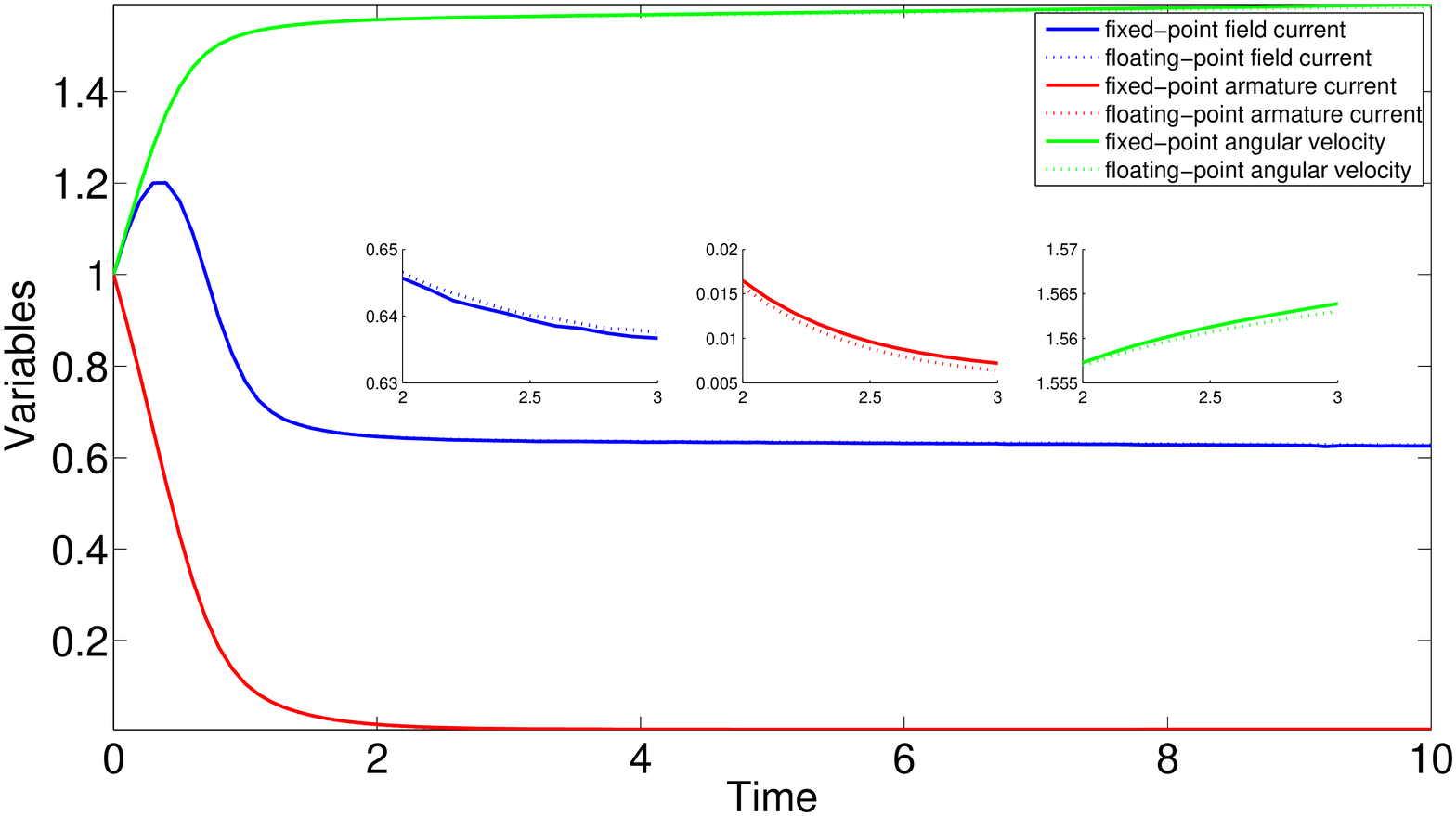}
\caption{DC Motor Using Floating-point and Fixed-point Controller. Fixedpoint and floatingpoint simulations almost overlap}
\label{fig:fx:dcmotorfx}
\end{figure}
\end{minipage}
\begin{minipage}[t]{0.5\textwidth}
\begin{figure}[H]
\centering
\includegraphics[width=2.3in]{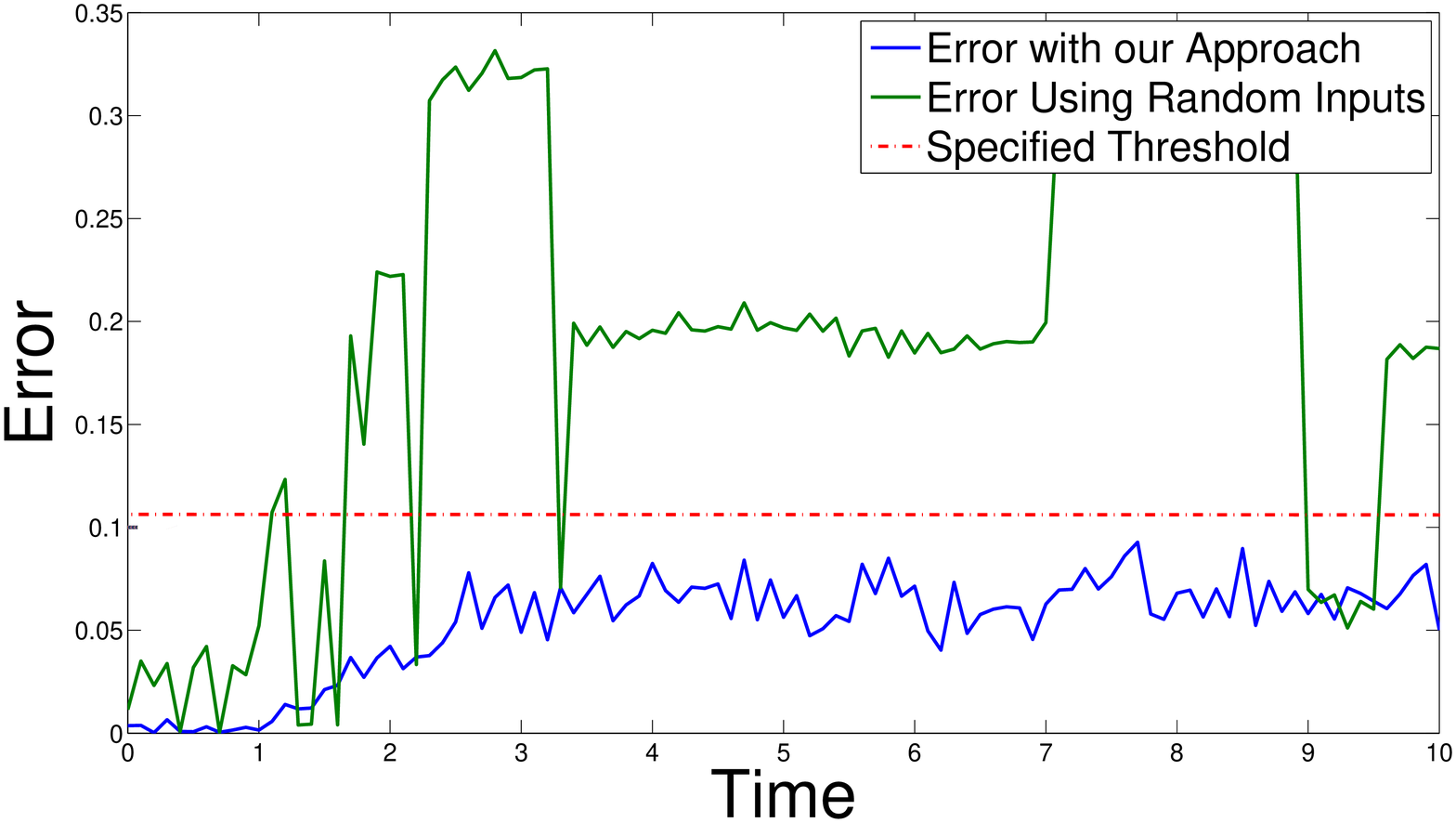}
\caption{DC Motor Error}
\label{fig:fx:motorerror}
\end{figure}
\end{minipage}
\end{minipage}

\noop{
\begin{minipage}[t]{\textwidth}
\noindent \begin{minipage}[t]{0.5\textwidth}
\begin{figure}[H]
\includegraphics[width=2.5in]{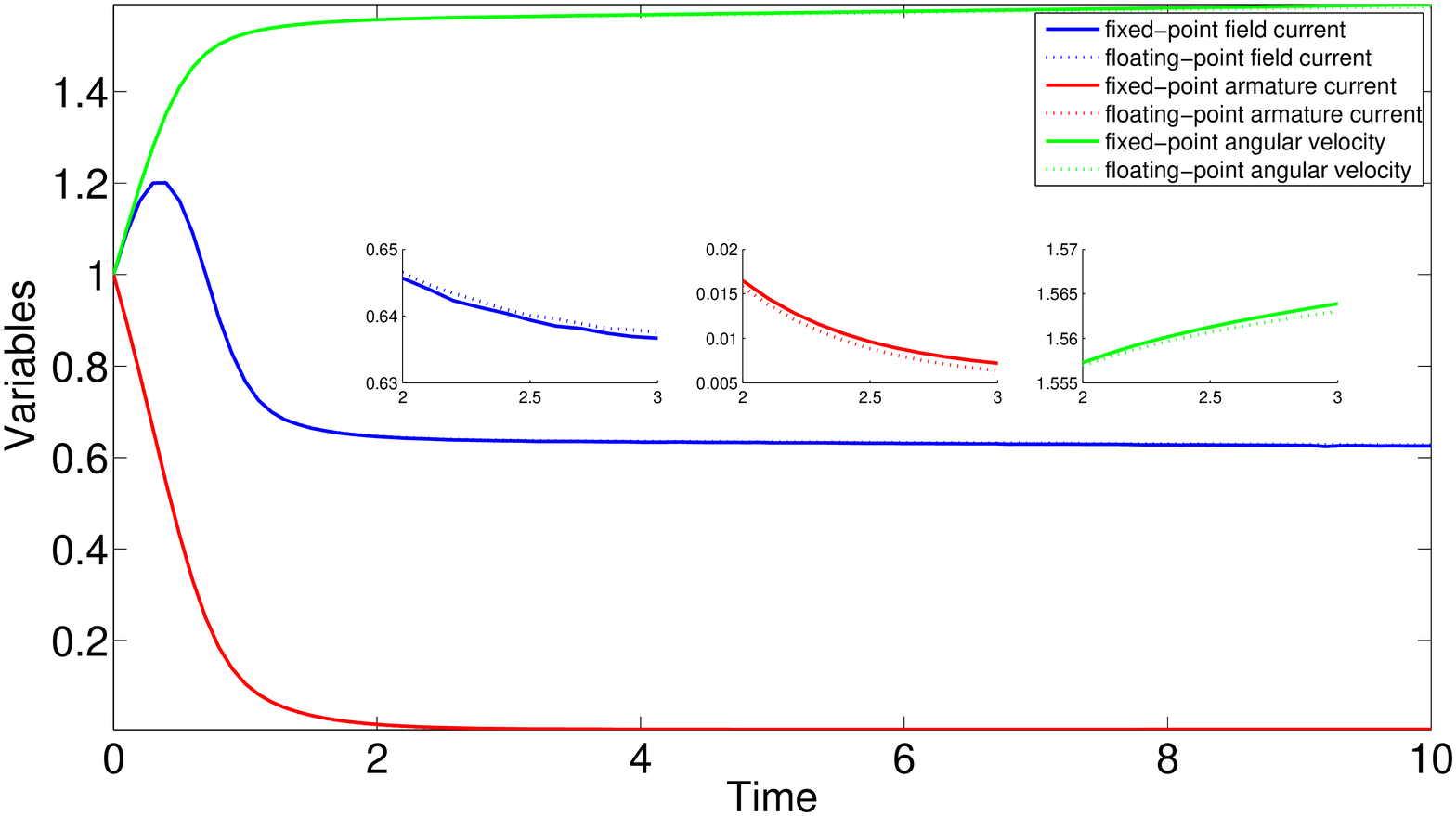}
\caption{DC Motor Using Floating-point and Fixed-point Controller}
\label{fig:fx:dcmotorfx}
\end{figure}
\end{minipage}[t]{0.5\textwidth}
\noindent \begin{minipage}[t]{0.5\textwidth}
\begin{figure}[H]
\includegraphics[width=2.5in]{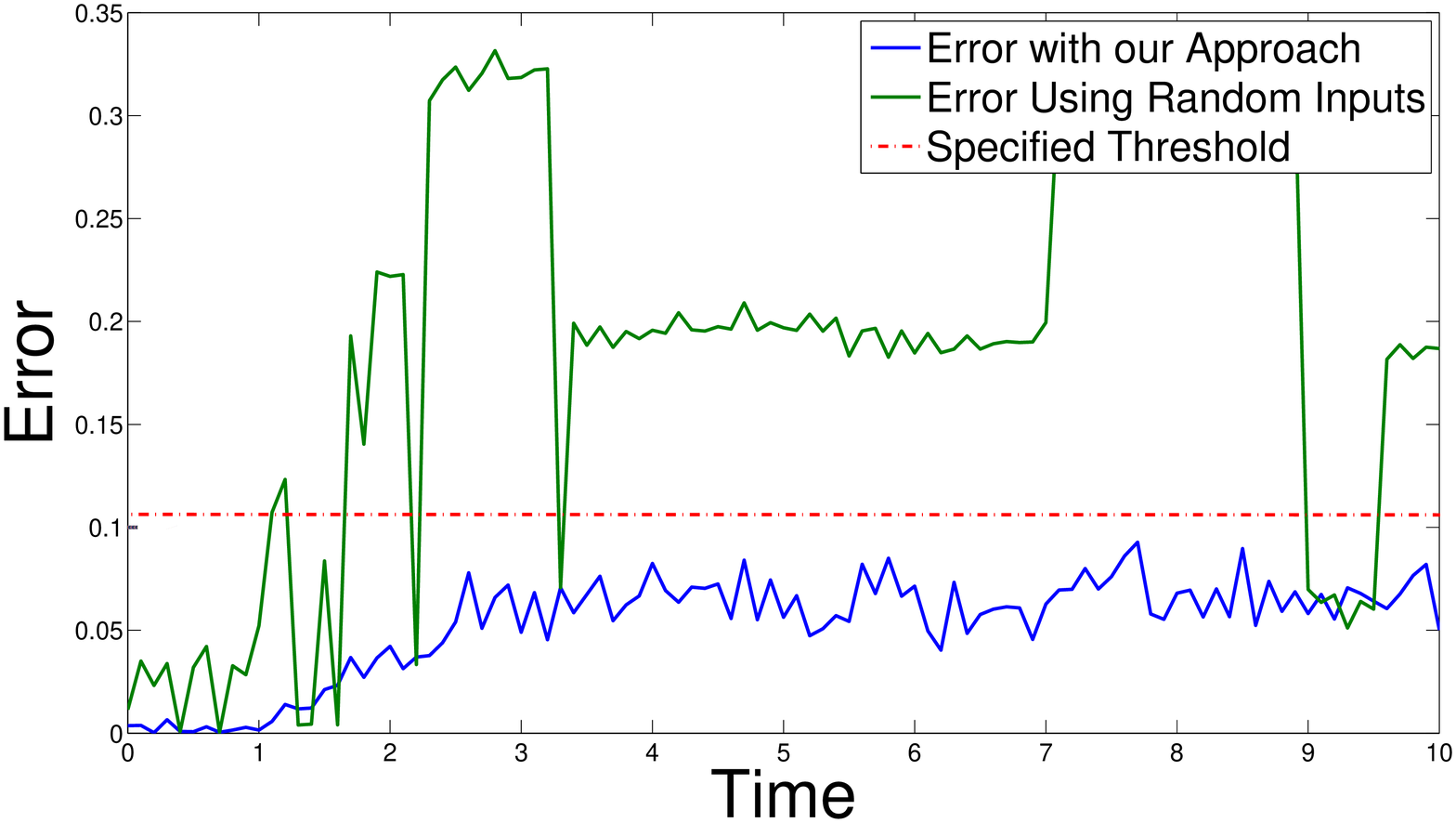}
\caption{DC Motor Error}
\label{fig:fx:motorerror}
\end{figure}
\end{minipage}
\end{minipage}
}

\subsection{Two-Wheeled Welding Mobile Robot}

The next case study is a nonlinear controller for a two-wheeled welding
mobile robot (WMR)~\cite{bui-jcas03}. 
$v$ and $\omega$ are the straight
and angular velocities of the WMR at its center point which
are the control parameters. 
Details of the robot model with equations of motion and the control law derivation
is presented in extended version~\cite{extended}.

\noop{
\begin{figure}[!ht]
\centering
\includegraphics[width=3.2in]{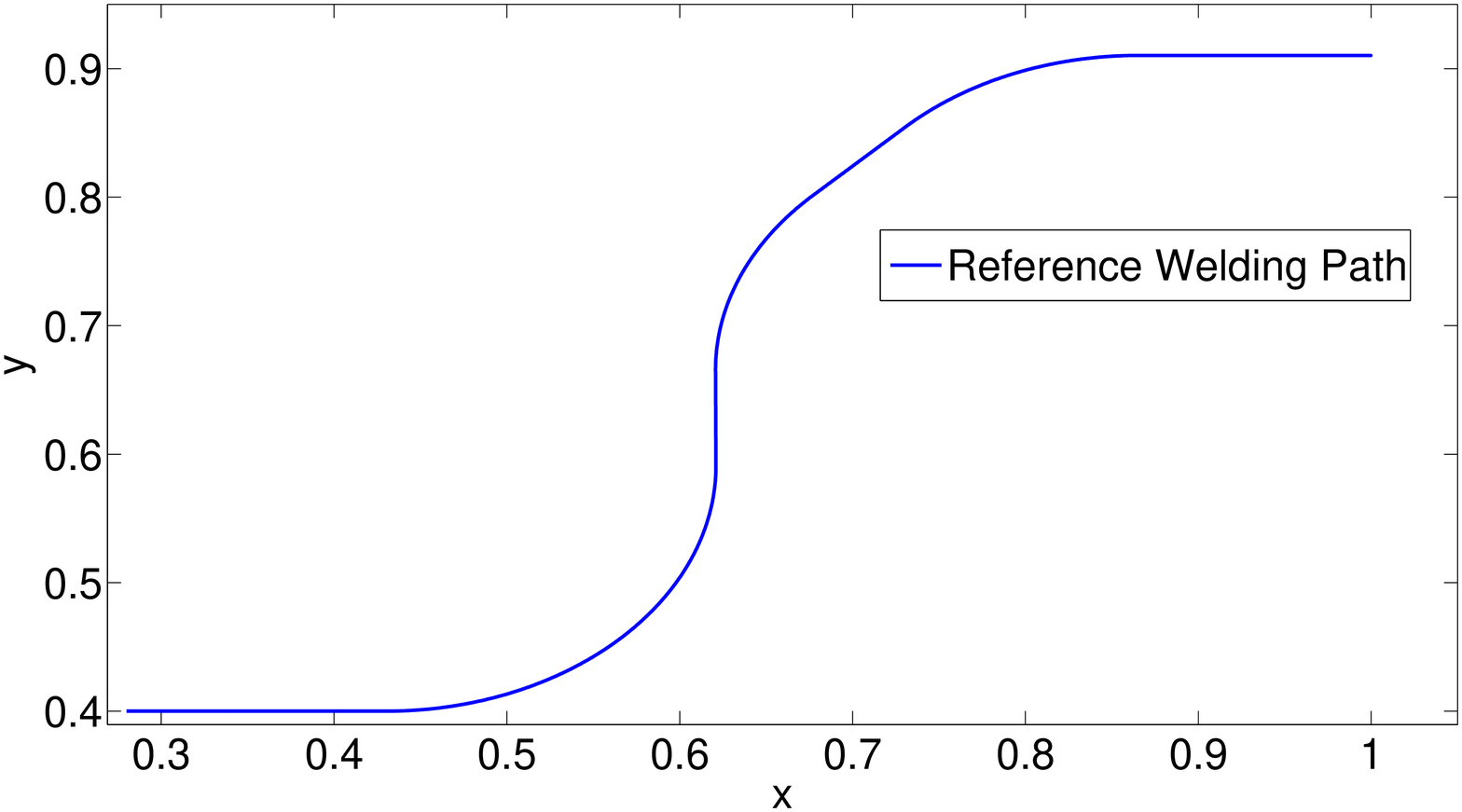}
\caption{Reference Welding Line}
\label{fig:fx:WMRref}
\end{figure}

\begin{figure}[!ht]
\centering
\includegraphics[width=3.2in]{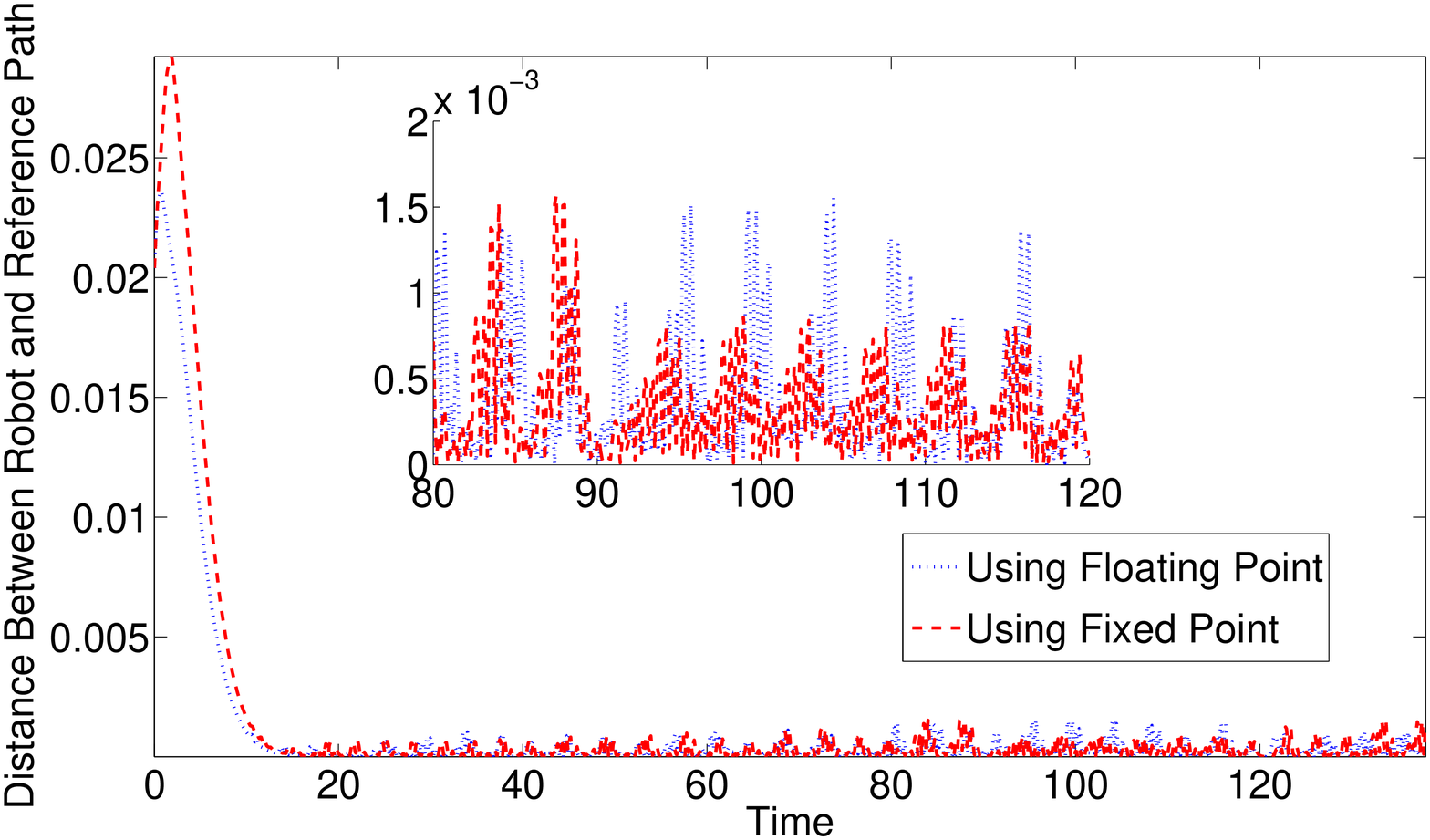}
\caption{Distance of WMR from Reference Line with zoomed-in view for $80$ to $120$ seconds.}
\label{fig:fx:WMRdist}
\end{figure}
}

\begin{figure}[!ht]
\centering
\subfigure[Reference Welding Line]
{
\includegraphics[width=0.48\textwidth]{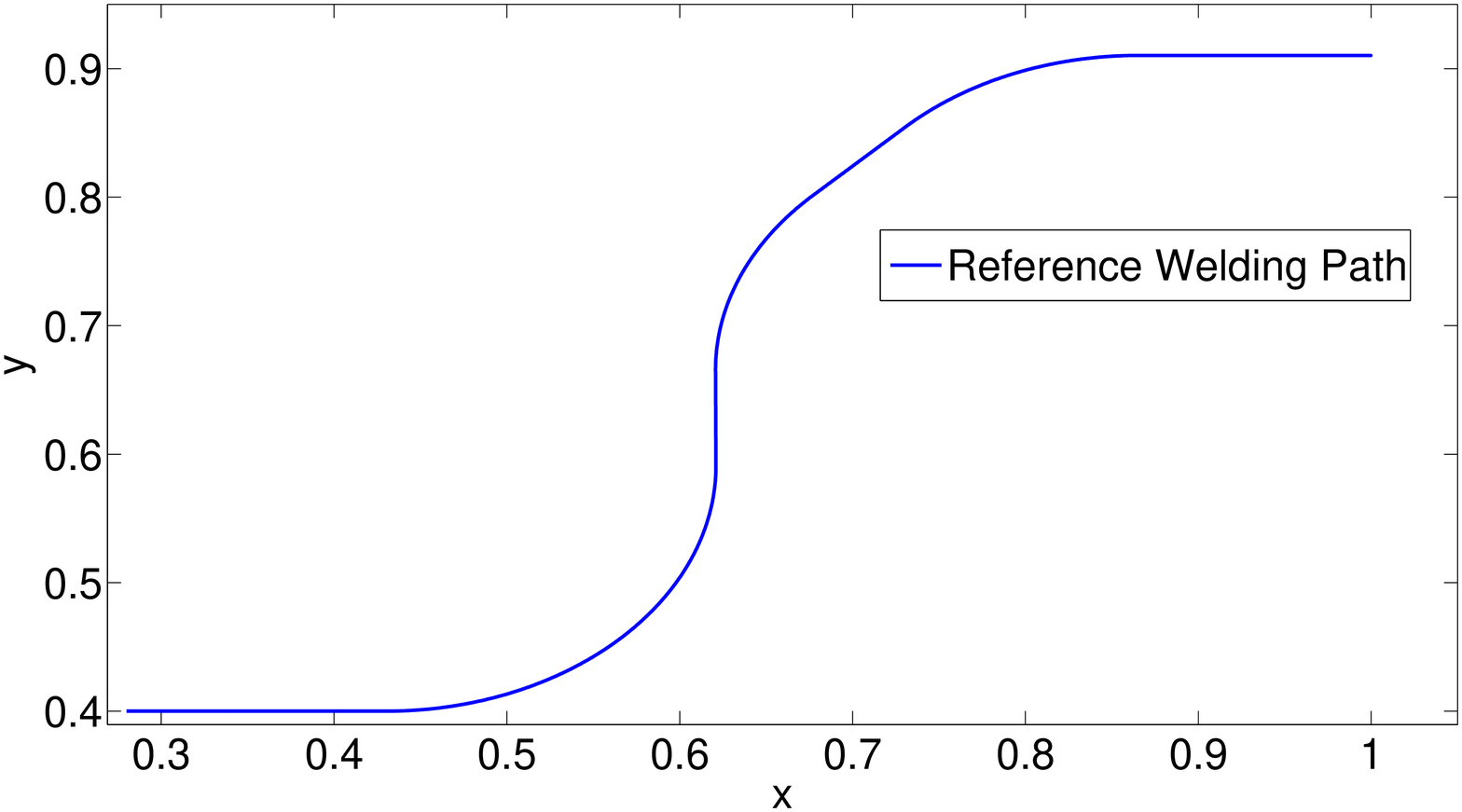}
\label{fig:fx:WMRref}
}
\subfigure[Distance of WMR from Reference Line]
{
\includegraphics[width=0.48\textwidth]{distance}
\label{fig:fx:WMRdist}
}
\caption{Welding Motor Robot}
\end{figure}

We use our synthesis technique to automatically synthesize fixed-point
program computing both control inputs: $v$ and $\omega$.
We require that the relative for both controllers ($v$ and $\omega$)
are less than $0.1$.
Figure~\ref{fig:fx:WMRref} shows the reference line for welding
and Figure~\ref{fig:fx:WMRdist} shows the distance of the WMR from the
reference line as a function of time for both cases: firstly, when
the controller is implemented as a floating-point program and secondly,
when the controller is implemented as a fixed-point program synthesized
using our technique.
The robot starts a little away
from the reference line but quickly starts tracking the line in both
cases. Figure~\ref{fig:fx:WMRerrorv} and Figure~\ref{fig:fx:WMRerroromega}
show the error between the floating-point
controller and fixed-point controller for both control inputs: $v$ and $\omega$,
respectively.
\noop{

\begin{figure}[!ht]
\centering
\includegraphics[width=3.2in]{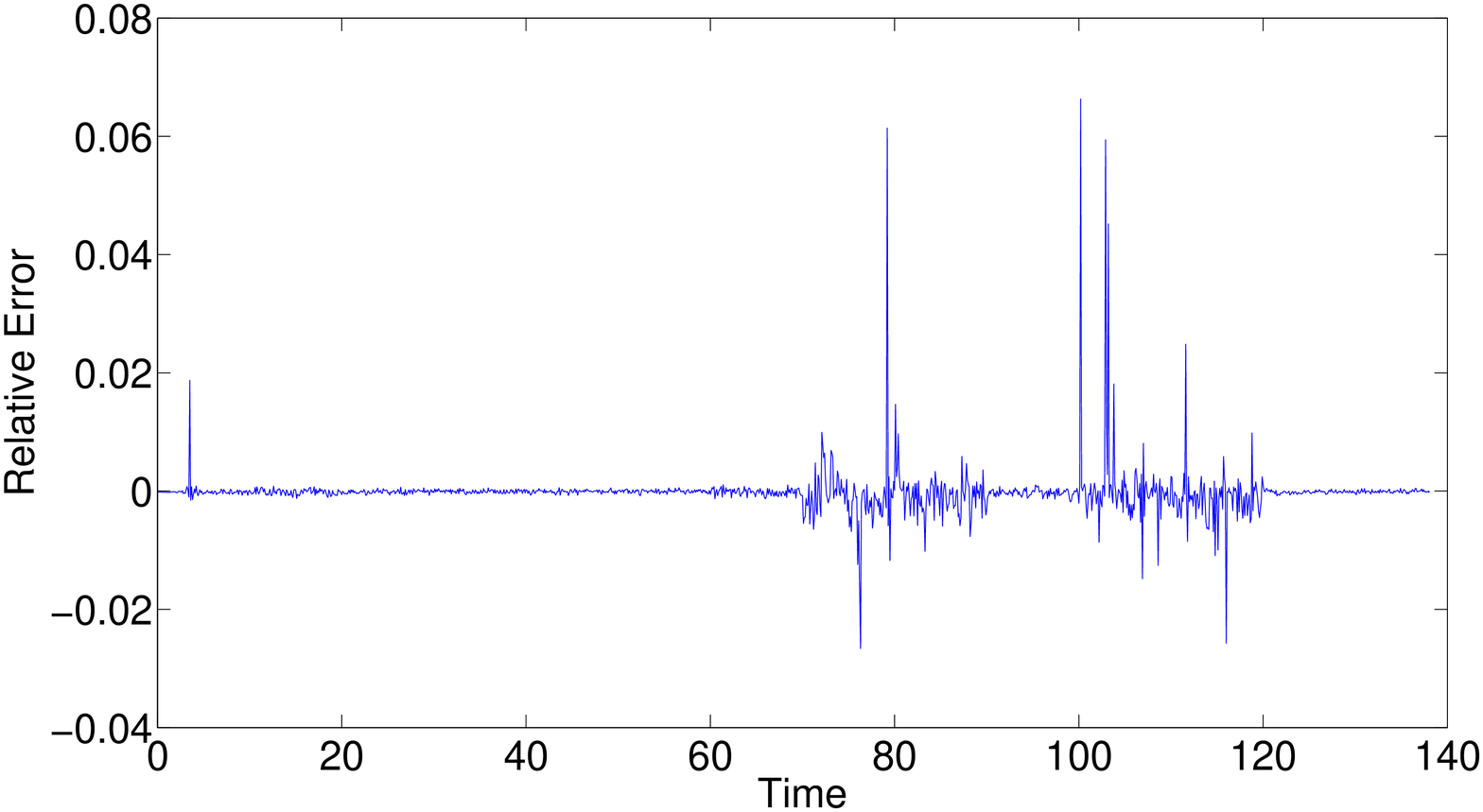}
\caption{Error in computing $v$}
\label{fig:fx:WMRerrorv}
\end{figure}

\begin{figure}[!ht]
\centering
\includegraphics[width=3.2in]{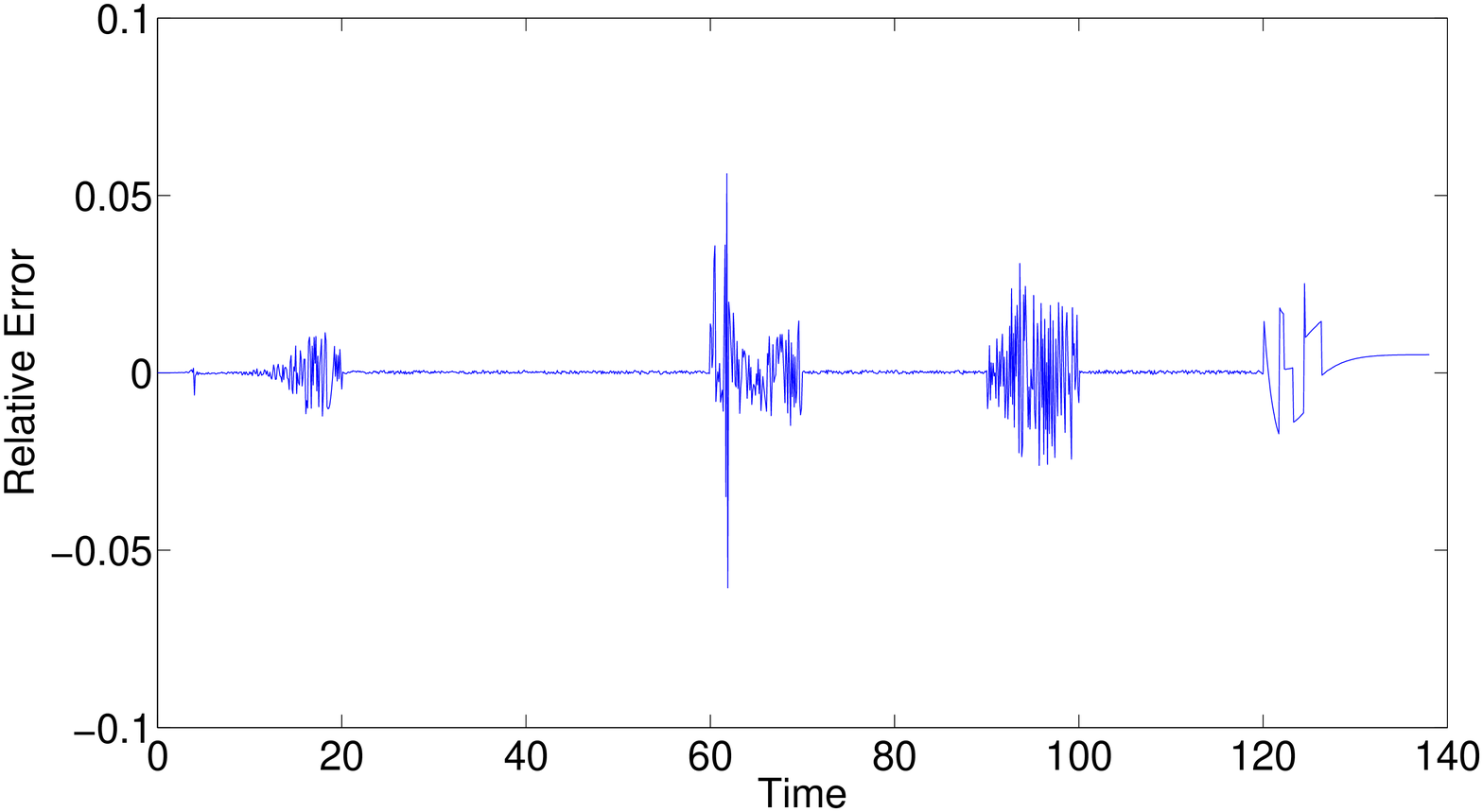}
\caption{Error in computing $\omega$}
\label{fig:fx:WMRerroromega}
\end{figure}
}

\begin{figure}[!ht]
\centering
\subfigure[Error in computing $v$]
{
\includegraphics[width=0.48\textwidth]{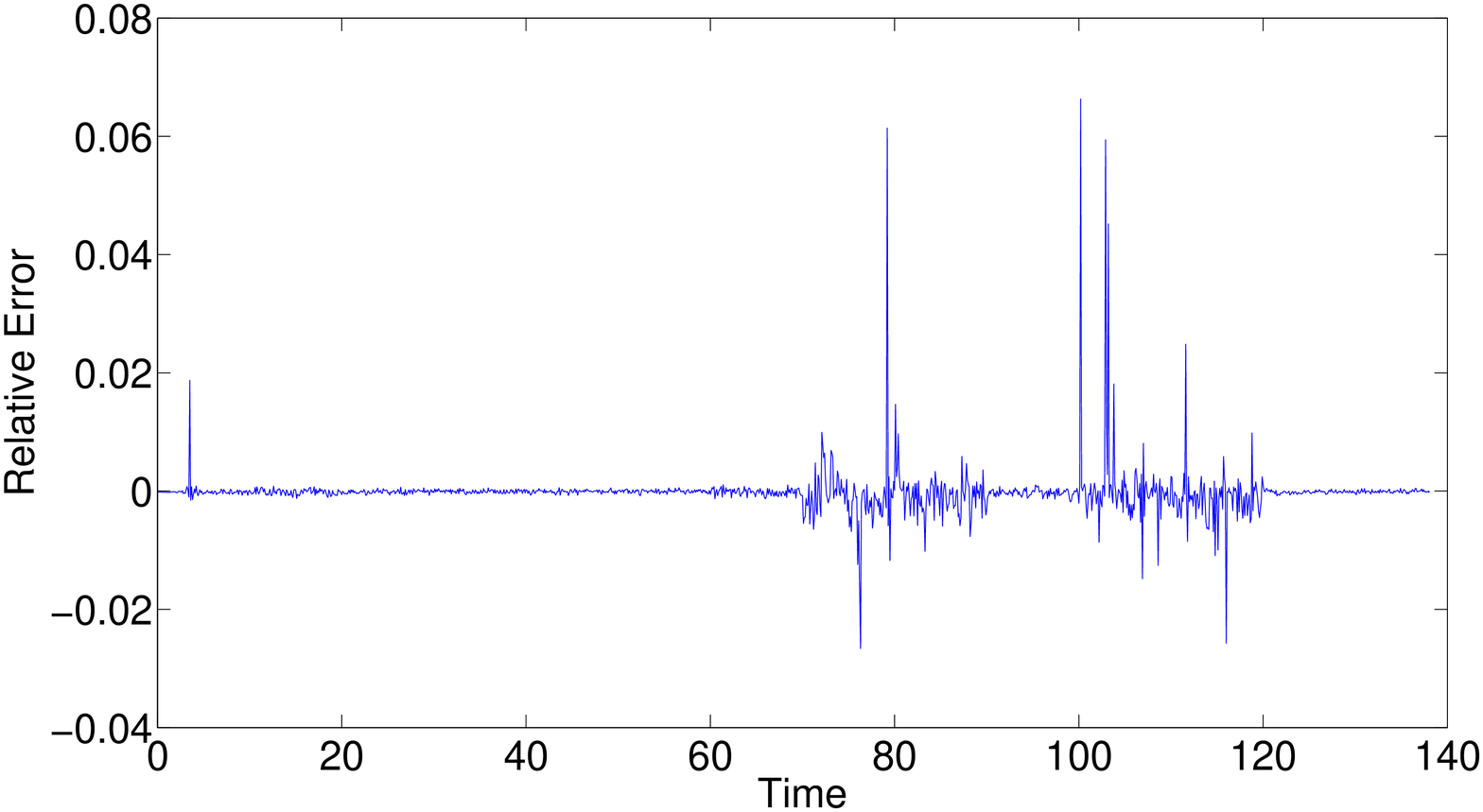}
\label{fig:fx:WMRerrorv}
}
\subfigure[Error in computing $\omega$]
{
\includegraphics[width=0.48\textwidth]{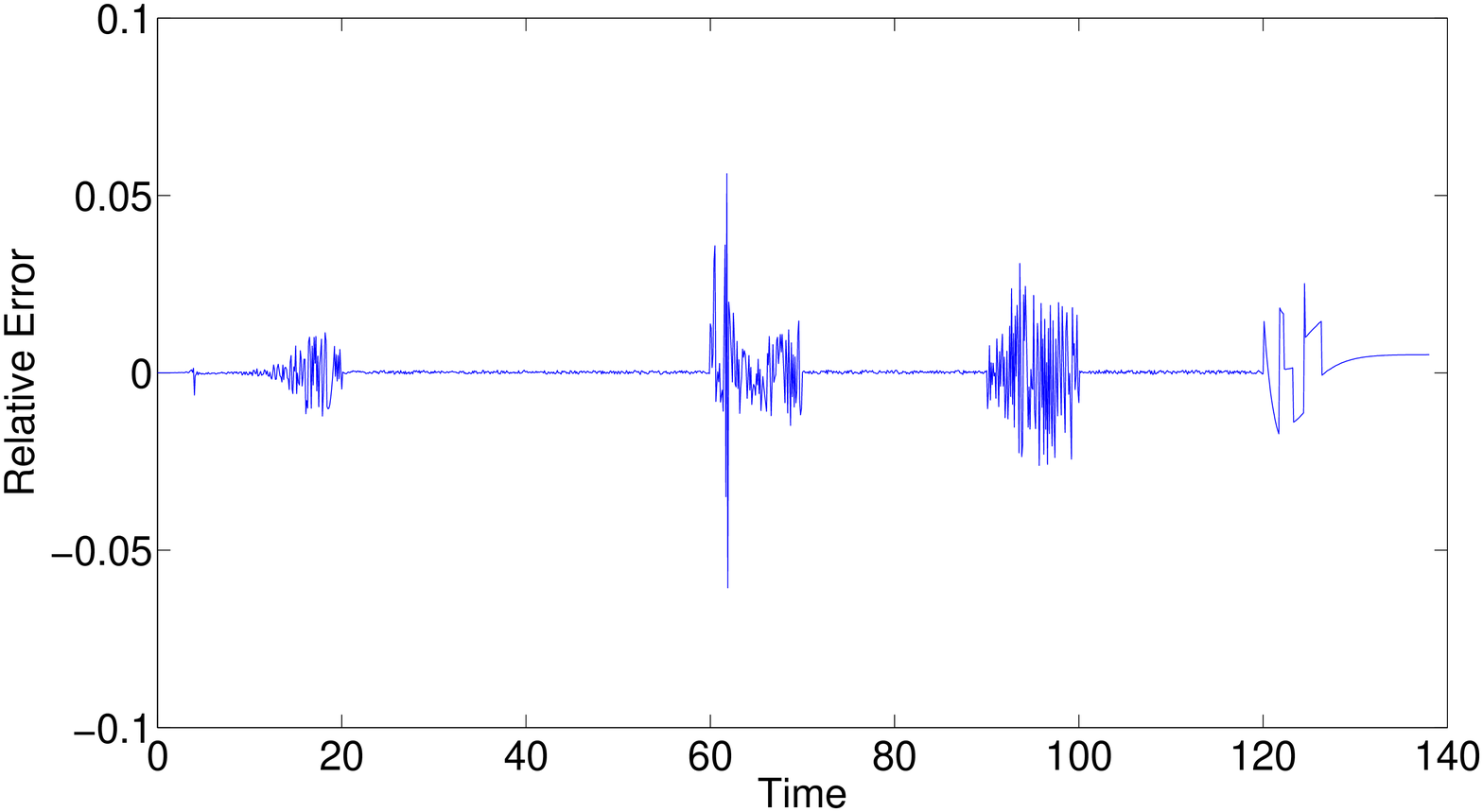}
\label{fig:fx:WMRerroromega}
}
\caption{Welding Motor Robot}
\end{figure}

}
{
\subsection{Infinite Impulse Response (IIR) Filter}

The first case study is an IIR filter which is used
in digital signal processing applications.
It is a first-order direct form-II IIR filter with
the schematic shown in Figure~\ref{fig:fx:iir}.
The constants are $a_1 = -0.5$, $b0 = 0.9$ and $b1 = 0.9$.
The fixed-point variables are identified in the schematic.
We use our synthesis technique to discover the appropriate
fixed-point types of these variables.
The input domain used in synthesis is $-2 < input < 2$.
The correctness condition for accuracy is to ensure that the relative
error between the floating-point and fixed-point program is less
than $0.1$.

\begin{figure}[!ht]
\centering
\includegraphics[width=4in]{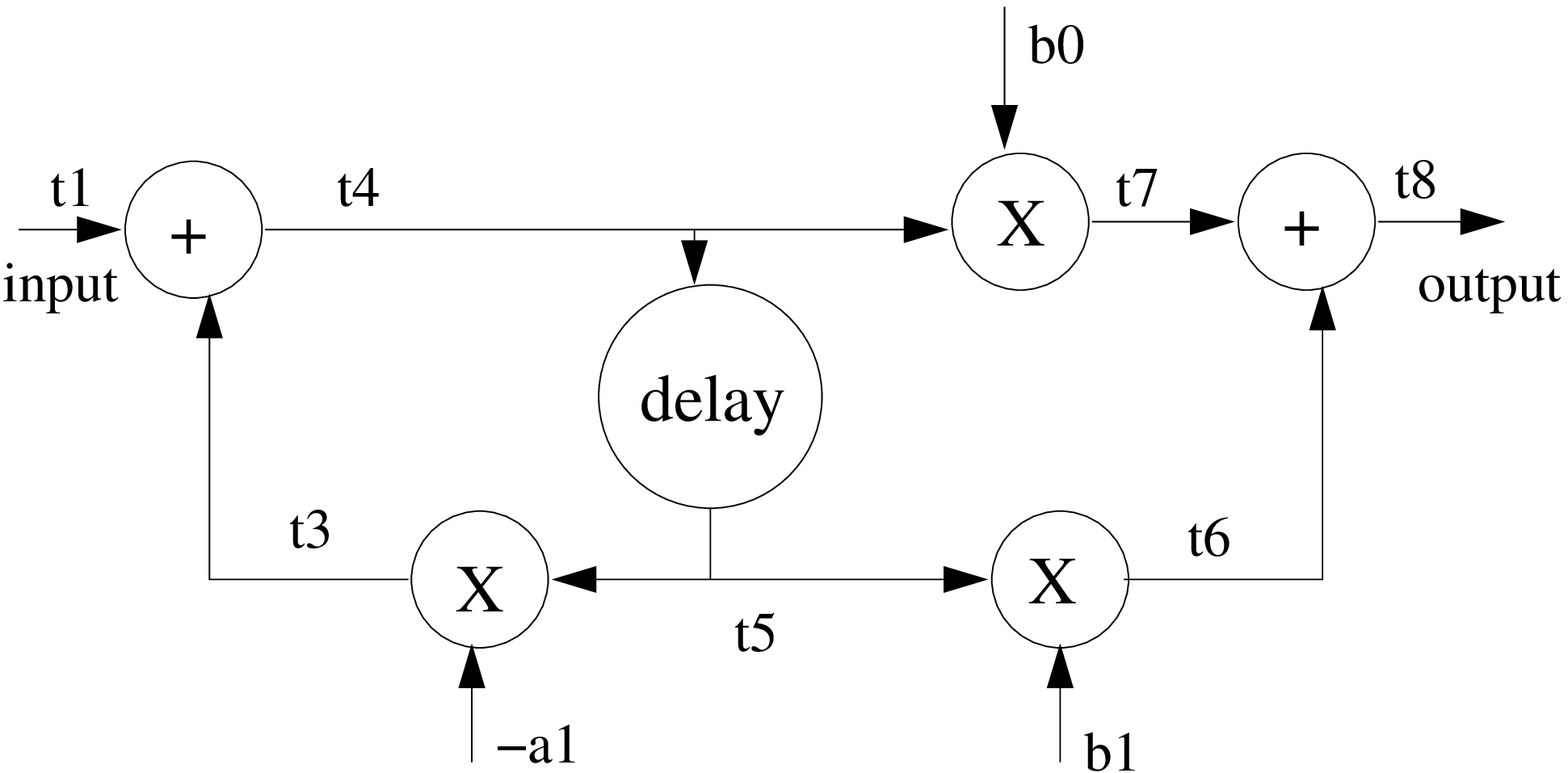}
\caption{IIR Filter Schematic}
\label{fig:fx:iir}
\end{figure}

\noop{
\begin{table}[!t]
\caption{Synthesized Fixed-point Types for IIR Filter}
\label{tbl:fx:iir}
\centering
\begin{tabular}{|c|c|c|c|c|c|}
\hline
Variables & a1 & b0 & b1 & t1 &  t3  \\
\hline
Fixed-point Types &  $\langle 1,2,1  \rangle$  &  $\langle 0,1,5  \rangle$ & $\langle 0,1,3  \rangle$ &  $\langle 1,2,6  \rangle$ &  $\langle 1,2,4  \rangle$  \\
\hline
Variables & t4 & t5 & t6 & t7 & t8 \\
\hline
Fixed-point Types &  $\langle 1,3,5  \rangle$ &  $\langle 1,3,5  \rangle$ &  $\langle 1,3,5  \rangle$ &  $\langle 1,3,6  \rangle$ &  $\langle 1,4,6  \rangle$ \\
\hline
\end{tabular}
\end{table}
}

\begin{figure}[!ht]
\centering
\includegraphics[width=4in]{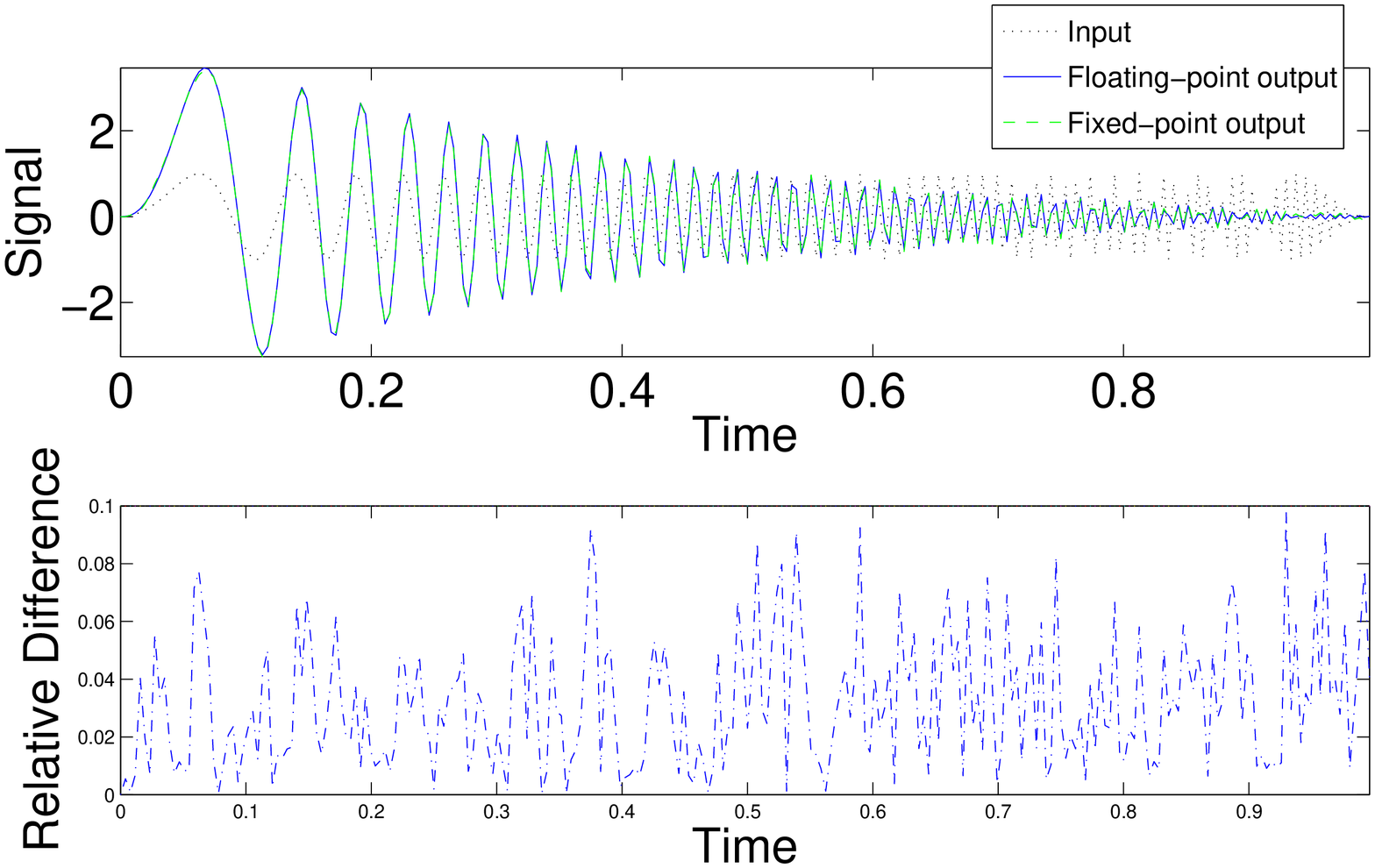}
\caption{IIR Filter Using Floating-point and Fixed-point.
{\footnotesize{In the top plot, the floating-point and fixed-point outputs
are virtually superimposed on each other.}}}
\label{fig:fx:iirsim}
\end{figure}

In order to test the correctness of our implementation,
we feed a common input signal to both the IIR filter implementations:
floating-point version and the fixed-point version obtained by our
synthesis technique.
The input signal is a linear chirp from $0$ to $\frac{Fs}{2}$ Hz in $1$ second.
$$input = (1-2^{-15}) \times sin(\pi \times \frac{Fs}{2}\times t^2)$$
where $Fs = 256$  and $t=0$ to $1-\frac{1}{Fs}$ and is
sampled at intervals of $\frac{1}{Fs}$.
Figure~\ref{fig:fx:iirsim} shows the
input,
outputs of both implementations and the relative error
between the two outputs. We observe that the implementation
satisfies the correctness condition and the relative error
remains below $0.1$ throughout the simulation.

\subsection{Finite Impulse Response (FIR) Filter}

The second case study is a low pass FIR filter of order 4
with tap coefficients $0.0346,0.2405, 0.4499, 0.2405$ and $0.0346$.
The input domain, correctness condition and input signal to test the
floating-point implementation and synthesized fixed-point program
are same as the previous case study.
Figure~\ref{fig:fx:firsim} shows the
input,
outputs of both implementations and the relative error
between the two outputs. We observe that the implementation
satisfies the correctness condition and the relative error
remains below $0.1$ throughout the simulation.

\begin{figure}[!ht]
\centering
\includegraphics[width=4in]{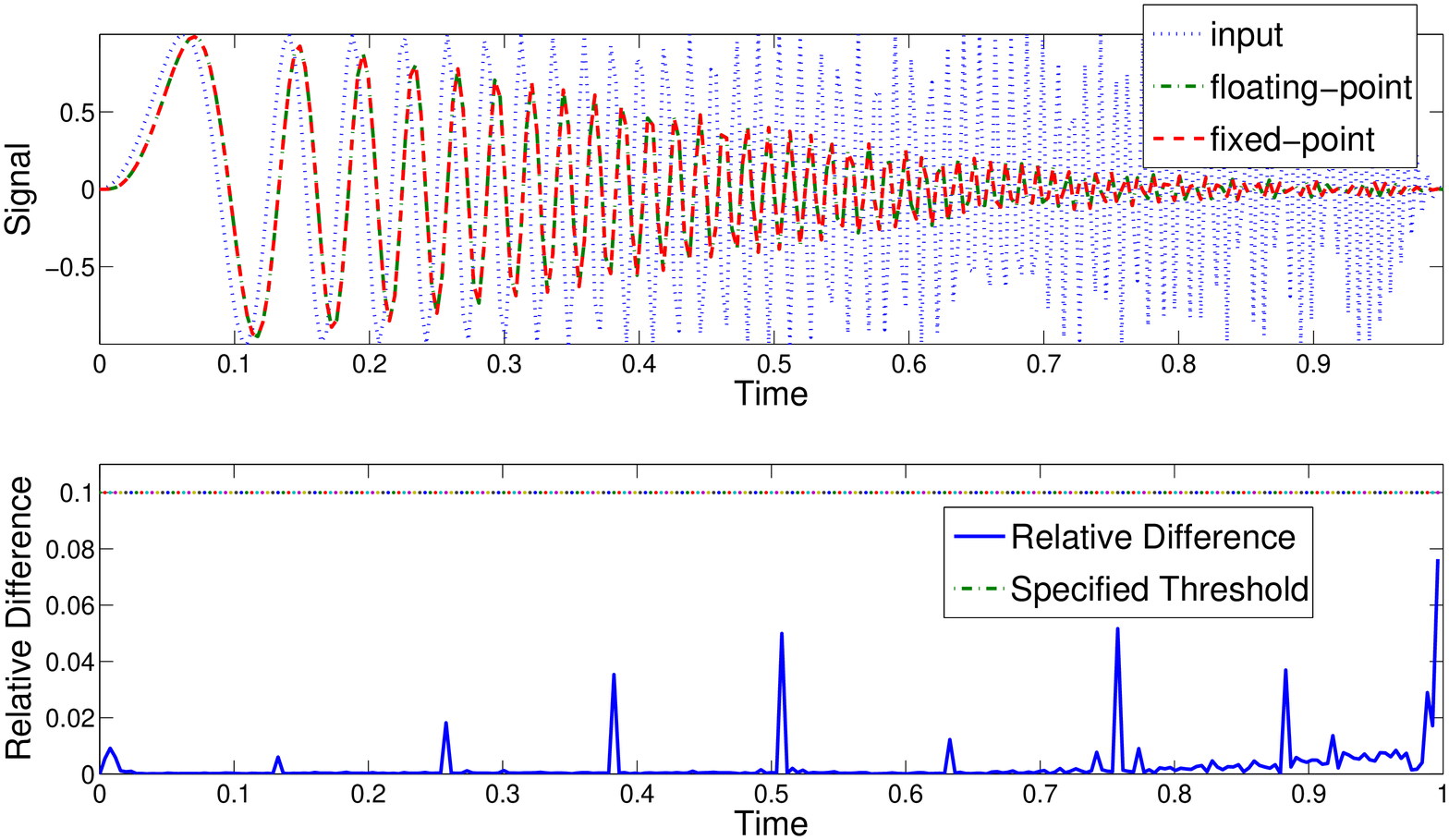}
\caption{FIR Filter Using Floating-point and Fixed-point.
{\footnotesize{The floating-point/fixed-point outputs
are virtually superimposed on each other.}}
}
\label{fig:fx:firsim}
\end{figure}

\subsection{Field Controlled DC Motor}

The next case study is a field controlled DC Motor.
It is a classic non-linear control example from
Khalil~\cite{khalil-book}. The example used here is an adaptation of
Khalil's example presented in \cite{claire-course}.
The system dynamics is described by the following ordinary differential
equations.
\begin{align*}
\dot v_f = R_fi_f + L_f \dot i_f\\
\dot v_a = c_1 i_f \omega + L_a \dot i_a + R_a i_a \\
\dot J \dot \omega = c_2 i_f i_a - c_3 \omega
\end{align*}

The first equation is for the field circuit with $v_f, i_f, R_f, L_f$
being its voltage, current, resistance, and inductance.
The variables $v_a, i_a, R_a, L_a$ are the corresponding
 voltage, current, resistance, and inductance
of the armature circuit described by the second equation.
The third equation is a torque equation for the shaft, with
$J$ as the rotor inertia and $c_3$ as a damping coefficient.
The term $c_1 i_f \omega$ is the back electromotive force induced in
the armature circuit, and $c_2 i_f i_a$ is the torque produced by the
interaction of the armature current with the field circuit flux.
In the field controlled DC motor,
field voltage $v_f$ is the control input while $v_a$ is held constant.
The purpose of the control is to drive the system
to the desired set point for the angular velocity $\omega$.

We can now rewrite the system dynamics in the following normal
form where $a = \frac{R_f}{L_f}$, $u = \frac{ v_f}{L_f}$, $b = \frac{R_a}{L_a}$,
$\rho =  \frac{v_a}{L_a}$, $c = \frac{c_1}{L_a}$,
$\theta = \frac{c_2}{J}$, $d = \frac{c_3}{J}$.
\begin{align*}
 \dot i_f = - a i_f + u \\
 \dot i_a =  - b i_a  + \rho - c i_f \omega \\
 \dot \omega = \theta i_f i_a - d \omega
\end{align*}
We assume no damping, that is, $c_3 = 0$ and set all the other
constants $a,b,c,\theta,\rho$ to $1$~\cite{claire-course}.
The state feedback law to control the system is given by
\noop{
\begin{tabular}{ c c }
\begin{minipage}{1.5in}
\begin{algorithmic} [style=tt]
\small
\REQUIRE $\mathtt{i_f, i_a, \omega, \theta, \rho, c, \epsilon}$
\ENSURE $\mathtt{u}$
\STATE $\mathtt{t1 = \theta \times i_a}$;
\STATE $\mathtt{t2 = \epsilon+t1}$;
\STATE $\mathtt{t3 = 1/t2}$;
\STATE $\mathtt{t31 = a+b}$;
\STATE $\mathtt{t32 = i_f \times i_a}$;
\STATE $\mathtt{t33 = t31 \times t32}$;
\STATE $\mathtt{t4 = \theta \times t33}$;
\STATE $\mathtt{t41 = \rho \times i_f}$;
\STATE $\mathtt{t5 = \theta \times t41}$;
\STATE $\mathtt{t6 = i_f \times i_f}$;
\STATE $\mathtt{t61 = t6 \times \omega}$;
\STATE $\mathtt{t62 = t61 \times \theta}$;
\STATE $\mathtt{t7 = c \times t62}$;
\STATE $\mathtt{t8 = t4+t5}$;
\STATE $\mathtt{t9 = t8 - t7}$;
\STATE $\mathtt{u = t3 \times t9}$;
\RETURN $\mathtt{u}$
\end{algorithmic}
Floating-point Code\\
\end{minipage}
&
\begin{minipage}{2in}
\large
$ u = \frac{\theta(a+b)i_f i_a + \theta \rho i_f - c \theta i_f^2 \omega}{\theta i_a + \epsilon}$\\
\normalsize
Control Law\\
\end{minipage}
\end{tabular}
}
$$ u = \frac{\theta(a+b)i_f i_a + \theta \rho i_f - c \theta i_f^2 \omega}{\theta i_a + \epsilon}$$
where $\epsilon = 0.01$ is added to denominator to avoid division by $0$. $i_a$
approaches $0$ at equilibrium. The corresponding floating-point code
is shown below.
\begin{algorithmic} [style=tt]
\small
\REQUIRE $\mathtt{i_f, i_a, \omega, \theta, \rho, c, \epsilon}$
\ENSURE $\mathtt{u}$
\STATE $\mathtt{t1 = \theta \times i_a}$; $\mathtt{t2 = \epsilon+t1}$;
 $\mathtt{t3 = 1/t2}$; $\mathtt{t31 = a+b}$;
\STATE $\mathtt{t32 = i_f \times i_a}$; $\mathtt{t33 = t31 \times t32}$;
 $\mathtt{t4 = \theta \times t33}$; $\mathtt{t41 = \rho \times i_f}$;
\STATE $\mathtt{t5 = \theta \times t41}$; $\mathtt{t6 = i_f \times i_f}$;
 $\mathtt{t61 = t6 \times \omega}$; $\mathtt{t62 = t61 \times \theta}$;
\STATE $\mathtt{t7 = c \times t62}$; $\mathtt{t8 = t4+t5}$;
 $\mathtt{t9 = t8 - t7}$; $\mathtt{u = t3 \times t9}$;
\RETURN $\mathtt{u}$
\end{algorithmic}
 The system is initialized with
field current $i_f = 1$, armature current $i_a = 1$
and angular velocity $\omega = 1$.

\begin{figure}[!ht]
\centering
\includegraphics[width=4in]{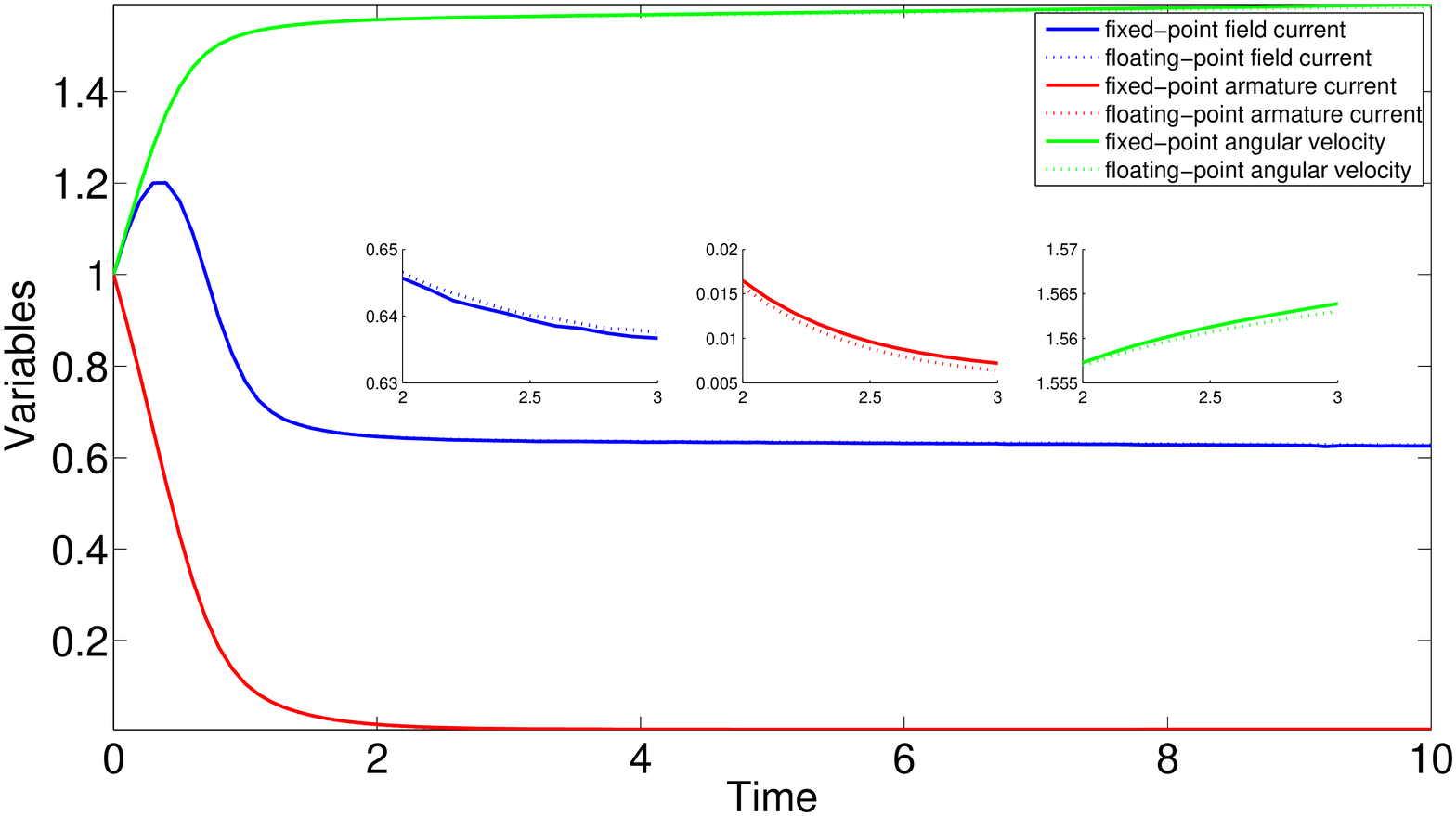}
\caption{DC Motor Using Floating-point and Fixed-point Controller with zoomed-in view for $2$ to $3$ seconds.}
\label{fig:fx:dcmotorfx}
\end{figure}

The computed control law can be mathematically shown to be
correct by designers who are more comfortable in reasoning with
real arithmetic but not with finite precision arithmetic.
Its implementation using floating-point computation
also closely mimics the arithmetic in reals
 but the control algorithms are often implemented using
fixed-point computation on embedded platforms.
We use our synthesis technique to automatically
derive a low cost fixed-point implementation of the control law
computing $u$.
The input domain is $0 \leq i_a,i_f,\omega \leq 1.5$.
The correctness condition for accuracy is that the absolute
difference between the $u$ computed by fixed-point program and the
floating-point program is less than $0.1$.

Figure~\ref{fig:fx:dcmotorfx} shows the simulation of the system
using the fixed-point implementation of the controller
and the floating-point implementation.
This end-to-end
simulation shows that fixed-point program generated by our technique
can be used to control the system as effectively as the floating-point program.
This illustrates the practical utility of our technique.
Figure~\ref{fig:fx:motorerror} plots the difference between
the control input computed by the fixed-point program and the floating-point
program. It
shows that the fixed-point types
synthesized using our approach
satisfy the correctness condition, and the
difference between the control input computed by
the fixed-point and floating-point program is within
the specified maximum error threshold of $0.1$.
The number of inputs needed in our approach was $127$.
In contrast, the fixed-point types found using $635$(5X our approach)
randomly selected inputs violate
the correctness condition for a large number of inputs.

\begin{figure}[!ht]
\centering
\centering
\includegraphics[width=4in]{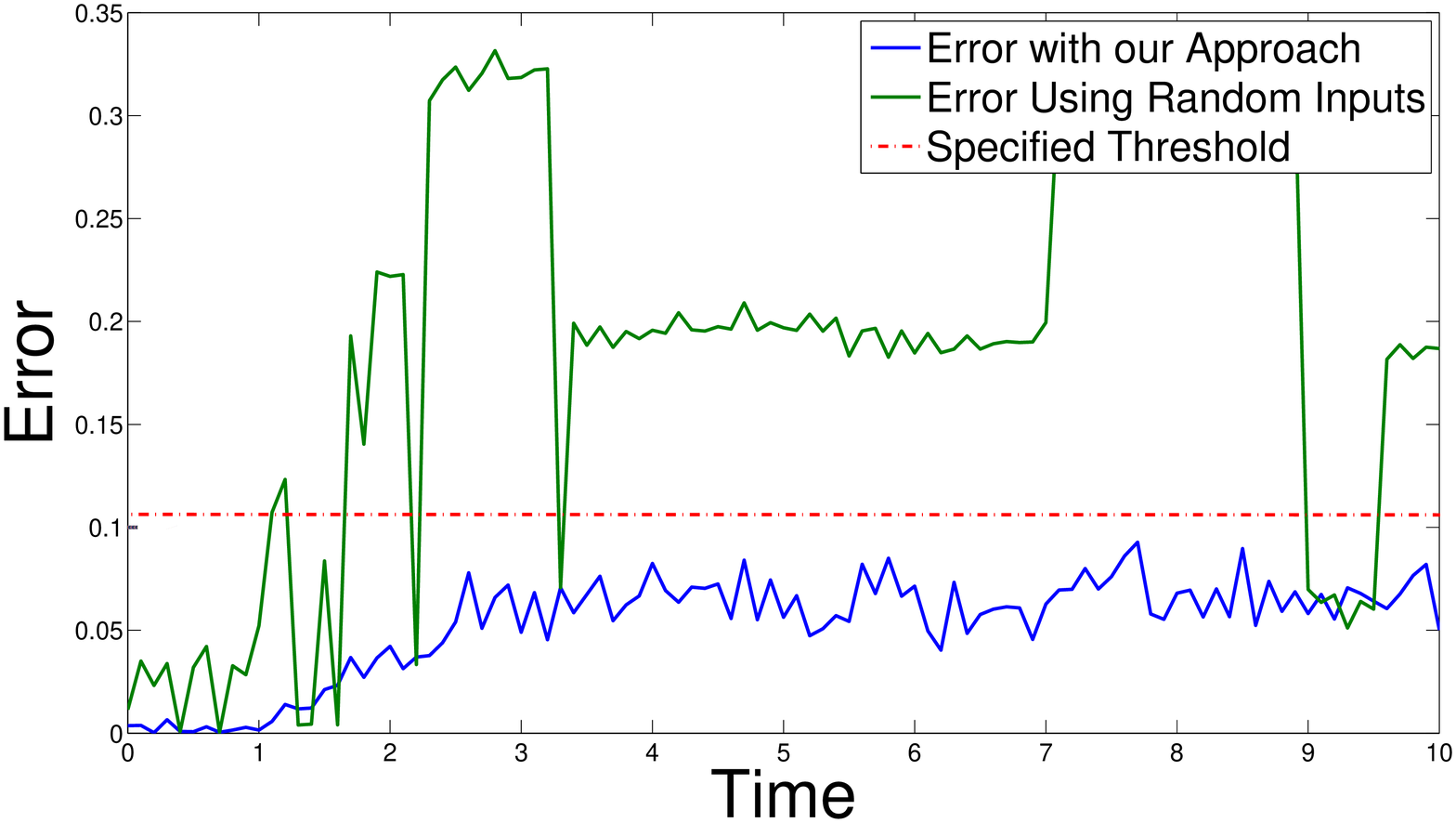}

\caption{Error in Control Input Using Fixed-point and Floating-point Program}
\label{fig:fx:motorerror}
\end{figure}

\subsection{Two-Wheeled Welding Mobile Robot}

The next case study is a nonlinear controller for a two-wheeled welding
mobile robot (WMR)~\cite{bui-jcas03}. The robot consists of
two wheels and a robotic arm. The wheels can roll and there is no slipping.
$(x,y)$ represents the Cartesian coordinate of the WMR's center point
and $\phi$ is the heading angle of the WMR. $v$ and $\omega$ are the straight
and angular velocities of the WMR at its center point.
The welding point coordinates $(x_w, y_w)$ and the heading angle $\phi_w$
can be calculated from the WMR's center point:
\begin{align*}
x_w = x - l \sin\phi\\
y_w = y + l \cos\phi\\
\phi_w = \phi
\end{align*}
So, the equation of motion for the welding point is as follows:
\begin{align*}
\dot x_w = v \cos\phi - l \omega \cos\phi - \dot l \sin\phi\\
\dot y_w = v \sin\phi - l \omega \sin\phi + \dot l \cos\phi\\
\dot \phi_w = \omega
\end{align*}

The objective of the WMR controller is to ensure that the robot
tracks a reference point R.
The reference point R moving with a constant velocity
of $v_r$ on the reference path has coordinates
$(x_r, y_r)$ and the heading angle $\phi_r$.
The tracking error is the difference between the location of
the robot and the reference point.

\[
\left[ {\begin{array}{c}
 e_1 \\
 e_2 \\
 e_3
\end{array} } \right]
=
\left[ {\begin{array}{ccc}
 \cos\phi & \sin\phi & 0 \\
 -\sin\phi & \cos\phi & 0 \\
 0 & 0 & 1 \\
\end{array} } \right]
\left[ {\begin{array}{c}
 x_r - x_w \\
 y_r - y_w \\
 \phi_r - \phi_w
\end{array} } \right]
\]

The two control parameters in the model are $v$ and $\omega$.
In order to ensure that the error quickly converges to $0$,
a nonlinear controller based on Lyapunov stability is
as follows:
\begin{align*}
v = l(\omega_r + k_2 e_2 v_r + k_3 \sin e_3) + v_r \cos e_3 + k_1 e_1\\
\omega = \omega_r + k_2 e_2 v_r + k_3 \sin e_3
\end{align*}
where $k_1, k_2$ and $k_3$ are positive constants.
Table~\ref{tbl:fx:mwrval} provides
the numerical values of constants and initial values of the
state variables from Bui et al~\cite{bui-jcas03}.
All lengths are in meters, angle in radians and time in seconds.

\begin{table}[!ht]
\small
\caption{Numerical and Constant Values}
\label{tbl:fx:mwrval}
\centering
\begin{tabular}{|c|c||c|c|}
\hline
Parameters & Values & Parameters & Values  \\
\hline
$k_1$ & $4.2$ & $l$ & $0.15$ \\
$k_2$ & $5000$ & $\dot l$ & $0$ \\
$k_3$ & $1$ & $v_r$ & $7.5e-3$\\
\hline
$x_r$ & $0.280$ & $x_w$ & $0.270$ \\
$y_r$ & $0.400$ & $y_w$ & $0.390$ \\
$\phi$ & $0$ & $\phi_w$ & $15$\\
\hline
\end{tabular}
\normalsize
\end{table}

\begin{figure}[!ht]
\centering
\includegraphics[width=4in]{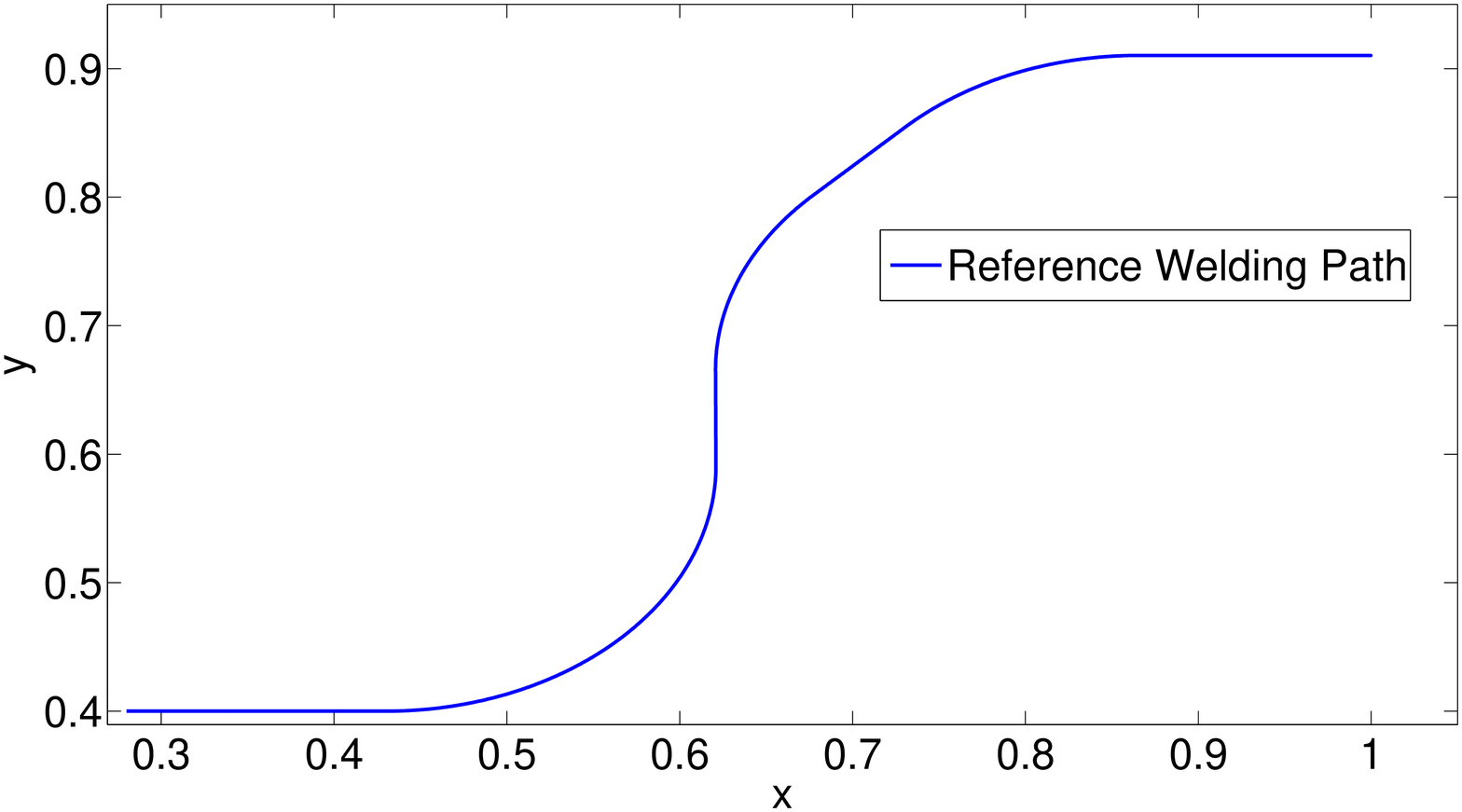}
\caption{Reference Welding Line}
\label{fig:fx:WMRref}
\end{figure}

\begin{figure}[!ht]
\centering
\includegraphics[width=4in]{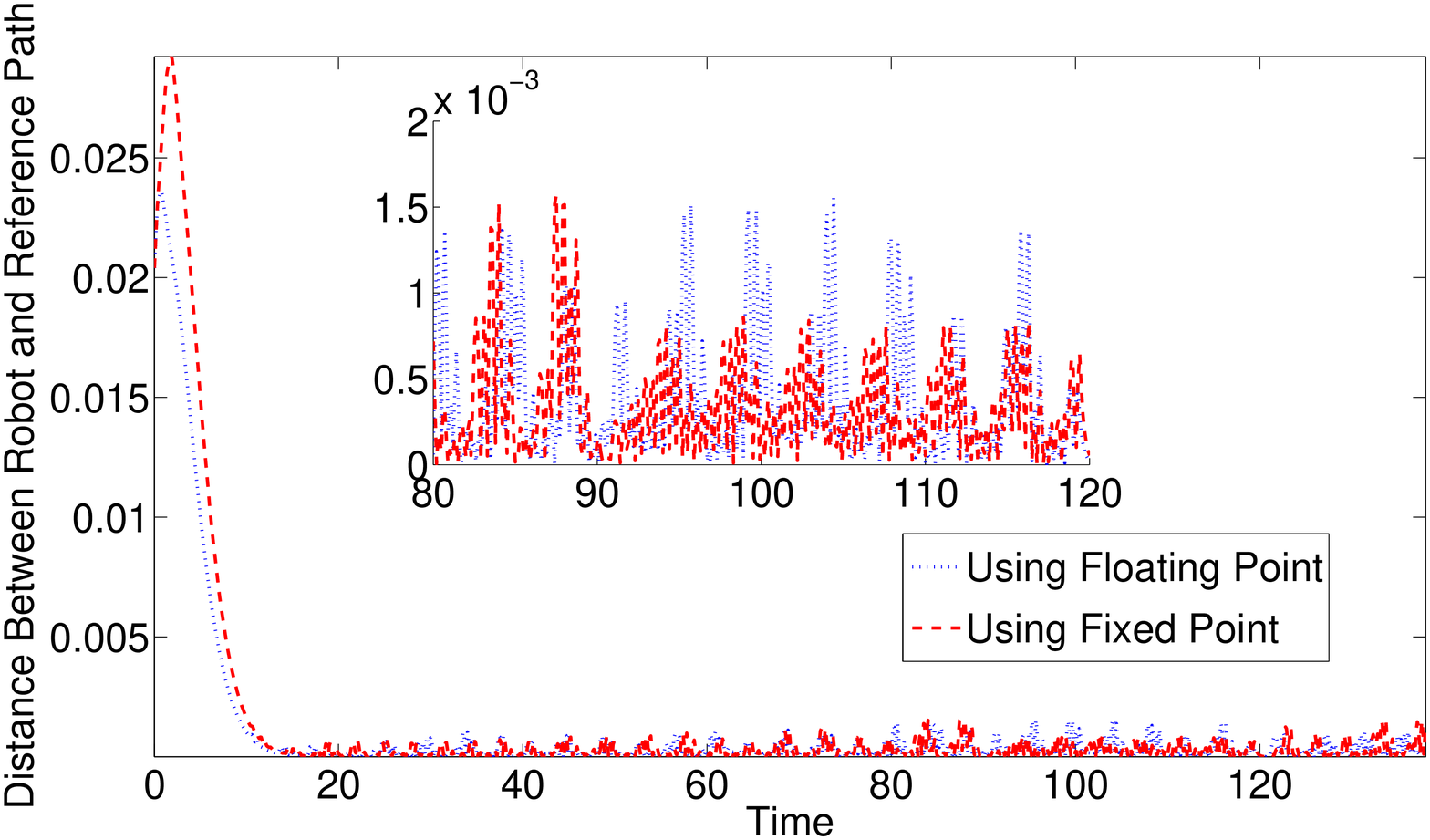}
\caption{Distance of WMR from Reference Line with zoomed-in view for $80$ to $120$ seconds.}
\label{fig:fx:WMRdist}
\end{figure}

\noop{
\begin{figure}[!ht]
\centering
\subfigure[Reference Welding Line]
{
\includegraphics[width=3.1in]{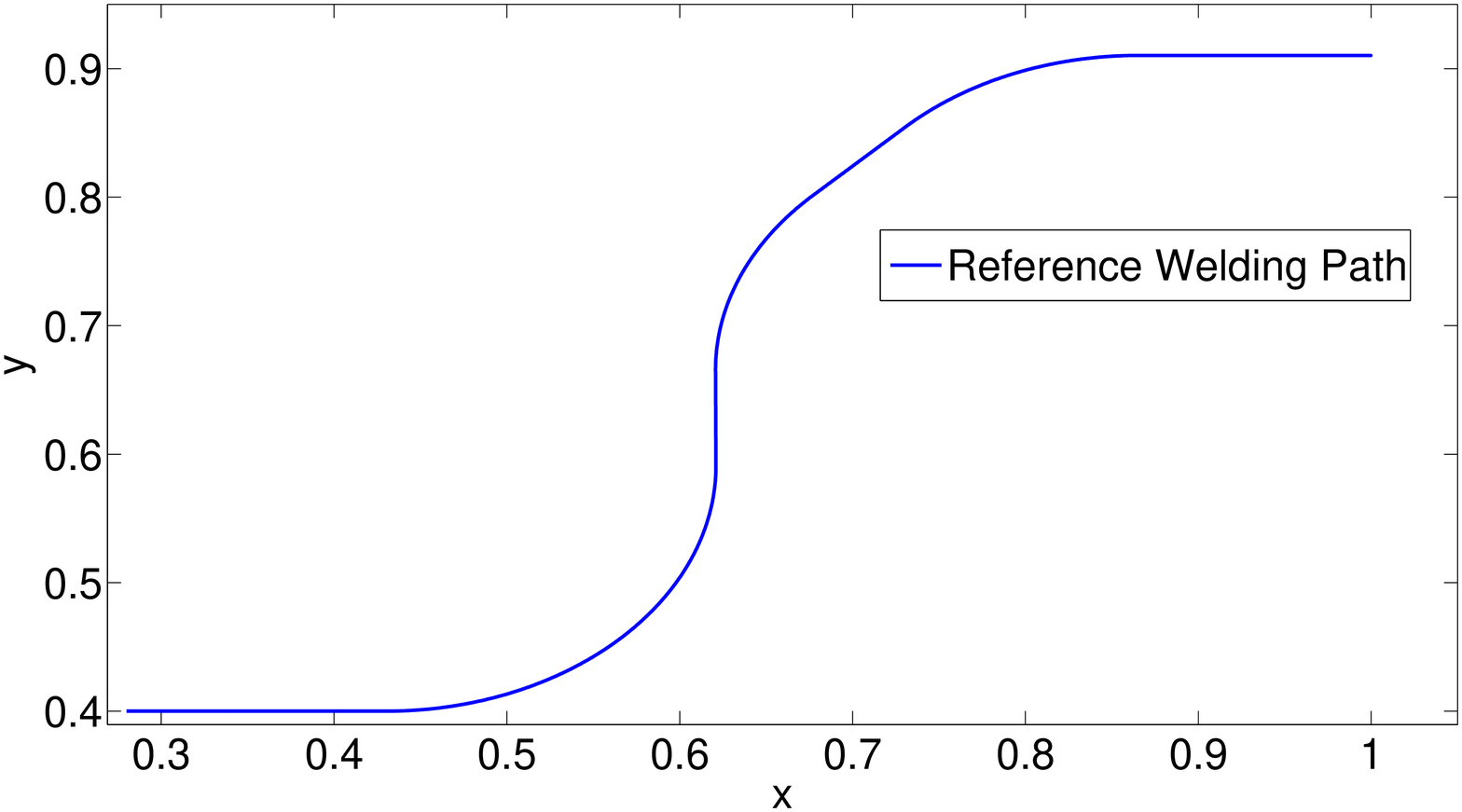}
\label{fig:fx:WMRref}
}

\subfigure[Distance of WMR from Reference Line]
{
\includegraphics[width=3.1in]{distance}
\label{fig:fx:WMRdist}
}
\caption{Welding Motor Robot}
\end{figure}
}

\begin{figure}[!ht]
\centering
\includegraphics[width=4in]{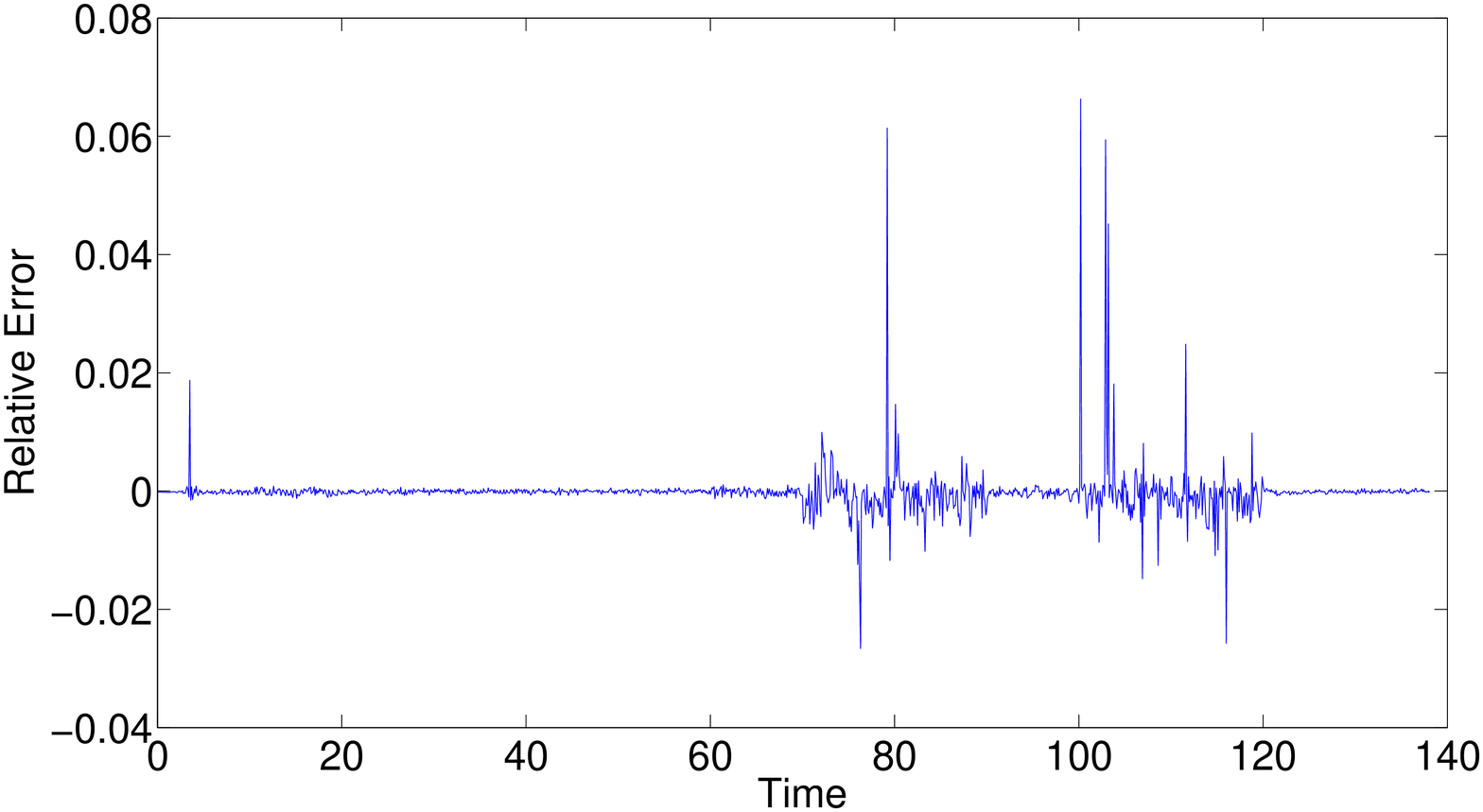}
\caption{Error in computing $v$}
\label{fig:fx:WMRerrorv}
\end{figure}

\begin{figure}[!ht]
\centering
\includegraphics[width=4in]{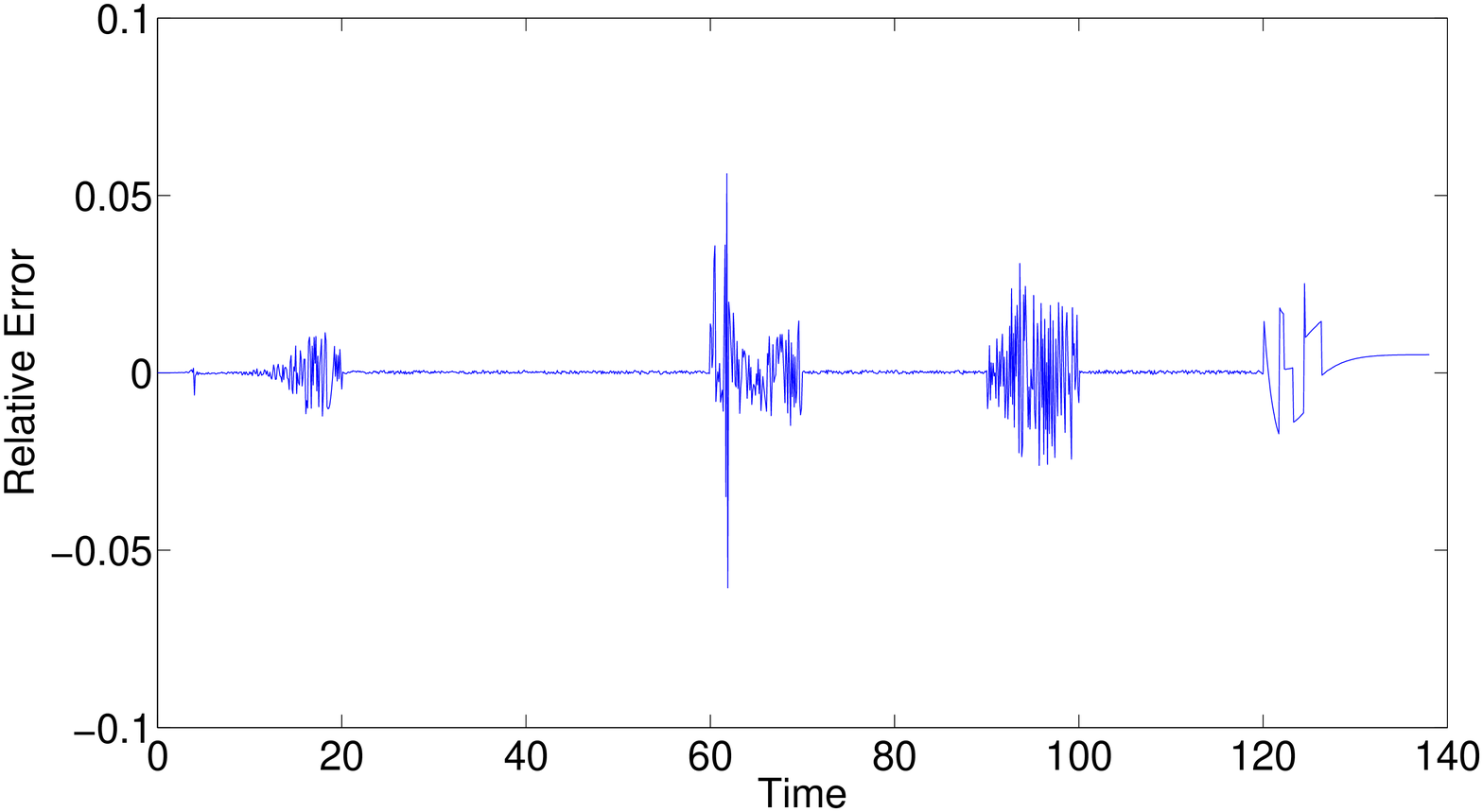}
\caption{Error in computing $\omega$}
\label{fig:fx:WMRerroromega}
\end{figure}

We use our synthesis technique to automatically synthesize fixed-point
program computing both control inputs: $v$ and $\omega$.
The error function used for $v$ is the relative difference
$$\frac{v_{floating-point} - v_{fixed-point}}{v_{floating-point}}$$
and the error function used for $\omega$ is the moderated
relative difference
$$\frac{\omega_{floating-point} - \omega_{fixed-point}}{\omega_{floating-point}+\delta}$$
where $\delta = 0.001$. The moderated relative difference is useful here
since $\omega_{floating-point}$ can be $0$.
We require that the difference values for both controllers
are less than $0.1$.
Figure~\ref{fig:fx:WMRref} shows the reference line for welding
and Figure~\ref{fig:fx:WMRdist} shows the distance of the WMR from the
reference line as a function of time for both cases: firstly, when
the controller is implemented as a floating-point program and secondly,
when the controller is implemented as a fixed-point program synthesized
using our technique.
The robot starts a little away
from the reference line but quickly starts tracking the line in both
cases. Figure~\ref{fig:fx:WMRerrorv} and Figure~\ref{fig:fx:WMRerroromega}
show the error between the floating-point
controller and fixed-point controller for both control inputs: $v$ and $\omega$,
respectively.
}

\ifthenelse {\boolean{IsFinal}}
{
\noindent \textbf{Performance: } 
Table~\ref{tbl:icse12:perf} summarizes the performance of our technique
in the four case-studies.

\begin{table}[!t]
\small
\caption{Performance}
\label{tbl:icse12:perf}
\centering
\begin{tabular}{|c|c|c|c|c|c|c|}
\hline
Perf & IIR Filter & FIR Filter & DC Motor & WMR $v$ & WMR $\omega$ \\
\hline
Runtime (s) & 268 & 379 & 4436 & 2218 & 1720 \\
\hline
\# Iterations & 5 & 4 & 8 & 7 & 4 \\
\hline
\end{tabular}
\normalsize
\end{table}
}
{
Table~\ref{tbl:icse12:perf} summarizes the performance of our technique
in the four case-studies.

\begin{table}[!t]
\small
\caption{Performance}
\label{tbl:icse12:perf}
\centering
\begin{tabular}{|c|c|c|}
\hline
Case-study & Runtime (seconds) & \# of Iterations \\
\hline
IIR Filter & 268 & 5 \\
\hline
FIR Filter & 379 & 4 \\
\hline
DC Motor $u$ & 4436 & 8 \\
\hline
WMR $v$ & 2218 & 7\\
$\omega$ & 1720 & 4\\
\hline
\end{tabular}
\normalsize
\end{table}
}

\section{Related Work}
\label{pldi12:sec:related}

\noop{
In this section, we discuss relevant related work and compare
our approach with existing techniques.
During the floating-point to fixed-point conversion process,
fixed-point wordlengths composed of the integer wordlength
part and the fractional wordlength part
need to be determined.
While the integer-part word-length are easily determined by
avoiding overflow and estimating the range of the variables,
the determination of the
fractional word-lengths is more challenging and is an area
of active research.
}
Previous techniques for optimizing fixed-point types
are based on statistical sampling of the input space.
These methods sample a large number
of inputs and heuristically solve an optimization problem
that minimizes implementation cost while ensuring that
some correctness specification is met over the sampled inputs.
The techniques differ in
in the heuristic search method employed,
in the measure of cost, or in how accuracy
of fixed-point implementation is determined.
Sung and Kum~\cite{sung-tsp95} use a heuristic
search technique which starts with the minimum
wordlength implementation as the initial guess.
The wordlengths are increased one by one
till the error falls below an acceptable threshold.
Han et al.~\cite{han-iscas01,han-icassp04}
use a gradient-based sequential search method which starts
with the minimum wordlength implementation as the initial guess.
The gradient (ratio of increase in accuracy and increase in wordlengths)
is computed for a set of wordlength changes at each step and the
search moves in the direction with maximum gradient.
Shi et al.~\cite{shi-ASSP03} propose a floating-point to fixed-point
conversion methodology for digital VLSI signal
processing systems.
Their approach is based on a perturbation theory
which shows that the change to the first order is a linear
combination of all the first- and second-order statistics of
the quantization noise sources.
Their technique works with general specification
critera, as long as these can be represented as large
ensemble averages of functions of the signal outputs.
For example,
they use mean-squared error (MSE) as the specification
function.
The cost of the implementation is a quadratic function.
Monte Carlo simulation of a large number
of input examples is used to formulate
a quadratic optimization problem based on perturbation theory.
In contrast,
our specification requires that the accuracy condition
holds for all inputs and not just on an average.
Further, the cost function can be any arbitrary function for
our technique and need not be quadratic.
Perhaps most importantly, our technique does not rely on apriori
random sampling of a large number of input values, instead using
optimization to discover a small set of
{\em interesting} examples which suffice to discover
optimal fixed-point implementation.

Purely analytical methods~\cite{Stephenson-sigplan00,kim-adsp98}
based on dataflow analysis
have also been proposed
for synthesizing fixed-point programs based on
forward and backward propagation in the program's dataflow graph.
The advantages of these techniques are that they
do not rely on picking the right inputs for simulation,
can handle arbitrary programs (with approximation),
and can provide correctness guarantees. However,
they tend to produce very conservative wordlength results.

Inductive synthesis based on satisfiability solving has
been previously used for synthesizing programs from
functional specifications.
These
approaches~\cite{jha-icse10,gulwani-pldi11}
rely on constraint solving in much the same way as we rely on
optimization routines. However, these approaches only seek to
find a correct program, without any notion of
cost and optimization.
In contrast, our technique is used to find a fixed-point program
which is not only correct with respect to a condition on accuracy
but is also of minimal cost.

\section{Conclusion}
\label{pldi12:sec:conc}
\label{icse12:sec:conc}

In this paper, we presented a 
novel approach to automated synthesis of fixed-point program from floating-point
program by discovering the fixed-point types of the variables.
The program is synthesized to satisfy the provided correctness condition
for accuracy and to have optimal cost with respect to the provided
cost model. 




\bibliographystyle{plain}
\bibliography{thesis}
\end{document}